\documentclass[11pt,a4paper]{article}
\usepackage{graphicx}
\usepackage{xcolor}
\usepackage{subcaption}
\usepackage{geometry}
\usepackage{amssymb}
\usepackage{amsmath}
\usepackage{amsthm}
\usepackage{textcomp}
\usepackage{xcolor}
\usepackage{adjustbox}  
\usepackage[english]{babel}
\usepackage[utf8]{inputenc}
\usepackage{physics}
\usepackage{siunitx}
\usepackage{csquotes}
\usepackage{bbold}
\usepackage{enumitem}
\usepackage{cite}
\usepackage{mathtools}
\usepackage{hyperref}
\hypersetup{
    colorlinks,
    citecolor=blue,
    filecolor=red!75!black,
    linkcolor=red!75!black,
    urlcolor=blue
}

\numberwithin{equation}{section}

\usepackage{tikz-feynman}
\tikzfeynmanset{compat=1.1.0}

\numberwithin{equation}{section}

\usepackage{appendix}
\usepackage{authblk}

\newcommand{\mpl}{M_P}

\theoremstyle{definition}

\newcommand{\la}{\lambda}
\newcommand{\lat}{\Tilde{\lambda}}
\newcommand{\rhot}{\Tilde{\rho}}

\newcommand{\M}{\mathcal{M}}
\newcommand{\A}{\mathcal{A}}

\newcommand{\angMM}[2]{\langle  \mathbf{#1} \mathbf{#2} \rangle}

\newcommand{\angmm}[2]{\langle  #1 #2 \rangle}
\newcommand{\sqrMM}[2]{[ \mathbf{#1} \mathbf{#2} ]}

\newcommand{\sqrmm}[2]{[ #1 #2 ]}

\newcommand{\angsqrMM}[3]{\langle \mathbf{#1}| #2 | \mathbf{#3} ]}

\newcommand{\angsqrmm}[3]{\langle #1| #2 | #3 ]}

\DeclareMathAlphabet{\mathmybb}{U}{bbold}{m}{n}

\DeclareMathOperator*{\SumInt}{%
\mathchoice%
  {\ooalign{$\displaystyle\sum$\cr\hidewidth$\displaystyle\int$\hidewidth\cr}}
  {\ooalign{\raisebox{.14\height}{\scalebox{.7}{$\textstyle\sum$}}\cr\hidewidth$\textstyle\int$\hidewidth\cr}}
  {\ooalign{\raisebox{.2\height}{\scalebox{.6}{$\scriptstyle\sum$}}\cr$\scriptstyle\int$\cr}}
  {\ooalign{\raisebox{.2\height}{\scalebox{.6}{$\scriptstyle\sum$}}\cr$\scriptstyle\int$\cr}}
}

\usetikzlibrary{decorations.markings,angles,quotes,positioning,decorations.pathreplacing,decorations.markings,snakes,arrows,calc}
\usetikzlibrary{fadings}
\usetikzlibrary{shapes.misc}
\tikzset{crossr/.style={cross out, draw=red, minimum size=4*(#1-\pgflinewidth), inner sep=0pt, outer sep=0pt},
crossr/.default={2pt}}
\tikzset{crossb/.style={cross out, draw=black, minimum size=4*(#1-\pgflinewidth), inner sep=0pt, outer sep=0pt},
crossb/.default={2pt}}
\tikzset{crossp/.style={cross out, draw=violet, minimum size=4*(#1-\pgflinewidth), inner sep=0pt, outer sep=0pt},
crossp/.default={2pt}}

\tikzfading 
[
  name=fade out,
  inner color=transparent!0,
  outer color=transparent!100
]

\tikzset{
upp/.style={postaction={decorate},
  decoration={markings,mark=at position .5 with  \arrow{>}}},
  dow/.style={postaction={decorate},
  decoration={markings,mark=at position .5 with  \arrow{<}}},
 }

\tikzset{
    ncbar angle/.initial=90,
    ncbar/.style={
        to path=(\tikztostart)
        -- ($(\tikztostart)!#1!\pgfkeysvalueof{/tikz/ncbar angle}:(\tikztotarget)$)
        -- ($(\tikztotarget)!($(\tikztostart)!#1!\pgfkeysvalueof{/tikz/ncbar angle}:(\tikztotarget)$)!\pgfkeysvalueof{/tikz/ncbar angle}:(\tikztostart)$)
        -- (\tikztotarget)
    },
    ncbar/.default=0.5cm,
}

\tikzset{round left paren/.style={ncbar=0.5cm,out=100,in=-100}}
\tikzset{round right paren/.style={ncbar=0.5cm,out=80,in=-80}}

\newcommand{\be}{\begin{equation}}
\newcommand{\ee}{\end{equation}}          
\newcommand{\bea}{\begin{eqnarray}}
\newcommand{\eea}{\end{eqnarray}}
\newcommand{\bc}{\begin{center}}
\newcommand{\ec}{\end{center}}

\begin{document} 


\begin{center} 

\vspace{2cm}
{\Large \bf (Super)$\,$Gravity from Positivity}

\vspace{1.4cm}

Brando Bellazzini$^{\sharp\,, \star}$, Alex Pomarol$^{\flat\,, \ast}$, Marcello Romano$^{\sharp}$, Francesco Sciotti$^{\flat}$
\vspace{0.5cm}

{\footnotesize{\textit{\noindent ${}^{\sharp}$ Université Paris-Saclay, CNRS, CEA, Institut de Physique Théorique, 91191, Gif-sur-Yvette, France \\ 
${}^{\star}$Kavli Institute for Theoretical Physics, Santa Barbara, California 93106, USA\\
 ${}^{\flat}$ IFAE and BIST, Universitat Aut\`onoma de Barcelona, 08193 Bellaterra, Barcelona \\
 ${}^{\ast}$ Departament de F\'isica, Universitat Aut\`onoma de Barcelona, 08193 Bellaterra, Barcelona
 }}}
 

\vskip 2cm
\begin{abstract}\noindent\normalsize

We investigate whether the effective theory for isolated, massive, and weakly interacting spin--$3/2$ particles is compatible with causality and unitarity---\textit{i.e.}, the positivity of scattering amplitudes. We find no solution to positivity constraints, except when gravitons are also present and couple in a (nearly) supersymmetric way. Gravity is thus bootstrapped from $S$--matrix consistency conditions for the longitudinal and transverse polarizations of massive spin--$3/2$ states. For two such particles forming a $U(1)$-charged state, a (gravi)photon gauging the symmetry is also required, with couplings characteristic of supergravity and consistent with both the no–global–symmetry and weak gravity conjectures. We further explore the EFT--hedron associated with the longitudinal polarizations, the Goldstinos, through novel $t$--$u$ symmetric dispersion relations. We identify the ‘extremal’ UV models that lie at the corners of the allowed parameter space, recovering familiar models of supersymmetry breaking and uncovering new ones.

\end{abstract}

\end{center}

\newpage

\renewcommand{\baselinestretch}{0.9}\normalsize
{ \hypersetup{hidelinks} 
 \tableofcontents }
\renewcommand{\baselinestretch}{1}\normalsize

\newpage

\section{Introduction}

No massive higher-spin particle ($J\geq 2$) has ever been found in isolation at the bottom of the mass spectrum, either in theoretical constructions or in experimental searches. Whenever higher-spin states appear, they are accompanied by a tower of additional states with parametrically similar masses, much like what occurs in bound-state spectra, regardless of how many theoretical knobs one adjusts. 
In other words, a single massive higher-spin particle inevitably comes with a host of companion states, suggesting that a rich spectrum is required at that mass scale.

Equivalently, one can say that an isolated massive higher-spin particle has a cutoff scale near its own mass, rendering any effective field theory (EFT) description invalid. In recent years, it has become increasingly clear that this feature is rooted in the stringent consistency conditions imposed by causality and unitarity—cornerstones of any physically viable theory—in the form of positivity bounds; see \textit{e.g.}~\cite{Adams:2006sv,Camanho:2014apa, Bellazzini:2014waa, Bellazzini:2015cra,  Bellazzini:2016xrt, deRham:2017avq, Bellazzini:2019bzh, Englert:2019zmt,  Bellazzini:2020cot,Arkani-Hamed:2020blm, Caron-Huot:2020cmc,  Tolley:2020gtv, Bern:2021ppb, Chiang:2021ziz, Caron-Huot:2021rmr, Davighi:2021osh, Serra:2022pzl,  Bellazzini:2022wzv, 
Caron-Huot:2022ugt, 
Guerrieri:2022sod, Henriksson:2022oeu, 
Haring:2022sdp, Bellazzini:2023nqj, Karateev:2023mrb, Haring:2023zwu, McCullough:2023szd,  Ma:2023vgc,EliasMiro:2023fqi, Berman:2024eid, Creminelli:2024lhd, Dong:2024omo,  Bertucci:2024qzt, Cremonini:2024lxn,  Cheung:2025krg, deRham:2025vaq, Peng:2025klv, Chang:2025cxc, Hui:2025aja, Correia:2025uvc}.\footnote{For a nearly comprehensive list, see the reference list in~\cite{Chang:2025cxc}.}

 In particular, the lesson from the massive $J=2$ case \cite{Bellazzini:2023nqj} is that massive higher-spin particles carry several polarizations  which inevitably cause scattering amplitudes to grow too rapidly with energy, scaling like $E^n$ with $n>4$ within the EFT.  This violates positivity bounds, which, among other constraints, require that the dominant term in an amplitude cannot grow with energy faster than $E^4$.

The central question we explore in this work is how the boundary case of massive spin-$3/2$ particles fits into this picture. On the one hand, explicit theories of massive Gravitinos can be constructed that are perfectly consistent and allow for a large separation of scales from other states in the theory---provided that a graviton is included in the low-energy spectrum. On the other hand, simple power-counting arguments suggest that amplitudes for lonely spin-$3/2$ particles grow as $E^6$ and thus violate positivity bounds, even at weak coupling, while respecting perturbative unitarity. \\

One of the main goals of this paper is to find an answer to the question: is gravity required by causality for consistency of the theory? And might even \textit{supergravity}---and hence supersymmetry (SUSY)---be similarly dictated by fundamental consistency conditions? As we will show, the answers to these questions are affirmative. In other words, any perturbative EFT (with finite number of degrees of freedom) of \textit{massive} spin-$3/2$ particles, valid at energies larger than $\sim 10m$, must necessarily be a (super)gravity theory.

The need for  gravity arises from the positivity requirement  of  canceling  the dominant $E^6$ term in the spin-3/2 amplitudes. Although this cancellation can also be achieved by scalars or vectors, in these cases the coefficient of the $E^4$ remaining term  becomes negative, in contradiction with positivity bounds.
Once gravity emerges from causality, several other hallmark features come along for the ride. In particular, the \textit{no global symmetries conjecture} and the \textit{Weak Gravity Conjecture} (WGC) appear as natural consequences of the same consistency conditions. Incidentally, we find that positivity bounds unambiguously fix the gyromagnetic factor to $g=2$.

In the limit in which the transverse modes $\lambda = \pm 3/2$ of the massive spin-3/2 particle, controlled by the Planck mass, 
decouple (\textit{i.e.}, $\mpl \to \infty$), 
the amplitudes are dominated by the longitudinal states of helicity 
$\lambda = \pm 1/2$, the Goldstinos.
These amplitudes are suppressed by $F^2= 3m^2 \mpl^2$ ($m$ is the mass of the spin-3/2 particle) that can remain finite in the limit $m\to 0$ and $\mpl\to\infty$ and define the EFT  of Goldstinos.
In this truncated sector, consistency conditions impose highly non-trivial constraints in the form of a novel class of $t$--$u$ symmetric dispersion relations, which we develop in detail.
We determine the allowed regions of the  parameter space when imposing  positivity, and at its boundaries we identify 
 well-known models of supersymmetry-breaking. We also find the regions corresponding to 
models with infinite higher-spin states  and study their properties.

%

The paper is structured as follows: in Section~\ref{sec:general_strategy} we state our assumptions and describe the strategy adopted to impose positivity bounds and build EFTs. In Section~\ref{sec:MajoranaGeneral} we study the consistency of a  Majorana spin-$3/2$ EFT, showing the necessary role of gravity, while in Section~\ref{sec:DiracGeneral} we extend the discussion to a Dirac spin-$3/2$ EFT with $U(1)$ symmetry. Section~\ref{sec:GoldstinoEFThedron} explores the properties of the Goldstino's EFT--hedron. Finally, we summarize our results and envisage future directions in Section~\ref{sec:conclusions}.\\
Conventions, togheter with a spinor-helicity reminder, are given in Appendix \ref{app:conventions}. In Appendix \ref{app:positivity_bounds} we review how positivity bounds used in the main text are derived from fundamental principles and discuss the effect of a finite mass, introducing further novel (technical)  results. In Appendix \ref{app:3_2eft} we summarise the on-shell three-points amplitudes for spin-$3/2$ particles. Finally, in Appendix \ref{app:diracscalarvector} we provide some additional details on scalar and vector extensions of the Dirac spin-$3/2$ EFT.

\section{General strategy}\label{sec:general_strategy}

We consider the effective field theory (EFT)  of spin $J ={3}/{2}$ particles of mass $m$ in four-dimensional Minkowski spacetime.
We will assume   
\begin{itemize}
    \item \textit{Large Scale Separation}: The EFT remains valid up to a cutoff scale $\Lambda \gg m$.
    \item \textit{Weak Coupling}: The EFT has a weakly-coupled description below $\Lambda$.\footnote{The weak coupling assumption does not restrict the generality of our analysis, as theories that are strongly coupled at the particle’s mass do not allow for a separation of scales in the first place. An example is $N_f = 1$ QCD, where the lightest baryon is a color-neutral spin-$3/2$ particle. It is already strongly coupled at $E \simeq m$, with the next excited state appearing at a comparable energy scale. } Loops of the EFT can thus be neglected. 
\end{itemize}
We then examine whether EFTs under these assumptions are compatible with unitarity and causality. We will find that EFTs of massive spin-3/2 states must be  implemented by a specific set of  light degrees of freedom with particular---gravitational---couplings.

We consider two main scenarios: a Majorana and a Dirac spin-$3/2$ particle. In the Dirac case, each helicity state has a degenerate partner which one can assign  opposite $U(1)$ charges. We will see that consistency with causality demands this $U(1)$ symmetry to be gauged.\\ 

In this section, we explain how positivity bounds can be used to constrain EFTs of massive higher-spin particles in a remarkably simple and powerful way, and outline the procedure to construct scattering amplitudes for these states.

\subsection{Positivity bounds}\label{subsec:positivitydecoupling}

The $2 \to 2$ scattering amplitude  for massive particles,
${\cal M}_{\lambda_1 \lambda_2}^{\lambda_3 \lambda_4}(s,t)$, where $\lambda_i$ corresponds to the helicity of the state $i$,
satisfies positivity properties that follow from causality and unitarity. These properties are summarized in detail in Appendix~\ref{app:positivity_bounds}, where we provide the derivations of the results stated in this section.

\begin{figure}[h!!]
    \centering
    \includegraphics[width=0.5\linewidth]{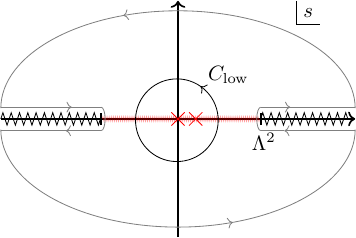}
    \caption{Integration contour $C_{\text{low}}$ in the complex-$s$ plane, enclosing subtraction, dynamical, kinematical poles (red crosses) and possibly kinematical branch cuts. We are neglecting IR branch-cuts (in red), based on the weak coupling assumption, so that the integration can be deformed to a UV contour (in grey).}
    \label{fig:dispersioncontours}
\end{figure}

In the complex-$s$ plane at fixed $-\Lambda^2 < t \leq 0$, 
we define the \emph{Arcs}~\cite{Bellazzini:2023nqj} as the contour integrals
\begin{equation} \label{eq:arcs_def}
    \A_{\lambda_1 \lambda_2}^{\lambda_3 \lambda_4}(t,n) = \frac{1}{2} \oint_{C_{\text{low}}} \frac{ds}{2\pi i} \frac{ \M_{\lambda_1 \lambda_2}^{\lambda_3 \lambda_4}(s,t) + \M_{\lambda_1 \bar{\lambda}_4}^{\lambda_3 \bar{\lambda}_2}(s,t) }{ \left(s - 2m^2 + \frac{t}{2} \right)^{3+n} }, \qquad n \geq 0\,,
\end{equation}
where $\M_{\lambda_1 \bar{\lambda}_4}^{\lambda_3 \bar{\lambda}_2}$ refers to the scattering amplitude with anti-particles $\bar{2}$ and $\bar{4}$ of helicities $\la_{\bar{i}}=\bar{\la}_{i}=-\la_i$.   
 The integration contour, illustrated in Fig.~\ref{fig:dispersioncontours}, is chosen with a radius large enough to enclose all kinematic singularities. 

On the one hand, Arcs can be computed within the EFT in terms of Wilson coefficients,   depending then on  infrared (IR) physics. 
On the other hand, 
by deforming the contour, the Arcs can be rewritten as integrals over physical discontinuities that run to arbitrary high energy scales, and then  becoming  sensitive to ultraviolet (UV) physics.
Under the weak-coupling assumption, we can choose the Arc radius such that the IR discontinuities (in red in Fig.~\ref{fig:dispersioncontours}) can be neglected as compared to the UV ones,\footnote{The impact of the IR discontinuities can be systematically accounted for, see \textit{e.g.}~\cite{Bellazzini:2020cot, Bellazzini:2021oaj, Chang:2025cxc, Beadle:2024hqg, Beadle:2025cdx, Chala:2021wpj}. 
At finite coupling, their contribution remains under theoretical control for a finite number of Arcs — with the lowest Arcs, as considered in this work, being the least sensitive to loop effects.} which start at the branch points $s = \Lambda^2$ and $s = -\Lambda^2 - t + 4m^2$.

 Extending the results of~\cite{Bellazzini:2023nqj}, we show in Appendix~\ref{app:positivity_bounds} that ratios of Arcs can be bounded exactly in terms of known, EFT-computable functions---see Eq.~\eqref{eq:fullcrossing_bound}. Even though these expressions are rather intricate, they simplify considerably in the regime where either $-t \ll \Lambda^2$, $m^2 \ll \Lambda^2$, or both, leading to the following inequality among Arcs\footnote{There are small corrections to \eqref{eq:positivity_bounds0} of order $\mathcal{O}\left( \frac{\sqrt{-t} \, m}{\Lambda^2} \right)$ times ratios of forward Arcs. These corrections originate from the non-trivial crossing symmetry of massive spinning particles and are discussed in detail in Appendix~\ref{app:positivity_bounds}. There, we show that they are always under control, being bounded by EFT-computable quantities.}
\begin{equation} \label{eq:positivity_bounds0}
\begin{split}
    0 &\leq \frac{\left| \A_{\lambda_1 \lambda_2}^{\lambda_3 \lambda_4}(t,n) \right| + \left| \A_{\lambda_1 \bar{\lambda}_2}^{\lambda_3 \bar{\lambda}_4}(t,n) \right|  }{ 
    \A_{\lambda_1 \lambda_2}^{\lambda_1 \lambda_2}(0,0) + 
    \A_{\lambda_3 \lambda_4}^{\lambda_3 \lambda_4}(0,0) + 
    \A_{\lambda_1 \lambda_4}^{\lambda_1 \lambda_4}(0,0) + 
    \A_{\lambda_3 \lambda_2}^{\lambda_3 \lambda_2}(0,0) 
    }
    \leq  \frac{1}{2} \frac{(\Lambda^2 - 2m^2)^3}{\left(\Lambda^2 - 2m^2 + \frac{t}{2} \right)^{n+3}} \,.
\end{split}
\end{equation}
This inequality constrains Arcs that may be non-forward, inelastic, or both, in terms of purely elastic and forward Arcs with minimal subtraction $n=0$, i.e., $\A_{\lambda_i \lambda_j}^{\lambda_i \lambda_j}(0,0)$, which are themselves positive
\begin{equation}
\label{eq:simplestIneq}
\A_{\lambda_i \lambda_j}^{\lambda_i \lambda_j}(0,0) \geq 0\,. 
\end{equation}
The Eq.~\eqref{eq:positivity_bounds0} is exact whenever $t=0$, $m=0$ or both. In particular, an even simpler expression follows in the  $m^2 \ll -t \ll\Lambda^2$ limit allowed by the large scale separation:
\begin{equation}\label{eq:positivity_bounds_dec}
\begin{split}
    0 &\leq \frac{\abs{\A_{\la_1\la_2}^{\la_3\la_4}(t,n)} + \abs{\A_{\la_1\bar\la_2}^{\la_3\bar\la_4}(t,n)}}{\Lambda^{-2n} \left[ \A_{\la_1\la_2}^{\la_1\la_2}(0,0) + \A_{\la_3\la_4}^{\la_3\la_4}(0,0) + \A_{\la_1\la_4}^{\la_1\la_4}(0,0) + \A_{\la_3\la_2}^{\la_3\la_2}(0,0) \right]}
      \leq \frac{1}{2}  \,.
\end{split}
\end{equation}
We can draw from Eqs.~\eqref{eq:positivity_bounds0},~\eqref{eq:positivity_bounds_dec} a very simple conclusion:
\begin{center}
    \textit{Every Arc is bounded by a linear combination of elastic Arcs with minimal subtractions.}
\end{center}

\textbf{Toy model example:} the practical implications of this statement can be easily explained by means of a toy-model example. Consider a shift-symmetric scalar, with $2\to2$ amplitude
\begin{equation}
    \M(s,t) = g_{2,0}(s^2 + t^2 + u^2) + g_{3,1}stu + g_{4,0}(s^2 + t^2 + u^2)^2 + \dots \,.
\end{equation}
In an EFT, one expects higher terms in the $s,t,u$ expansion to be smaller than previous ones, \textit{e.g.} $g_{2,0} \gg  g_{4,0}(s^2 + t^2 + u^2)$. Nevertheless, these expectations can in principle be violated by symmetry arguments or accidental cancellations in the Wilson coefficients. Therefore, it is pertinent to ask the following question:
can the amplitude at some center-of-mass (CoM) energy $E$ be dominated by the $g_{4,0}$ term? Or, in other words, can it scale as $E^8$ within the EFT? This would mean $g_{2,0} \ll \abs{g_{4,0}} E^4$ (along with $\abs{g_{4,0}} E^4 \ll 16\pi^2$ such that the EFT is still within perturbation theory). The answer is given by Eq.~\eqref{eq:positivity_bounds_dec}, which in our case implies that this is not possible:
\begin{equation}
    0 \leq \abs{\A(t,2)} = 4 \abs{g_{4,0}} \leq \A(0,0) = \frac{2g_{2,0}}{\Lambda^4} \,.
\end{equation}
The same is true for $g_{3,1}$, as we also have
\begin{equation}
    0 \leq \abs{\A(t,0)} = \abs{2g_{2,0} - t g_{3,1} + 6t^2 g_{4,0}} \leq \A(0,0) = 2g_{2,0} \,,
\end{equation}
which, if we try to impose $g_{2,0} \ll t g_{3,1}, t^2 g_{4,0}$, would just tell us $\abs{-t g_{3,1} + 6t^2 g_{4,0}} \leq 0$, \textit{i.e.}, $g_{3,1} = g_{4,0} = 0$. Thus, the $g_{2,0}$ contribution to the amplitude can never be suppressed compared to higher-order Wilson contributions.

\noindent \textbf{Take-home message:}
\begin{center}
    \textit{EFT's amplitudes consistent with positivity cannot be dominated by $E^n$ terms with $n>4$.}
\end{center}
Moreover, elastic EFT's amplitudes must have positive $E^4$-term coefficients.

It turns out that inequalities~\eqref{eq:positivity_bounds0} and \eqref{eq:simplestIneq}  represent just the simplest instances in an infinite tower of positivity bounds, each targeting more precisely individual terms in a Mandelstam expansion---none of which can dominate lower-order ones, beginning with the $E^4$ term.  For example, in Section~\ref{sec:GoldstinoEFThedron}, we derive the full (infinite) set of positivity constraints relevant for helicity $\pm1/2$ amplitudes, where our goal is to delineate sharp boundaries for the EFT-hedron of Goldstinos---crucially, a class of theories whose leading energy scaling saturates the $E^4$ bound. In that case, extracting meaningful constraints requires imposing additional conditions, which we do.

\noindent
However, for the purpose of excluding theories with a gross violation of the $E^4$ scaling rule, the minimal bounds in Eq.~\eqref{eq:positivity_bounds0} are already sufficient.\\

\noindent These bounds turn out to be extremely powerful for theories with isolated massive states with $J>1$ spin, as generic amplitudes---\textit{without any tuning of the parameters}---typically grow faster than $E^4$ at high energies. For instance, the $\lambda = 0$ (longitudinal) polarization of an integer-spin particle behaves as $(E/m)^J$ for $E \gg m$, leading to a naïve scaling of the amplitude $\M^{00}_{00}(E \gg m) \sim (E/m)^{4J}$. Of course, cancellations may occur and the energy scaling can be reduced, but crucially, not enough to meet the $E^4$ threshold for $J> 1$.~\footnote{At the Lagrangian level, the difference between $J=1$ and $J\geq3/2$ can be tracked back to the healthy (finite) kinetic term for the longitudinal modes, which are thus ordinary NGBs, in the former case, in contrast with the higher-spin case, where the kinetic term is proportional to the mass and arises from mixing with the transverse modes. In particular, for $J=3/2$, separating the transverse and longitudinal modes in the mass term, $\psi_{\mu}+\partial_{\mu}\chi$, generates a mixing $\sim m \bar\psi_{\mu}\gamma^{\mu\nu}\partial_{\nu}\chi+h.c.$, which can be diagonalized providing the mass--suppressed kinetic term for $\chi$.}

As we show below, achieving that requires the introduction of new light degrees of freedom,  on top of specific tunings.

The hypothesis of large scale separation allows us to expand the amplitude in powers of the CoM energy $E$, in the regime $m^2\ll E^2\ll\Lambda^2$, and obtain the following schematic scaling for the largest term of the amplitude
\begin{equation}\label{eq:generichescaling}
    \frac{E^n}{M^n}\left(1 + \mathcal{O}\left(\frac{m}{E}\right) \right) \,,
\end{equation}
where  $n$ is a positive integer and $M$ sets the strong coupling scale. Generically, $\Lambda$---the scale where the EFT breaks down---is bounded by $M$, since new physics may be weakly coupled. 
If one finds $n>4$, with the $E^4$ contribution suppressed by powers of $m/E$,  positivity will tell us, as in the toy model example, that the coefficient in front of the term \eqref{eq:generichescaling} must be effectively set to $0$.

In practice, one can define a \textit{decoupling limit} in which we formally send $m \to 0$ while holding $M$ fixed.  
This effectively yields massless scattering amplitudes, which remarkably simplifies the application of positivity bounds while keeping their physical implications fully transparent. In such a limit, Eq.~\eqref{eq:positivity_bounds_dec} applies.

\noindent Our strategy is essentially iterative and by contradiction:
\begin{enumerate}
    \item Determine which is  the dominant term  $(E/M)^n$ of an amplitude of a given EFT --see Section~\ref{subsec:EFTLogic}.  
    \item Impose the  positivity bound  Eq.~\eqref{eq:positivity_bounds_dec}.  If $n > 4$, this bound  will typically require tuning among parameters of the EFT in order to reduce the  energy scaling of the dominant term:  $(E/M)^n\to (E/M)^{n-1}$. 
    \item Repeat the process until
either:  (i) all  bounds are satisfied and the EFT is thus admissible. In particular,  the dominant term of the amplitudes scales at most as $E^4$, or (ii) only the trivial solution is found. 
    \item If only the trivial solution is found, the EFT must be modified by adding new light degrees of freedom, and the process is repeated until a non-trivial theory is achieved or eventually ruled out.  In order to work with a finite number of degrees of freedom, we focus on EFTs which include spin-0, spin-1,  and massless spin-2 particles, on top of the massive spin-3/2 states.  
\end{enumerate}

One should keep in mind that these are necessary conditions, as further constraints come from the full set of positivity bounds, as we show in detail for the Goldstino theory in Section~\ref{sec:GoldstinoEFThedron}. 

\noindent Clearly, the bounds obtained in the decoupling limit are always to be understood up to corrections of order $\mathcal{O}\left(m/\Lambda\right)$. In Appendix~\ref{app:positivity_bounds}, we show explicitly, through an example, how finite-mass effects affect the bounds. In fact, one can use them to place upper bounds on $\Lambda/m$.

\medskip

\noindent A comment is due on the forward limit $t \to 0$. In this work we include gravitons in the EFT, which notoriously make this limit singular due to the pole at $t = 0$. Crucially, in the decoupling limit this term is instead subleading, so that one can formally take $t \to 0$. In other words, the two limits do not commute, see also \cite{Chang:2025cxc}.  Let us remind, however, that the decoupling limit is just a computational expedient: we are never really taking the mass to zero, but rather zoom in regions of energy where $m\ll E \ll \Lambda$. 
%

\subsection{Constructing EFT Amplitudes}
\label{subsec:EFTLogic}

We construct tree-level $2\to2$ scattering amplitudes for spin-$3/2$ particles  using on-shell methods. A summary of our conventions can be found in Appendix~\ref{app:conventions}. Given the spectrum, the amplitude is built according to the following steps:
\begin{enumerate}
    \item Write down all on-shell three-point amplitudes allowed by symmetries.
    \item Fix the factorizable contribution to the four-point by gluing three-point amplitudes.
    \item Include all possible on-shell contact four-point contributions allowed by symmetries, with at most the same energy growth as the factorizable amplitude.
\end{enumerate}

\par Moreover, we restrict our attention to amplitudes that are $CP$-invariant and, when applicable, separately charge-conjugation and parity invariant. Extending our analysis to $CP$-violating interactions is straightforward,  and we don't expect them to affect our conclusions.

\par The classification of on-shell three-point amplitudes is well known~\cite{Arkani-Hamed:2017jhn}, as are the methods to glue them into higher-point functions. In this work, we do not use BCFW-like recursion relations~\cite{Britto:2005fq}, as we wish to avoid any additional assumptions about the UV behavior of the amplitude or invoke supersymmetry (see~\cite{Gherghetta:2024tob,Ema:2025qgd} for an analysis of spin-$3/2$ amplitudes from that point of view).

\par Instead, the factorizable part of the amplitude is obtained by constructing an ansatz consistent with factorization. To this, we add a general contact term contribution. The on-shell classification of contact terms for both massless and massive particles has been studied in~\cite{Durieux:2020gip,Dong:2021vxo,Dong:2022mcv,DeAngelis:2022qco}. In this work, we employ the algorithm developed in~\cite{DeAngelis:2022qco}, implemented in the public \texttt{Mathematica} package \href{https://github.com/StefanoDeAngelis/MassiveEFT-Operators}{\texttt{MassiveGraphs}}.

\section{Majorana spin-3/2 particle}
\label{sec:MajoranaGeneral}

In this section, we study the EFT of a massive Majorana spin-$3/2$ particle. 

We begin by analyzing the spin-3/2 state in isolation, and then we show how consistency forces the inclusion of additional light degrees of freedom. We start  focusing on a simplified version of the theory, concentrating solely on the longitudinal sector ($\lambda = \pm 1/2$) with minimal interactions. The conclusions  for a more general theory will be analogous and are discussed at the end of the section.

\subsection{Isolated spin-3/2 particle (longitudinal sector)}
\label{sec:isolated_majorana}

In this simplified case, the only interactions contributing to the scattering amplitude of four spin-$3/2$ particles are four-point contact terms. At lowest order in the EFT expansion, there are three independent $CP$-invariant contributions, which in spinor-helicity variables (see Appendix \ref{app:conventions} for a reminder) read
\begin{equation}
\label{eq:contactMajorana}
    \begin{split}
        \M_4^{contact}&=\frac{h_1}{M^6}\angMM{1}{4}\angMM{2}{3}\sqrMM{1}{4}\sqrMM{2}{3}(\angMM{1}{4}\angMM{2}{3}+\sqrMM{1}{4}\sqrMM{2}{3})\\
        &+\frac{h_2}{M^6}\angMM{1}{4}\angMM{2}{3}\sqrMM{1}{4}\sqrMM{2}{3}(\angMM{1}{4}\sqrMM{2}{3}+\sqrMM{1}{4}\angMM{2}{3})\\
        &+\frac{h_3}{2M^6}(\angMM{1}{4}\angMM{2}{3}\angMM{3}{4}\sqrMM{1}{2}\sqrMM{1}{4}\sqrMM{2}{3}+\sqrMM{1}{4}\sqrMM{2}{3}\sqrMM{3}{4}\angMM{1}{2}\angMM{1}{4}\angMM{2}{3})\\
        &+\text{antisymm.} \,.
    \end{split}
\end{equation}
The $M$ represents the mass scale of the interactions, while $m$ is the spin-$3/2$ mass. 
We assume $M\gtrsim \Lambda$ such that the interactions are always weak for $E\lesssim \Lambda$.
The amplitude \eqref{eq:contactMajorana} corresponds to (linear combinations of) the following four-fermion operators in the spin-$3/2$ Lagrangian
\begin{equation}
    (\bar\psi_{\mu}\gamma_{\rho}\psi^{\mu})(\bar\psi_{\nu}\gamma^{\rho}\psi^{\nu})\;\;,\;\;\;(\bar\psi_{\mu}\psi^{\mu})^2\;\;,\;\;\;(\bar\psi_{\mu}\gamma_5\psi^{\mu})^2\,,
\end{equation}
    where $\psi_{\mu}$ is a (Majorana) Rarita-Schwinger field and $\gamma^{\mu}$ are the standard gamma matrices. We can compute the amplitude in the center-of-mass frame and expand it in the  limit $m\to0$ with $M$ fixed, with the CoM energy $m\ll E\ll M$. For the $\lambda_i=\pm 1/2$ components we obtain
\begin{equation}\label{eq:amplideclong}
\begin{split}
\M_{-\frac{1}{2}-\frac{1}{2}}^{-\frac{1}{2}-\frac{1}{2}}(s,t)&\to-\frac{2h_2s^3+h_3s[s^2+2t(s+t)]}{18M^6}+\mathcal{O}\left(\frac{m^2E^4}{M^6}\right)\,,\\
\M_{-\frac{1}{2}-\frac{1}{2}}^{+\frac{1}{2}+\frac{1}{2}}(s,t)&\to-\frac{h_1st(s+t)}{3M^6}+\mathcal{O}\left(\frac{m^2E^4}{M^6}\right)\,,\\
\M_{-\frac{1}{2}+\frac{1}{2}}^{+\frac{1}{2}-\frac{1}{2}}(s,t)&\to-\frac{2h_2t^3+h_3(s+t)[t^2+2s(s+t)]}{18M^6}+\mathcal{O}\left(\frac{m^2E^4}{M^6}\right) \,.
\end{split}
\end{equation}
Notice that they grow as $E^6$, with $E^4$ terms suppressed by $(m/E)^2$,  and therefore will be subject to severe bootstrap constraints, as described above for the toy model. Let us take a closer look at them.

The corresponding Arcs are
\begin{equation}
\begin{split}
\A_{-\frac{1}{2}-\frac{1}{2}}^{-\frac{1}{2}-\frac{1}{2}}(t,0)&\to t\frac{6h_2-h_3}{36M^6}+\mathcal{O}\left(\frac{m^2}{M^6}\right)\,,\\
\A_{-\frac{1}{2}-\frac{1}{2}}^{+\frac{1}{2}+\frac{1}{2}}(t,0)&\to-t\frac{h_1}{3M^6}+\mathcal{O}\left(\frac{m^2}{M^6}\right)\,,\\
\A_{-\frac{1}{2}+\frac{1}{2}}^{+\frac{1}{2}-\frac{1}{2}}(t,0)&\to-t\frac{h_3}{9M^6}+\mathcal{O}\left(\frac{m^2}{M^6}\right) \,.
\end{split}
\end{equation}
The key observation is that the leading terms of the Arcs at $t\to 0$ are subleading in $m$:
\begin{equation}
    \A_{-\frac{1}{2}-\frac{1}{2}}^{-\frac{1}{2}-\frac{1}{2}}(0,0)=\A_{+\frac{1}{2}+\frac{1}{2}}^{+\frac{1}{2}+\frac{1}{2}}(0,0)=\frac{2m^2(2h_1-h_3)}{9M^6}  \,.
\end{equation}
This creates a tension, as Eq.~\eqref{eq:positivity_bounds_dec} requires all inelastic Arcs to be bounded by elastic ones
\begin{equation}\label{eq:finitemboundsmajorana_approx}
\begin{split}
\abs{\frac{t}{m^2}\frac{(6h_2-h_3)}{36}+\mathcal{O}(1)}&\leq \frac{2(2h_1-h_3)}{9} \,,\\
\abs{\frac{t}{m^2}\left(-\frac{h_1}{3}\mp \frac{h_3}{9}\right)+\mathcal{O}(1)}&\leq\frac{4(2h_1-h_3)}{9}\,.
\end{split}
\end{equation}
The leading effect in $m^2\ll-t\ll\Lambda^2$ can be captured by working in the $m/M\to 0$ limit, so that the bounds just read
\begin{equation}
    \abs{6h_2-h_3}\leq0\;,\;\;\abs{-3h_1\mp h_3}\leq0 \,,
\end{equation}
whose only solution is the trivial one
\begin{equation}
    h_1=h_2=h_3=0 \,.
\end{equation}

 Therefore, we reach the conclusion that
\begin{center}
    \textit{Isolated Majorana spin-$3/2$ particles are either free, or $\Lambda\simeq m$.}
\end{center}

As it should be clear already from Eq.~\eqref{eq:finitemboundsmajorana_approx}, the bounds constrain ratios of $h_i$ and therefore are independent of their absolute size. In particular, a deformation of the free theory, schematically $\mathcal{M} \sim \frac{m^2}{M^8}(m^2 E^4 + E^6)$---which vanishes as $m \to 0$---is still not acceptable. Positivity is powerful in that it constrains even arbitrarily weakly coupled theories. This is possible because the free theory lies right at the boundary of exclusion by positivity, and the direction in which it is approached---\textit{i.e.}, the ratios of couplings---is crucial. A suppressed ratio of the $E^4$ term relative to the $E^6$ term, for instance, is inconsistent with positivity---regardless of how small the overall amplitude may be.

In  Ref.~\cite{Melville:2019tdc} part of this analysis was performed but
inelastic channels were not included. 
As we have shown, the latter play an essential role to reach the above conclusion.\footnote{In particular, Ref.~\cite{Melville:2019tdc} does not constrain $h_1$, seemingly leaving room for a consistent solution.}\\

\begin{figure}[h!!]
    \begin{subfigure}[h]{0.33\linewidth}
    \includegraphics[width=\textwidth]{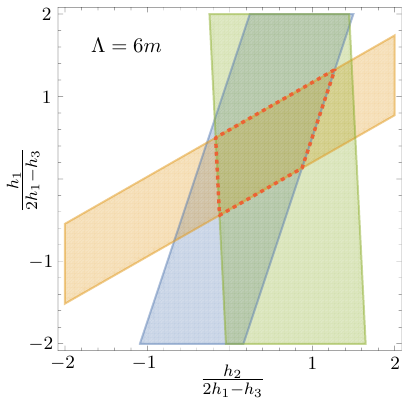}
    \end{subfigure}
    \begin{subfigure}[h]{0.33\linewidth}
    \includegraphics[width=\textwidth]{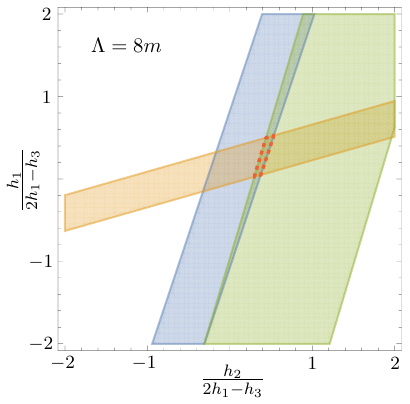}
    \end{subfigure}
    \begin{subfigure}[h]{0.33\linewidth}
    \includegraphics[width=\textwidth]{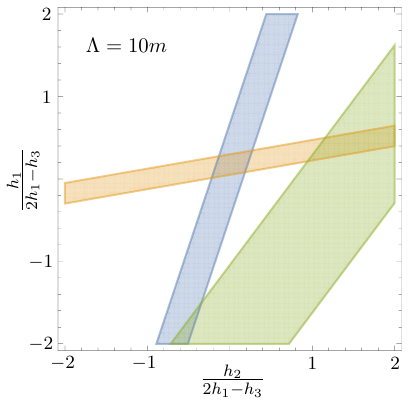}
    \end{subfigure}
    \caption{Constraints on $h_1$, $h_2$ and $h_3$ from positivity bounds at finite mass $m$, evaluated at~$t=-\Lambda^2/10$ with $\Lambda=6m,8m,10m$ (from left to right). The blue, orange and green bands correspond to bounding $|\A_{-\frac{1}{2}-\frac{1}{2}}^{-\frac{1}{2}-\frac{1}{2}}(t,0)|$, $|\A_{-\frac{1}{2}-\frac{1}{2}}^{+\frac{1}{2}+\frac{1}{2}}(t,0)+\A_{-\frac{1}{2}+\frac{1}{2}}^{+\frac{1}{2}-\frac{1}{2}}(t,0)|$ and $|\A_{-\frac{1}{2}-\frac{1}{2}}^{+\frac{1}{2}+\frac{1}{2}}(t,0)-\A_{-\frac{1}{2}+\frac{1}{2}}^{+\frac{1}{2}-\frac{1}{2}}(t,0)|$. The intersection region, when present, is indicated by a red dashed contour. Further details are given in Appendix~\ref{app:positivity_bounds}.}\label{fig:finitem_main}
\end{figure}

Working at finite $m/\Lambda$, we can extract a quantitative upper bound  on $\Lambda$ for nonzero $h_i$, as explained in Appendix~\ref{app:positivity_bounds}. 
The area of parameter space allowed by positivity bounds for different values of $m/\Lambda$  is shown in Fig.~\ref{fig:finitem_main}, within the red dashed contour.   We observe that  
\begin{equation}
\label{eq:boundMajorana8}
\Lambda \lesssim 9 m\,,
\end{equation}
otherwise the allowed area  shrinks to zero, see also Fig.~\ref{fig:finitmboundsintersectionarea}.
Eq.~\eqref{eq:boundMajorana8}  quantifies the limited range of validity of the EFT for massive spin-$3/2$ particles. The bound comes with a $|t|/\Lambda^2=10\%$ accuracy; a seemingly  more stringent bound, $\Lambda\simeq m$, obtained with $|t|\simeq \Lambda^2$, would only be an ${\cal O}(1)$ estimate, while \eqref{eq:boundMajorana8} is an accurate---conservative---upper bound.

The next question to address is whether the inclusion of additional light degrees of freedom can enlarge the range of validity of the EFT of massive spin-3/2 particles.

\subsection{
The need of Gravity}

We will only consider additional degrees of freedom that are themselves well-behaved. 
This restricts us to particles with spin $J \leq 1$ and the graviton. Specifically, the only allowed exchanges are scalars, vectors, and a massless graviton.

Scalars may be massless, but vector bosons must be massive, since the on-shell coupling of a massless photon to a massive Majorana fermion vanishes due to antisymmetry.

\par The generic $CP$-invariant three-point amplitudes are (the normalizations of the amplitudes do not affect our results, so they are fixed by convenience):\footnote{The two spin-$3/2$ particles in the vertices are labeled by $\bf 1$ and $\bf 2$.  We refer to the spin-$0$ and spin-$1$ states respectively as  $S_{CP}$ and $V_{CP}$ with $CP=\pm 1$ being the $CP$ quantum number.}
\begin{itemize}
    \item \underline{Scalar and pseudoscalar:}
    \begin{equation}\label{eq:3ptmajorana_scalars}
        \begin{split}
            \M(\mathbf{1}\mathbf{2}\mathbf{3}^{S_{+}})&=\frac{a_1^{+}}{M_S^2}(\angMM{1}{2}^3+\sqrMM{1}{2}^3)+\frac{a_2^{+}}{M_S^2}\angMM{1}{2}\sqrMM{1}{2}(\angMM{1}{2}+\sqrMM{1}{2})\,,\\
            \M(\mathbf{1}\mathbf{2}\mathbf{3}^{S_{-}})&=i\frac{a_1^{-}}{M_S^2}(\angMM{1}{2}^3-\sqrMM{1}{2}^3)+i\frac{a_2^{-}}{M_S^2}\angMM{1}{2}\sqrMM{1}{2}(\angMM{1}{2}-\sqrMM{1}{2})\,,
        \end{split}
    \end{equation}
    where $M_S$ controls the mass dimension of the three-point vertex, and $m_{S_\pm}$ is the spin-0 mass. 
    \item \underline{Vector and pseudovector:}\footnote{The powers of $m_{V_\pm}$ are chosen so to have a $1/m_{V_\pm}$ factor when projecting into  the longitudinal mode of the vector. We also notice that the pseudovector couples through the $b_2^{-}$ term to a $U(1)$ current built with the  spin-3/2 Weyl fermion. Conservation of this current is explicitly broken by the Majorana mass term. }
    \begin{equation}\label{eq:3ptmajorana_vectors}
        \begin{split}
            \M(\mathbf{1}\mathbf{2}\mathbf{3}^{V_{+}})&=  i\frac{b_1^{+}}{m_{V_+} M_V^2}(\angMM{1}{3}\sqrMM{2}{3}-\angMM{2}{3}\sqrMM{1}{3})(\angMM{1}{2}^2-\sqrMM{1}{2}^2)\,,\\
            \M(\mathbf{1}\mathbf{2}\mathbf{3}^{V_{-}})&=\frac{b_1^{-}}{m_{V_-} M_V^2}(\angMM{1}{3}\sqrMM{2}{3}-\angMM{2}{3}\sqrMM{1}{3})(\angMM{1}{2}^2+\sqrMM{1}{2}^2)\\
            &+\frac{b_2^{-}}{m_{V_-} M_V^2}(\angMM{1}{3}\sqrMM{2}{3}-\angMM{2}{3}\sqrMM{1}{3})\angMM{1}{2}\sqrMM{1}{2} \,,
        \end{split}
    \end{equation}
    where  $m_{V_\pm}$ is the mass of the vector, and $M_V$ the scale controlling the overall mass dimension of the vertex.
     \item \underline{Graviton:}
    \begin{equation}\label{eq:3ptmajorana_graviton}
        \begin{split}
            \M(\mathbf{1}\mathbf{2}3^{++})&=\frac{x_{12}^2}{\mpl^2}\left[\angMM{2}{1}^3+\frac{g_{4}}{m^2}\angMM{2}{1}\angsqrMM{1}{3}{2}^2+\frac{g_{8}}{m^3}\angsqrMM{1}{3}{2}^3\right]\,,\\
            \M(\mathbf{1}\mathbf{2}3^{--})&=\frac{x_{12}^{-2}}{\mpl^2}\left[\sqrMM{2}{1}^3+\frac{g_{4}}{m^2}\sqrMM{2}{1}\angsqrMM{2}{3}{1}^2+\frac{g_{8}}{m^3}\angsqrMM{2}{3}{1}^3\right] \,.
        \end{split}
    \end{equation}
where $\mpl$ is the Planck mass and the three terms correspond, respectively, to minimal coupling, the gravitational quadrupole ($g_4$), and the octupole ($g_8$).\footnote{A priori, an additional three-point amplitude is possible, corresponding to a gravitational dipole. However, this coupling is inconsistent with factorisation and soft theorems~\cite{Chung:2018kqs,Falkowski:2020aso}. In fact, we explicitly checked that positivity forces it to vanish as well.} The $x_{12}$-factor~\cite{Arkani-Hamed:2017jhn} is defined as $x_{12}=\frac{\angsqrmm{\xi}{p_1}{3}}{m\angmm{3}{\xi}}$, with $\xi$ and arbitrary reference spinor.
\end{itemize}

To keep the discussion simple to show the main result, we start considering first only the case of spin-0 states and the graviton with minimal gravitational coupling, {\it i.e.}, setting $g_4 = g_8 = a_{1}^{\pm} = b_{i}^{\pm} = 0$.  In this case we will only need to analyze the amplitudes of the longitudinal  spin-$3/2$ particle. 
The fully general case, where all couplings are turned on, is  discussed in Section~\ref{sec:generalcase}.
The  interactions that we are considering correspond to the following Lagrangian operators:
\begin{equation}
    \frac{1}{\mpl}h_{\mu\nu}T^{\mu\nu}\;,\;\;\frac{m^2}{M_S^2}\phi_+\bar\psi_{\mu}\psi^{\mu}\;,\;\;i\frac{m^2}{M_S^2}\phi_{-}\bar\psi_{\mu}\gamma_5\psi^{\mu}\,.
\end{equation}
The amplitudes\footnote{For this simplified setting, we add contact terms to reduce the energy growth of the gravitational contribution from $E^8$, as obtained from the on-shell computation, to $E^6$.} in Eq.~\eqref{eq:amplideclong} become now 

\begin{equation}\label{eq:amplideclong2}
\begin{split}
\M_{-\frac{1}{2}-\frac{1}{2}}^{-\frac{1}{2}-\frac{1}{2}}(s,t)&\to-\frac{s^2[(a_2^+)^2 +(a_2^-)^2]}{9M_S^4}+\frac{s[s^2-6t(s+t)]}{72f_G^6}-\frac{2h_2s^3+h_3s[s^2+2t(s+t)]}{18M^6}
\,,\\
\M_{-\frac{1}{2}-\frac{1}{2}}^{+\frac{1}{2}+\frac{1}{2}}(s,t)&\to\frac{2[s^2+s(s+t)][(a_2^+)^2 -(a_2^-)^2]}{9M_S^4}-\frac{h_1st(s+t)}{3M^6}
\,,\\
\M_{-\frac{1}{2}+\frac{1}{2}}^{+\frac{1}{2}-\frac{1}{2}}(s,t)&\to-\frac{t^2[(a_2^+)^2 +(a_2^-)^2]}{9M_S^4}+\frac{t[t^2-6s(s+t)]}{72f_G^6}-\frac{2h_2t^3+h_3(s+t)[t^2+2s(s+t)]}{18M^6}
\,,
\end{split}
\end{equation}
where we have taken the limit
\begin{equation}
\label{eq:limit}
m\to0\;,\quad \mpl \to \infty\;,\quad f_G^3\equiv m^2 \mpl=\mbox{finite} \,,
\end{equation}
and  $M$ is the scale of the contact terms in \eqref{eq:contactMajorana}. In principle, scalar exchanges could give the dominant contribution to the elastic amplitude. However, let us observe that this would give a negative-definite contribution to the elastic Arcs: 
\begin{equation}
    0<\A_{-\frac{1}{2}-\frac{1}{2}}^{-\frac{1}{2}-\frac{1}{2}}(0,0)\big|_{S}=\A_{+\frac{1}{2}+\frac{1}{2}}^{+\frac{1}{2}+\frac{1}{2}}(0,0)\big|_{S}=-\frac{(a_2^+)^2 +(a_2^-)^2}{9M_S^4}+\mathcal{O}\left(\frac{m^2}{M^6}\right) \,.
\end{equation}
Therefore, scalar exchanges cannot provide the dominant contribution, and is eventually bounded by subleading terms in the limit~\eqref{eq:limit}, see later Eq.~\eqref{eq:scalarcouplingsbounds}. We can thus neglect them at order $E^6$.

Positivity of the remaining terms, under the form of Eq.~\eqref{eq:positivity_bounds_dec}, requires
\begin{equation}
\label{eq:tuninMaj0}
    \abs{t\frac{6h_2-h_3}{36M^6}-t\frac{5}{24f_G^6}}\leq0\;,\;\;\abs{-t\frac{h_1}{3M^6}}\leq0\;,\;\;\abs{-t\frac{h_3}{9M^6}-t\frac{3}{18f_G^6}}\leq0 \,,
\end{equation}
where the RHS zeros should be understood as quantities that vanish in the limit~\eqref{eq:limit}, similarly to the previous case Sec.~\ref{sec:isolated_majorana}. 
The only non-trivial solution is
\begin{equation}
\label{eq:tuninMaj0b}
    h_1=0\;,\;\;\frac{h_2}{M^6}=\frac{1}{m^4\mpl^2}\;,\;\;\frac{h_3}{M^6}=-\frac{3}{2m^4\mpl^2} \,. 
\end{equation}
We can then conclude that
 \begin{center}
    \textit{Gravity is necessary for getting a valid  EFT of a Majorana spin-$3/2$ state.}
\end{center}
The next step is to understand what kind of theory we have gotten after fixing Eq.~\eqref{eq:tuninMaj0b}. Now, the four-point amplitude  grows as $E^4$ and can be studied in the limit
\begin{equation}
m\to0\;, \quad  \mpl\to\infty\;, \quad  m^2\mpl^2=\mbox{finite} \,,
\end{equation}
yielding
\begin{equation}\label{eq:amplidec_goldstinos}
\begin{split}
\M_{-\frac{1}{2}-\frac{1}{2}}^{-\frac{1}{2}-\frac{1}{2}}(s,t)&\to\frac{s^2}{3m^2\mpl^2}-\frac{s^2[(a_2^+)^2 +(a_2^-)^2]}{9M_S^4}+\mathcal{O}\left(\frac{E^2}{\mpl^2}\right)\,,\\
\M_{-\frac{1}{2}-\frac{1}{2}}^{+\frac{1}{2}+\frac{1}{2}}(s,t)&\to\frac{2[s^2+s(s+t)][(a_2^+)^2 -(a_2^-)^2]}{9M_S^4}+\mathcal{O}\left(\frac{E^2}{\mpl^2}\right) \,.
\end{split}
\end{equation}
We recognize Eq.~\eqref{eq:amplidec_goldstinos} as describing the scattering of \textit{Goldstinos}, with decay constant
\begin{equation}
    F^2=3m^2\mpl^2 \,.
\end{equation}
Remarkably, the relation between $\sqrt{F}$,  that can be associated to a supersymmetry-breaking scale, the spin-$3/2$ mass,  and the Planck mass emerges purely from causality and unitarity, with no reference to supersymmetry.

The positivity of the $E^4$ term, together with bounds on inelastic Arcs, imposes
\begin{equation}\label{eq:scalarcouplingsbounds}
	\begin{split}
		\frac{(a_2^+)^2 +(a_2^-)^2}{M_S^4}&\leq\frac{9}{F^2} \,,\\
		\frac{2}{9}\frac{\abs{(a_2^+)^2 -(a_2^-)^2}}{M_S^4}&\leq\frac{1}{F^2}-\frac{(a_2^+)^2 +(a_2^-)^2}{9M_S^4} \,.
	\end{split}
\end{equation}
Thus, all couplings of the spin-$3/2$ particle are controlled by $1/F^2=1/(3m^2\mpl^2)$ and they must vanish if gravity is removed at finite $m$. In full analogy, the transverse sector—comprising a massless spin-$3/2$—is found to couple minimally to gravity when positivity bounds are imposed.

The picture that emerges is none other than spontaneously broken $\mathcal{N}=1$ supergravity: the spin-$3/2$ particle is the Gravitino, which acquires mass by \textit{eating} the longitudinal (Goldstino) modes. The decoupling limit $\mpl\to\infty$ with $F$ fixed corresponds to the rigid supersymmetric limit.

It is amusing that the same conclusion can be reached from a completely different direction: by demanding amplitudes to be constructible via BCFW-like recursion relations, and using the Ward identities of the transverse amplitudes, as done in~\cite{Gherghetta:2024tob}. However, such an approach misses the key insight into what goes fundamentally wrong if constructibility fails. After all, one can write down a seemingly consistent EFT for a spin-$3/2$ particle that violates the recursion relations and still enjoys a large mass gap—yet its pathology lies in causality violations.\\

While the scalars do not play a central role in ensuring consistency, it is instructive to examine the special case where the bounds in Eq.~\eqref{eq:scalarcouplingsbounds} are saturated:
\begin{equation}
	\frac{(a_2^+)^2}{M_S^4}=\frac{(a_2^-)^2}{M_S^4}=\frac{9}{2F^2} \,.
\end{equation}
In this case the $E^4$ terms cancel out (except for the graviton $t$-channel pole). The theory can now be studied in two relevant limits:
\begin{itemize}
    \item[a)] \textbf{Rigid SUSY limit} ($m\to0$, $\mpl\to\infty$, $F,m_{S_\pm}$ fixed): The graviton and the transverse Gravitinos decouple, and we find the simplest supersymmetry-breaking model corresponding to a scalar and its super-partner, the Goldstino. In Section~\ref{sec:GoldstinoEFThedron} we will  more generically explore  the space of consistent supersymmetry-breaking theories by deriving the constraints on the EFT of Goldstinos.\\

    \item[b)] \textbf{Unbroken SUGRA limit} ($m,m_{S_\pm}\to0$, $\mpl$ fixed): This corresponds to the Polonyi model of supergravity \cite{Polonyi:1977pj}, in which supersymmetry-breaking is sourced by a scalar. In this  limit we recover a $\mathcal{N}=1$ chiral supermultiplet (made of the scalars $S_\pm$ and the longitudinal mode of the spin-3/2 state) coupled to the graviton supermultiplet (the graviton and the transverse modes of the spin-3/2 states).
\end{itemize}

Let us pause to consider a general observation. As already mentioned, the tunings which define a consistent solution, aside from the free theory, should not be taken as exact. Positivity bounds, imposed in successive decoupling limits, provide in fact a power-counting scheme, determining the theory around which a consistent EFT expansion is possible. Deviations from the tunings are allowed, as long as the resulting amplitudes do not dominate the $E^4$ contribution or violate basic positivity of the Arcs. These corrections can be systematically bounded, exploiting the full set of consistency conditions, either working at finite mass or in a scaling limit in which the $E^4$ term is not vanishing. An example is given, in the Goldstino limit, in Section~\ref{sec:GoldstinoEFThedron}.\\

\subsection{General case}
\label{sec:generalcase}
The conclusions reached in the simplified case discussed above  remain valid in the most general case, as we discuss in the present section. In particular, we find eventually that again gravity must be present, with vanishing multipoles.

\par The relevant three-point amplitudes are those involving scalars, vectors, and gravitons, as given in Eqs.~\eqref{eq:3ptmajorana_scalars}, \eqref{eq:3ptmajorana_vectors}, and \eqref{eq:3ptmajorana_graviton}. The factorizable contributions to the four-point amplitudes, arising from scalar, vector, and graviton exchange, grow respectively as $E^4$, $E^6$, and $E^8$ in the high-energy limit $E\gg m$. Therefore we have to include  all 27 $CP$-invariant independent contact terms with up to $E^8$ energy growth. The full expressions for the four-point amplitudes—including both exchange and contact contributions—can be found in the ancillary file \texttt{Amplitudes.nb}.

\par We proceed as in the previous section: we first expand the amplitudes in the limit $E\gg m$,  and  apply the positivity bounds on the largest term,  the $E^8$ term. These bounds will (non-trivially) require  this term to cancel.   After this is guaranteed by adjusting the parameters of  the EFT,  we repeat the process for the subleading contributions $E^7,E^6,...$.  Positivity bounds will  effectively force the elimination of all growing terms down to the $E^4$ level. 

In particular, we find that gravitational multipoles must vanish 
\begin{equation}\label{eq:fullmajorana_tuning1}
g_4=g_8=0 \,,
\end{equation}
while Wilson coefficients $h_i/M^8$ must be tuned to be either vanishing or $\mathcal{O}((m^6\mpl^2)^{-1},(m^2m_V^2M_V^4)^{-1})$  to cancel the terms of the amplitudes growing faster than $E^4$. From now on, we will assume these  tunings  for the $h_i$.

\par At order $E^4$ we find further  constraints from positivity of elastic Arcs. Working in the limit 
\begin{equation}
m \to 0\;,\quad m_{V_{\pm}} \to 0\;,\quad \mpl \to \infty\;,\quad F^2 = 3m^2\mpl^2=\mbox{finite}\;,\quad \mu_{V_{\pm}} = \frac{m_{V_\pm}}{m}=\mbox{finite} \,,
\end{equation}
we have in the transverse sector
\begin{equation}
	\begin{split}
		\M_{-\frac{3}{2}-\frac{3}{2}}^{-\frac{3}{2}-\frac{3}{2}}(s,t)\to -s^2\frac{(a_1^+)^2 +(a_1^-)^2}{M_S^4} - \frac{2s^2}{M_V^4}\left(\frac{(b_1^+)^2}{\mu_{V_+}^2} +\frac{(b_1^-)^2}{\mu_{V_{-}}^2}\right)\,.
	\end{split}
\end{equation}
Positivity of  the $s^2$ coefficient thus requires 
\begin{equation}
a_1^+ = a_1^- = b_1^+ = b_1^- = 0\,. 
\end{equation} For the longitudinal modes we have:
\begin{equation}
	\begin{split}
		\M_{-\frac{1}{2}-\frac{1}{2}}^{-\frac{1}{2}-\frac{1}{2}}(s,t) &\to \frac{s^2}{F^2} - \frac{s^2[(a_2^+)^2 +(a_2^-)^2]}{9M_S^4} - s^2\frac{(2+\mu_{V_-}^2)(b_2^-)^2}{9\mu_{V_-}^2 M_V^4} 
        \,,
	\end{split}
\end{equation}
and positivity tells us
\begin{equation}\label{eq:finalboundmajorana}
	\begin{split}
        \frac{(a_2^+)^2 +(a_2^-)^2}{M_S^4} + \left(\frac{2}{\mu_{V_-}^2 M_V^4}+\frac{1}{M_V^4}\right)(b_2^-)^2 &\leq \frac{9}{F^2} \,,
	\end{split}
\end{equation}
where we have split the massive vector contribution into its longitudinal component---singular in $\mu_{V_-}^2$ and equivalent to the pseudoscalar---and a transverse component.\footnote{Although in this amplitude the longitudinal and the transverse modes appear with the same functional dependence, they are distinguished by other helicity configurations. The longitudinal mode can be decoupled, and the pseudovector made massless, only in the limit $m,m_{V_-}\to0$ with $\mu_{V_-}\to+\infty$, so that the $U(1)$ symmetry is restored.}

The Eq.~\eqref{eq:finalboundmajorana} shows that all contributions are bounded by the term $1/F^2$, which is intrinsically gravitational. Moreover, it makes evident the reason for the need of gravity: a theory of scalars and vectors contribute negatively to the $s^2$ term of the amplitude, violating positivity.\\

It is possible to derive further bounds, \textit{e.g.} considering inelastic Arcs as in the second line of Eq.~\eqref{eq:scalarcouplingsbounds}. We leave this  for Section~\ref{sec:GoldstinoEFThedron}, where we study the Goldstino EFT.

\section{Dirac spin-3/2 particle}
\label{sec:DiracGeneral}

We consider now a scenario in which the spin-$3/2$ particle is a Dirac fermion and the theory enjoys a  global $U(1)$ symmetry, under which the former is charged. As in the Majorana case, we will proceed step-by-step, starting from the isolated spin-3/2 theory and gradually introducing new matter content and interactions.

\par Since the external states are charged, we will have different amplitudes for different charge configurations. Restricting to charge-conjugation invariant interactions, we have the following three channels for spin-$3/2$ scattering:
\begin{equation}
\psi_q\psi_{-q} \to \psi_q\psi_{-q}\ , \ \ \psi_q\psi_{-q} \to \psi_{-q}\psi_{q}\ , \ \ \psi_q\psi_{q} \to \psi_q\psi_{q}\,,
\end{equation}
where $q$ denotes the $U(1)$ charge. Notice that under crossing the first and the third channels are exchanged, while the second maps to itself. Unless otherwise specified, the amplitudes discussed below refer to the first channel, and the Arcs are elastic in the charges.

\par Based on the results of the previous section, we can already build some intuition on this case. Indeed, we can think of the Dirac particles as a pair of Majorana modes $\xi_1$ and $\xi_2$, transforming under an $SO(2)$ symmetry. In this basis,  we have three types of scattering amplitudes, those involving only  $\xi_1$, those involving only  $\xi_2$, and those involving both $\xi_1$ and $\xi_2$. Notice that the $SO(2)$ symmetry relates the three types of amplitudes, so they are not independent. For the first two types,  we can use the  result of the previous section and claim that  gravity is needed for consistency of the EFTs. The novelty of the  case discussed here resides therefore in the amplitudes involving the two different Majorana states. As we will see, consistency will require the gauging of the $SO(2)$ symmetry. To understand the positivity constraints, it is simpler to work with a Dirac fermion of definite $U(1)$ charge.

\subsection{Isolated spin-3/2 particle}

At the lowest order, the EFT of a massive Dirac spin-3/2 particle  consists of  contact terms  that  grow as $E^6$. There are 27 separately $C$- and $P$-invariant terms with  Wilson coefficients  $h_i$ ($i=1,...,27$). We repeat the procedure already applied to the Majorana case. Positivity requires the cancellation of the $E^6$ terms, and we find that actually there is a non-trivial solution:
\begin{equation}
\begin{split}
    \frac{h_{i}}{M^8}&=0\;,\;i=1,...,19,23,...,27 \,,\\
    h_{20} &= -\frac{1}{6} h_{21} \,,\\
    h_{22} &= \frac{2}{3} h_{21} \,.
\end{split}
\end{equation}
The reason is that the  $h_{21}$ contribution is degenerate with a linear combination of the $h_{20}$ and $h_{22}$ ones at order $E^6$. However, as we move to order $E^5$,  one finds that this is term is proportional to $h_{21}$,  and its cancellation, required by positivity, leads to   $h_{21} = 0$. Thus, we conclude that 
\begin{center}
    \textit{Isolated Dirac spin-$3/2$ particles are either free, or $\Lambda\simeq m$.}
\end{center}

A quantitative bound on $\Lambda/m$ can be obtained by working at finite $m/\Lambda$, analogously to the analysis of Section~\ref{sec:isolated_majorana}.

\par We can now swiftly move on and introduce additional light degrees of freedom, in the hope of finding a non-trivial solution  consistent with positivity bounds.

\subsection{Including light degrees of freedom}

The possible matter content that can be added to the low-energy spectrum is richer in this case. Beyond the graviton, we can now have also a massless photon, associated to the $U(1)$ symmetry.\footnote{A priori the $U(1)$ symmetry could also not be gauged, with the photon coupling only via $F^{\mu\nu}$ interactions. We will see shortly that this is not consistent with positivity bounds.} Moreover, one can exchange scalars and massive vectors, both neutral and charged, both even and odd under $C$- and $P$-transformations. In order to keep the discussion simple, we will consider separately different scenarios.

\subsubsection{Massless graviton and photon}\label{subsec:diracgravitonphoton}
The three-point amplitudes are
\begin{itemize}
    \item \underline{Photon}:
    \begin{equation}\label{eq:threepoint_photon}
        \begin{split}
            \M(\mathbf{1}_q\mathbf{2}_{-q}3^{+})&=\sqrt{2}qe\frac{x_{12}}{m^2}\left[\angMM{2}{1}^3+\frac{3c_{2}}{2m}\angMM{2}{1}^2\angsqrMM{1}{3}{2}+\frac{c_{4}}{m^2}\angMM{2}{1}\angsqrMM{1}{3}{2}^2+\frac{c_{8}}{m^3}\angsqrMM{1}{3}{2}^3\right] \,,\\
            \M(\mathbf{1}_q\mathbf{2}_{-q}3^{-})&=\sqrt{2}qe\frac{x_{12}^{-1}}{m^2}\left[\sqrMM{2}{1}^3+\frac{3c_{2}}{2m}\sqrMM{2}{1}^2\angsqrMM{2}{3}{1}+\frac{c_{4}}{m^2}\sqrMM{2}{1}\angsqrMM{2}{3}{1}^2+\frac{c_{8}}{m^3}\angsqrMM{2}{3}{1}^3\right] \,.
        \end{split}
    \end{equation}
    \item \underline{Graviton}:
    \begin{equation}
        \begin{split}
            \M(\mathbf{1}_q\mathbf{2}_{-q}3^{++})&=\frac{x_{12}^2}{\mpl^2}\left[\angMM{2}{1}^3+\frac{g_{4}}{m^2}\angMM{2}{1}\angsqrMM{1}{3}{2}^2+\frac{g_{8}}{m^3}\angsqrMM{1}{3}{2}^3\right] \,,\\
            \M(\mathbf{1}_q\mathbf{2}_{-q}3^{--})&=\frac{x_{12}^{-2}}{\mpl^2}\left[\sqrMM{2}{1}^3+\frac{g_{4}}{m^2}\sqrMM{2}{1}\angsqrMM{2}{3}{1}^2+\frac{g_{8}}{m^3}\angsqrMM{2}{3}{1}^3\right] \,.
        \end{split}
    \end{equation}
\end{itemize}
The couplings $c_2$, $c_4$ and $c_8$ correspond respectively to the electromagnetic dipole, quadrupole and octupole moments, see \textit{e.g.} \cite{Chung:2018kqs} for a discussion. In particular, $c_2$ is directly connected to the gyromagnetic ratio $g$ 
\begin{equation}
c_2 = \frac{g - 2}{2}\,.
\end{equation}
The choice to normalize the multipole couplings with $m$ is purely conventional--- and does not affect our argument.  Here we follow the traditional choice for the $g-2$.

It can be useful to match these interaction to the following Lagrangian operators often considered in the literature
\begin{equation}
    \mathcal{L}_{int}=qe\bar\psi^{\mu}\gamma_{\mu\nu\rho}A^{\nu}\psi^{\rho}-i\frac{qe}{m}l_1\bar\psi^{\mu}F_{\mu\nu}\psi^{\nu}-i\frac{qe}{m}l_2\bar\psi^{\mu}F_{\nu\rho}\gamma^{\nu\rho}\psi_{\mu}\,,
\end{equation}
up to terms that vanish on shell, with the choice $3c_2/2=l_1+2l_2+1$, $c_4=l_1/2+2l_2$, $c_8=0$.

In Section~\ref{sec:g-2_causality} we comment on the relation between positivity bounds for charged spin-$3/2$ particles and their causal propagation in classical backgrounds, which has been extensively studied in previous works.

As the spin-$3/2$  amplitudes mediated by gravitons and photons grow as $E^8$, we must include all contact terms that can give contributions up to   this  order.
There are 79 possible $C$- and $P$-invariant contact terms.  The full amplitudes are provided in the ancillary files. Following the strategy of the previous section, positivity bounds non-trivially require the cancellation of the terms growing as  $E^8$ and $E^7$, which yields the condition
\begin{equation}
g_8=0 \,,
\end{equation}
combined with the tuning of the Wilson coefficients $h_i$, with $i>27$.

At the $E^6$ order also the photon exchanges enter the amplitudes and positivity bounds require
\begin{equation}\label{eq:tuningdiracE6}
g_4=0\ ,\ \ \ \   c_8=0\,.
\end{equation}
Similarly as in the previous section, the remaining $h_i$ are also fixed, with the exception of the Wilson coefficient $h_{21}$, as it is degenerate with $h_{20}$ and $h_{22}$ at this order.
\par After the above tuning of parameters, the amplitude is dominated by  $E^5$ terms. Now, crucially, we have the  couplings of the graviton and photon, beyond $h_{21}$, so one can hope to find  non-trivial solutions consistent with positivity bounds. 
Indeed, the latter impose
\begin{equation}\label{eq:constraintsdiracE5}
    \begin{split}
        0&=\frac{q^2e^2}{m^5}(2+3c_2)(3c_2+4c_4) \,,\\
        \frac{8}{3}\frac{h_{21}m^3}{M^8}&=\frac{q^2e^2}{m^5}\left[2+\frac{16}{3}c_4+c_2(8+9c_2+10c_4)\right] \,,\\
        \frac{1}{\mpl^2m^3}&=\frac{q^2e^2}{m^5}(2+3c_2) \,.
    \end{split}
\end{equation}
The system has formally three solutions
\begin{equation}
\begin{split}
    \text{a)}\;\;\;\;\;\frac{h_{21}}{M^8}&=\frac{q^2e^2c_4^2}{m^8}\;,\;\;
    c_4=-\frac{3}{4}c_2\;,\;\;qe\to0\;,\;\; \,\frac{1}{\mpl}\to0\;,\;\;qec_4\;\text{finite} \,,\\
    \\
    \text{b)}\;\;\;\;\;\frac{h_{21}}{M^8}&=\frac{q^2e^2(1-2c_4)}{m^8}\;,\;\;\frac{1}{\mpl}\to0\;,\;\;c_2=-\frac{2}{3} \,,\\
    \\
    \text{c)}\;\;\;\;\;\frac{h_{21}}{M^8}&=\frac{1}{16q^2e^2m^4\mpl^4}+\frac{1}{4m^6\mpl^2}\;,\;\;
    \frac{q^2e^2c_2}{m^2}=\frac{1}{3\mpl^2}-\frac{2}{3}\frac{q^2e^2}{m^2}\;,\;\;
    c_4=-\frac{3}{4}c_2 \,.
\end{split}
\end{equation}
The first two, in which gravity and/or the minimal EM coupling are removed, turn out to be inconsistent at order $E^4$, as they violate the positivity of the elastic Arcs, unless all remaining couplings are set to zero.\footnote{Notice that the solution $b)$ has a gyromagnetic factor $g=\frac{2}{3}$ and corresponds to a spin-$3/2$ coupled to the photon only through its $U(1)$ current (with gravity decoupled).}
We are then left with the last solution.

At order $E^4$,  we have the constraint
\begin{equation}
    \A_{-\frac{3}{2}-\frac{1}{2}}^{-\frac{3}{2}-\frac{1}{2}}(0,0)=-\frac{(m^2-2q^2e^2\mpl^2)^2}{24q^2e^2m^6\mpl^2}\geq0 \,,
\end{equation}
which yields the condition
\begin{equation}\label{eq:charge_tuning}
    \frac{q^2e^2}{m^2}=\frac{1}{2\mpl^2} \,.
\end{equation}
In summary, we find that the tunings required by consistency are
\begin{equation}\label{eq:tuningdirac_simp}
\begin{split}
    \frac{h_{i}}{M^8}&=\mathcal{O}\left(\frac{1}{m^6\mpl^2}\right)\;,\;i=1,...,79 \,,\\
    g_8&=g_4=0 \,,\\
    c_8&=c_4=c_2=0 \,,\\
    \frac{q^2e^2}{m^2}&=\frac{1}{2\mpl^2} \,.
\end{split}
\end{equation}

\noindent Some comments are in order. 
Firstly, an immediate conclusion we can draw  is that   
\begin{center}
    \textit{Graviton and photon are both necessary for the consistency of  Dirac spin-$3/2$  EFTs.}
\end{center}
In particular, we see that causality and unitarity demand that the photon necessarily gauges the  global $U(1)$ symmetry, providing a bottom-up example of the \textit{no-global symmetries} conjecture in gravitational theories \cite{Vafa:2005ui,Ooguri:2006in,Banks:2010zn,Harlow:2018tng} (see also \cite{Palti:2019pca} for a review of the topic).
Moreover, it is quite remarkable that Eq.~\eqref{eq:charge_tuning} precisely saturates the weak gravity conjecture bound, $|q|e\geq\frac{m}{\sqrt{2}\mpl}$ originally proposed in \cite{Arkani-Hamed:2006emk}. 

Secondly, the strength of all non-zero couplings is set by $\mpl$, \textit{i.e.}, the theory is intrinsically gravitational. The EFT with the conditions Eq.~\eqref{eq:tuningdirac_simp} corresponds to a truncated version of $\mathcal{N}=2$ AdS gauged supergravity  \cite{Freedman:1976aw}
with a  positive cosmological constant added $V=\frac{3}{2}m^2\mpl^2$  to have the model in flat space instead of AdS. In flat space supersymmetry is broken and the Gravitino remains massive.
On the other hand,  in the limit $e\to0$, $m\to0$, and $\mpl$ fixed,
the transverse sector\footnote{To make sense of this limit in the longitudinal sector, since $F\to0$, it is necessary to include the superpartners, \textit{e.g.} as discussed for the Majorana case.}
reproduces the original $\mathcal{N}=2$ supergravity model of Ref.~\cite{Ferrara:1976fu} corresponding to a massless graviton supermultiplet that consists of states with helicities $(\pm 2,\pm\frac{3}{2},\pm\frac{3}{2},\pm 1)$.

Finally, we note that the tuning $c_2=0$ implies a gyromagnetic factor $g=2$. The "naturalness" of this value has been widely discussed in the literature, both as satisfying the criterion of best high-energy behaviour of photon Compton scattering with spin-$3/2$ particles and  as a solution to the causality constraints of the latter. In particular, our result completes and extends the findings of Ref.~\cite{Deser:2001dt}, where it was shown that, assuming $c_i = g_i = 0$, spin-$3/2$ particles can propagate causally on electromagnetic and gravitational backgrounds provided $\frac{q^2 e^2}{m^2} = \frac{1}{2 \mpl^2}$. Other scenarios permitted by that analysis—such as minimal electromagnetic coupling—are shown instead to be inconsistent by our results. We defer further comparison with these works to Section~\ref{sec:g-2_causality}.\\

 In the next section, we will explore some extensions of this framework with the inclusion of further light degrees of freedom: spin-$0$ and spin-$1$ states. We present below only the main results, deferring   the details  to  Appendix~\ref{app:diracscalarvector}.

\subsubsection{Extra spin-0 and spin-1 states}
\label{subsec:diracscalarvector}

The inclusion of scalars and massive vectors in the Dirac spin-$3/2$ EFT allows for several scenarios which we analyze below,  highlighting the  implications of the positivity bounds. Let us anticipate that in all cases the presence of gravity and the gauging of the $U(1)$ symmetry will remain a necessary condition, but the conditions on the parameters of the EFT obtained in the previous section will be modified.
For the extra spin-0 and spin-1 states we follow the notation  $S_{P,C}$ and $V_{P,C}$ respectively.

\subsubsection*{Spontaneously broken $U(1)$ symmetry:}

Let us consider the scenario where the photon acquires a mass, possibly due to spontaneous breaking of the $U(1)$ symmetry. Although charge is strictly speaking no longer a good quantum number, in this case, to label states, we can still work within the regime of an approximate symmetry and consider scattering of definite-charge eigenstates. Alternatively, the charged scalar responsible for the photon’s longitudinal mode---by acquiring a vacuum expectation value---may have very suppressed couplings to the spin-$3/2$ particle.

In any case, this scenario effectively amounts to replacing the massless photon on-shell three-point amplitudes with those of a massive vector ($V_{-,-}$), shown in Eqs.~\eqref{eq:threepoint_massivephoton},~\eqref{eq:relabeling_massivephoton}. In the limit in which the mass of the vector vanishes $m_V\to0$, the four-point amplitudes match the massless photon ones up to contact terms.
\par In the decoupling limit analysis we do not find any qualitative difference with the case of a massless photon. We arrive indeed to the same final tunings of Eq.~\eqref{eq:tuningdirac_simp}.

Interestingly, from the positivity of elastic Arcs in the longitudinal sector we get the following bound on the photon mass\footnote{In the presence of the scalars or massive vectors discussed later in this section this upper bound  only gets more stringent.}
\begin{equation}\label{eq:massbound}
    m_V\leq\sqrt{6}m \,.
\end{equation}
This is nicely consistent with the analysis of the previous section: if the limit $m_V \gg m$ were allowed, we could integrate out the massive vector and effectively flow to the EFT without the photon—precisely the scenario we showed to be inconsistent. Moreover, if we suppose the vector mass to be generated by some scalar vacuum expectation value (\textit{VEV}) $m_V=e\langle\phi\rangle$, one can interpret Eq.~\eqref{eq:massbound} as a bound on the \textit{VEV} itself, in the spirit of the Swampland \textit{distance conjecture} \cite{Ooguri:2006in}.

\subsubsection*{Massless photon and massive $V_{-,-}$ vector:}\label{sec:twophotons}

 Let us consider the exotic case in which there are two vectors with the same quantum numbers: a massless photon, coupled as in Eq.~\eqref{eq:threepoint_photon}, and a massive $V_{-,-}$ vector, coupled as in Eq.~\eqref{eq:threepoint_massivephoton}, \eqref{eq:relabeling_massivephoton}. As long as the masses and kinetic mixings are negligible, as we assume, we can always take linear combinations of the two vectors so that only one of the two has a minimal coupling.

\par In this case, the decoupling limit analysis presents non-trivial features compared to the one without the vector. Details are provided in Appendix~\ref{app:diracscalarvector}. Notably, we find that the charge "quantization" condition Eq.~\eqref{eq:charge_tuning} can be detuned, and is replaced by the weaker condition
\begin{equation}\label{eq:chargetuning_weak}
\begin{split}
    \frac{1}{2\mpl^2}&\leq \frac{q^2e^2}{m^2}\leq\frac{2}{\mpl^2} \,,\\
\end{split}
\end{equation}
in addition to further conditions in Eq.~\eqref{eq:D9}. Nonetheless, the ratio $qe/m$ can still vary only in a range of $\mathcal{O}(1)$ in units of $\mpl^{-1}$, which also implies $g-2=\mathcal{O}(m/\mpl)$. All other multipoles are instead required to vanish. Therefore, we can conclude that the theory is still gravitational. Moreover, the statement that both gravity and the gauging of the $U(1)$ symmetry are necessary for consistency still holds true.

 In this case as well positivity of the elastic Arcs in the longitudinal sector provides a bound on the vector mass
\begin{equation}\label{eq:doublevector_massbound}
    m_V^2\leq\frac{6m^2}{q^2e^2-\frac{m^2}{2\mpl^2}} \,.
\end{equation}
The upper bound is controlled by the degree of de-saturation of Eq.~\eqref{eq:charge_tuning}. Consistently, achieving $m_V \gg m$ and integrating out the vector requires returning to saturation of the weak gravity bound, $q^2 e^2 \to \frac{m^2}{2\mpl^2}$.

\subsubsection*{Massless photon and massive $V_{+,-}$ pseudovector:}\label{sec:axialpseudovector}

\par Another extension of the EFT via spin-$1$ particles is to include, in addition to the massless photon, a neutral pseudovector ($V_{+,-}$) with same charge-conjugation transformation of a photon, but opposite parity. The three-point amplitudes ar given in Eq.~\eqref{eq:3ptaxialpseudovector}.

\par Applying positivity bounds we find a picture qualitatively similar to the previous case, and the same condition Eq.~\eqref{eq:chargetuning_weak} is required. The ratio $qe/m$ can be detuned from the WGC lower bound, but only of order $\mathcal{O}(\mpl^{-1})$. Moreover, we obtain the same bound on the vector mass
\begin{equation}\label{eq:pseudovector_massbound}
    m_V^2\leq\frac{6m^2}{q^2e^2-\frac{m^2}{2\mpl^2}} \,.
\end{equation}

\subsubsection*{Spin-0 and the other spin-1 states:}

The last modification of the EFT that we consider is to include scalars and the remaining massive vectors, either neutral ($S_{\pm,+}$ and $V_{\pm,+}$) or with charge $\pm 2q$, or both. The neutral vectors have opposite charge-conjugation phase compared to a photon. 
The on-shell three-point amplitudes are given in Eqs.~\eqref{eq:diracneutralscalar},~\eqref{eq:diracneutralvectorgb},~\eqref{eq:3ptdiracchargedscalars} and~\eqref{eq:3ptdiracchargedvector}.
\par Imposing consistency conditions as in the previous sections, we eventually find Eq.~\eqref{eq:tuningdirac_simp} to still hold. In addition, as in the Majorana case, the scalar and vector contributions alone at order $E^4$ violate positivity of the elastic Arcs---see later Eqs.~\eqref{eq:boundsdirac01nc} and~\eqref{eq:scalarvectbound_dirac}---, hence the presence of gravitons and photons is still necessary. Accordingly, at this order the scalar and vector couplings are either required to vanish (when coupled mostly to the transverse modes) or bounded by $1/F^2=1/(3m^2\mpl^2)$.

\subsection{Gyromagnetic factor and causality of spin-3/2 particles}\label{sec:g-2_causality}
\par The results of the previous sections show that consistency with causality requires a very specific coupling of the spin-$3/2$ particle to the photon. In minimal cases, this translates into the requirement $g = 2$ for the gyromagnetic factor. The naturalness of this value has been widely discussed in the literature, mainly based on two different lines of reasoning.
\par A first class of arguments can be collectively described as based on \textit{best high-energy behaviour}. In a seminal work, Weinberg \cite{Deser:1970spa} showed via dispersive arguments on soft photon amplitudes that, assuming good high-energy behaviour \footnote{In particular, assuming that the amplitude admits a once-subtracted dispersive representation.} for particles of any spin $J$ the deviation $g - 2$ is proportional to an integral over the amplitude's discontinuity. In a theory weakly coupled all the way to the UV, this implies $g \approx 2$. Further works, starting from \cite{Ferrara:1992yc} for general spin and later \cite{Deser:2000dz} specifically for spin-$3/2$, argued that $g = 2$ is necessary to ensure the best high-energy behaviour of the photon Compton amplitude. While in a Lagrangian setting this choice corresponds to a non-trivial tuning of the coupling to the $U(1)$ current (which in itself would give $g = 1/J$) and of non-minimal higher-derivative operators, on-shell methods developed in recent years naturally organize interactions according to their high-energy behaviour \cite{Arkani-Hamed:2017jhn,Chung:2018kqs,Chiodaroli:2021eug,Ema:2025qgd}.
\par A second more specific line of reasoning comes from attempts to address the causality problems of spin-$3/2$ electrodynamics \cite{Velo:1969bt}. The works \cite{Ferrara:1992yc,Deser:2000dz} and later \cite{Deser:2001dt,Porrati:2009bs,Benakli:2024aes} point to $g = 2$ as the preferred value ensuring causal propagation on external fixed electromagnetic backgrounds without introducing spurious degrees of freedom. These analyses focus on pure electrodynamics without gravity (except for \cite{Deser:2001dt}). 

The results of the previous sections show that, when causality is imposed in the form of positivity bounds, gravity becomes necessary for the theory’s consistency, under the assumptions of weak coupling and large scale separation. Nonetheless, $g = 2$ arises in minimal solutions to the causality constraints and deviations are allowed only at the level $g - 2 = \mathcal{O}(m/\mpl)$.

\section{The Goldstino EFT-hedron}\label{sec:GoldstinoEFThedron}

In this section we turn our investigations to the longitudinal $\lambda = \pm1/2$ polarizations of the spin-3/2 state in the limit $\mpl\to\infty$, $m\to 0$, and $F=\sqrt{3}m \mpl$ held fixed, after all the tunings discussed in previous sections, needed to ensure $E^4$ scaling as the dominant term in the amplitudes,  have been performed.
The resulting theory describes the EFT of a spin-$1/2$ fermion, the \textit{Goldstino}, associated with (rigid) supersymmetry-breaking.

This section is constructed to be self-contained, allowing readers interested only in spin-$1/2$ positivity bounds---for Goldstinos or other massless fermions---to follow without needing to refer to earlier sections. Accordingly, the notation is adapted to the more specific case under study and made lighter (\textit{e.g.},  using the all-incoming momenta convention and  spinor-helicity variables), while the full power of positivity constraints at finite transverse momentum will be deployed. 

We consider  the case where the Goldstino is of Majorana type and uncharged under any global symmetry. We review the $2 \to 2$ scattering amplitudes involving external Goldstinos and construct a novel class of $t-u$ symmetric dispersion relations. These allow for an elegant organization of the ultraviolet constraints imposed by positivity and crossing symmetry. We then explore how these constraints bound the space of Wilson coefficients---defining what we refer to as the \textit{Goldstino EFT-hedron}---and identify particularly interesting (extremal) regions of this space. In doing so, we recover known scenarios of  supersymmetry-breaking but also find new UV models.  

\subsection{Goldstino Scattering Amplitude}

The four Goldstino scattering amplitude  (all incoming) 
can be written using spinor-helicity variables as
\be
\M(1^+2^+3^-4^-) = [12] \langle 34 \rangle f(s,t)\, ,
\label{ampdef}
\ee
where the super-indices $\pm$ refer to the helicity of the fermions.
The little group factor $[12] \langle 34 \rangle$ reduces to $-s$ in the center-of-mass frame, and $f(s,t)$ is a Lorentz-invariant scalar function of the Mandelstam variables $s$ and $t$. 

Crossing symmetry and fermion statistics  requires invariance under $t \leftrightarrow u$ exchange,
\be
 f(s,t) = f(s,u)\, ,
\ee
and allows us to generate other amplitudes from the one above. For example,
\begin{eqnarray}
\M(1^+2^-3^-4^+) &=& [14] \langle 23 \rangle f(u,t)\, ,\\
\M(1^+2^-3^+4^-) &=& [13] \langle 42 \rangle f(t,s)\, \label{inamp},
\end{eqnarray}
which correspond to elastic and inelastic processes, respectively.

With no other light states in the EFT, the scalar function's discontinuities in the IR arise entirely from multi-fermion intermediate exchanges, \textit{i.e.}, loops. Therefore, whenever scattering amplitudes are well-approximated by tree-level physics in the infrared, $|s|, |t| \ll \Lambda^2$, the scalar function is analytic and admits an expansion in terms of an infinite set of Wilson coefficients $g_{n,k}$  
\be
f(s,t) = g_{0,0} + g_{1,0}s + g_{2,0}s^2 + g_{2,1}tu + g_{3,0}s^3 + g_{3,1}stu + \cdots = \sum_{n \geq 2k}^{\infty} g_{n,k} s^{n - 2k} (tu)^k\,,
\label{iramp}
\ee
where $s + t + u = 0$ for massless external states.

In the case of Goldstinos, there is an additional constraint: $g_{0,0} = 0$. However, this does not affect the tree-level positivity bounds derived in what follows, except that—by dimensional analysis—the coefficients $g_{1,0}$, $g_{2,0}$, and $g_{2,1}$ do not run, as $g_{0,0} = 0$. That is, any non-analyticity from loops starts at $\mathcal{O}(s^3)$,  rendering  the tree-level positivity bounds on $g_{1,0}$, $g_{2,0}$, and $g_{2,1}$ more robust to (part of the effect of) loop corrections \cite{Bellazzini:2020cot, Bellazzini:2021oaj, Beadle:2024hqg,Chang:2025cxc}.

\subsection{$t$--$u$ symmetric dispersion relations}

It has become clear in recent years that $s$--$t$--$u$ crossing-symmetric dispersion relations for scattering of identical particles (see \textit{e.g.}~\cite{Sinha:2020win, Li:2023qzs, Berman:2024kdh}) are more manageable and stable under loop corrections than fixed-$t$ dispersions \cite{Chang:2025cxc, Beadle:2025cdx}. This is because the extraction of \textit{null constraints}—relations on amplitude discontinuities due to crossing symmetry—is systematic, clean, and does not rely on specific kinematical configurations such as $t=0$, which may be singular for massless particles \textit{e.g.} when loops are considered \cite{Bellazzini:2021oaj, Beadle:2024hqg, Beadle:2025cdx,Peng:2025klv}. 
 
We now develop a novel class of $t$--$u$ crossing-symmetric dispersion relations, tailored for distinguishable particles, as is the case here where particles and antiparticles carry different helicities. We will see that this approach yields a new family of UV constraints, which can be interpreted as generating functionals for all the null constraints that might otherwise be extracted at fixed $t$.

For $t$--$u$ symmetric amplitudes, a natural choice is to define the variable
\be
p^2 = \frac{tu}{s}\,,
\ee
so that the low-energy amplitude given in Eq.~\eqref{iramp} becomes
\be
f(s,p^2)= \sum_{n\geq2k}^\infty g_{n,k} s^{n-k}p^{2k}= \sum_{a\geq k}^\infty g_{k+a,k} s^{a}p^{2k}\, .
\ee 
Below we compute $s=0$ residues of the subtracted amplitude $f(s,p^2)/s^m$; only a finite number of terms with $k \leq m - 1$ contribute. \\

To implement the dispersion relations, we must first derive the inverse transformations $t(s,p^2)$ and $u(s,p^2)$, which are not injective (\textit{i.e.}, they are multivalued):\footnote{Notice that this inverse transformation has two branch points (at $p^2 = s/4$ and $p^2 = \infty$), and a branch cut running between them. It is thus defined on a two-sheeted Riemann surface.}
\be
t_{\pm}(s,p^2)=-\frac{s}{2} \pm \frac{s}{2} \sqrt{1-4p^2/s}  \,.
\ee
If we fix $p^2 \geq 0$, then for all real values of $s$, we find that $t_+ < 0$. The sign of $t_-$ depends on the sign of $s$: it is positive when $s < 0$ and negative when $s > 0$. 
We adopt the convention in which we identify
\be
t_+(s,p^2) = t(s,p^2)\ \ ;\ \  t_-(s,p^2) = u(s,p^2) \,,
\label{prescription}
\ee
although, due to the amplitude’s $t$--$u$ symmetry, the opposite choice would lead to equivalent results.

In this prescription, when taking the limit $|s| \to +\infty$ at $p$ fixed, we find
\be
\lim_{|s| \to +\infty} t(s,p^2) \sim -p^2 \quad \text{and} \quad \lim_{|s| \to +\infty} u(s,p^2) \sim -s + p^2 \,.
\label{regscal}
\ee
Finally, in order for $t(s,p^2)$ and $u(s,p^2)$ to remain real, we must restrict to the region $0 \leq p^2 \leq M_s^2/4$, where $M_s^2$ is the mass squared of the lightest particle exchanged in the $s$-channel.\\

\begin{figure}[t]
\begin{center}
\includegraphics[width=0.85\textwidth]{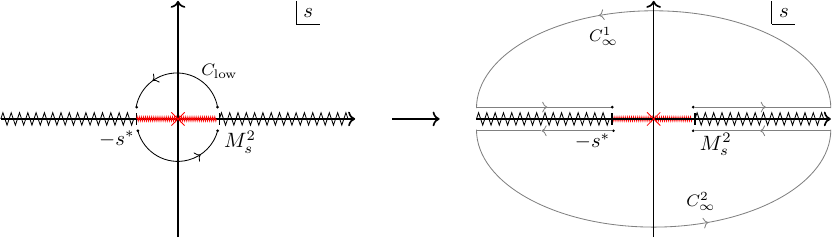}
\end{center}
  \caption{Scheme of the contour integrals appearing in the definition of the Arcs used in this section, and of the analytic structure of the integrated function.}
  \label{arcdef}
\end{figure}

We are now ready to perform the dispersion relations in the $s$-plane at fixed $p^2 \geq 0$. Following the contour depicted in Fig.~\ref{arcdef}, we define the Arc $\A_n(p^2)$ as the contour integral
\be
\A_n(p^2) = \int_{C_\text{low}}  \frac{\M(1^+2^+3^-4^-)}{[12]\langle 34\rangle}\frac{1}{s^{n}}\frac{ds}{s} = \int_{C_\text{low}}  \frac{f\big(s,t(s,p^2)\big)}{s^{n}}\frac{ds}{s}   \, .
\label{arcir}
\ee
Due to the IR discontinuity, $\A_n(p^2)$ generally depends on the Arc radius even when the Mandelstam variables lie within the EFT’s regime of validity. However, within the tree-level approximation---where $f$ is treated as analytic---the radius can be extended all the way to the first UV discontinuity.

As shown in Fig.~\ref{arcdef}, we can deform the integration contour $C_\text{low}$---using Cauchy's theorem---into the sum of two integrals running along the discontinuities on the real $s$-axis, and two arcs at infinity, $C_\infty^{1}$ and $C_\infty^{2}$. Assuming further that
\begin{equation}
\lim_{|s|\to 0}\frac{\M(1^+2^+3^-4^-)}{s^2}\to 0\,,
\end{equation}
we can  drop the contributions from $C_\infty^{1}$ and $C_\infty^{2}$ for  $n \geq 1$. Therefore, the Arc becomes equal to
\begin{align}
\A_n(p^2) 
=& -\frac{1}{2\pi i}\int_{M_s^2}^\infty \frac{\text{Disc} \, \mathcal{M}^{++--}\big(s+i\epsilon,t(s,p^2)\big)}{s^{n+2}}\,ds  -\frac{1}{2\pi i}\int^{s^*}_{-\infty} \frac{\text{Disc} \, \mathcal{M}^{++--}\big(s+i\epsilon,t(s,p^2)\big)}{s^{n+2}}\,ds \nonumber \\
=& -\frac{1}{2\pi i}\int_{M_s^2}^\infty \frac{\text{Disc} \, \mathcal{M}^{++--}\big(s+i\epsilon,t(s,p^2)\big)}{s^{n+2}}\,ds   \nonumber \\ 
& -\frac{1}{2\pi i}\int^{s^*}_{-\infty} \frac{\text{Disc} \, \mathcal{M}^{+--+}\big(-s-i\epsilon - t(s,p^2), t(s,p^2)\big)}{s^{n+2}}\,ds\, ,
\label{arcuv}
\end{align}
where the second equality follows from  crossing symmetry  and we have defined ${\cal M}^{\lambda_1 \lambda_2 \lambda_3 \lambda_4 }(s,t)$ the amplitude $\M(1^{\lambda_1} 2^{\lambda_2} 3^{\lambda_3} 4^{\lambda_4})$ in the CoM frame as function of the Mandelstam variables.  

The prescription chosen in Eq.~\eqref{prescription} is such that for positive $p^2$ and negative real $s$, $u(s,p^2)\geq0$, and therefore the second integral in Eq.~\eqref{arcuv} picks up the discontinuity in the $u$-channel. On the other hand, the $t$-channel discontinuity is never picked up, because on the real $s$-axis, $t(s,p^2)\leq 0$ always. We can therefore fix $s^* = -M_u^4/(M_u^2 + p^2)$. Finally, we can perform the change of variables
\be
u\equiv -s-t(s,p^2) \ \rightarrow\ s=-\frac{u^2}{u+p^2} \quad \text{and}\quad  t(s=-\frac{u^2}{u+p^2},p^2)  = -\frac{p^2u}{p^2+u}\equiv t_u(u,p^2) \,,
\ee
in the second integral, yielding 
\begin{align}
\A_n(p^2) 
=& -\frac{1}{2\pi i}\int_{M_s^2}^\infty \frac{\text{Disc} \, \mathcal{M}^{++--}\big(s+i\epsilon,t(s,p^2)\big)}{s^{n+2}}\,ds \nonumber \\
& +\frac{1}{2\pi i}\int_{M_u^2}^\infty \bigg(1-\frac{p^4}{(p^2+u)^2}\bigg) \bigg(\frac{u+p^2}{-u^2}\bigg)^{n+2}\text{Disc} \, \mathcal{M}^{+--+}\big(u+i\epsilon,t_u(u,p^2)\big)\,du \,.
\label{arcuv2}
\end{align}
Notice that both angles
\be
\cos \theta_s = \sqrt{1-4p^2/s} \quad , \quad \cos\theta_{u}=\frac{u-p^2}{u+p^2}\,,
\ee
are physical within their respective regions of integration, and therefore we can expand the discontinuities in partial waves and rewrite the definition of the Arc in a more compact form
\be
-\A_n(p^2) = \bigg\langle \frac{P_J\big(\sqrt{1-4p^2/s}\big)}{s^{n-1}}\bigg \rangle_{++}+(-1)^{n+1}\bigg\langle \frac{\big(1+\frac{p^2}{u}\big)^{n-1} \big(1+\frac{2p^2}{u}\big)P_{J-1}^{(0,2)}\big(\frac{u-p^2}{u+p^2}\big)}{u^{n-1}}\bigg \rangle_{+-}\, .
\label{xsymDR}
\ee
Here we defined the \textit{high energy averages}
\be
\big\langle (\cdots)  \big\rangle_{++} =  \SumInt_{M_s^2\,,J\in\mathrm{even} } (2J+1) \frac{ds}{s^{3}}  \mathrm{Im}\mathcal{M}^{++--}_{J}  (\cdots)\ \ , \ \ 
\big\langle (\cdots)  \big\rangle_{+-} =  \SumInt_{M_u^2\,,J\geq 1 } (2J+1) \frac{du}{u^{3}}  \mathrm{Im}\mathcal{M}^{+--+}_{J}  (\cdots)\,,
\ee
and $P_J$ and $P_{J}^{(a,b)}$ are, respectively, the Legendre and Jacobi polynomials that arise from the ratio of Wigner $d$-functions and the little group factors, and we absorbed the state normalization factor into the imaginary parts.

Notice that helicity selection rules force the first measure $\langle(\cdots)\rangle_{++}$ to project only on even spin states, while the second measure $\langle(\cdots)\rangle_{+-}$ projects on all states with $J\geq1$.

 \subsubsection{Sum rules and null constraints}

Let us display explicitly the implications of the first few subtractions of the crossing symmetric dispersion relation in Eq.~\eqref{xsymDR}. 
%
Working in the weak coupling limit, where the IR cuts depicted in Fig.~\ref{arcdef} can be neglected, we have for $n=1$
\be
-\A_{1}(p^2)=-g_{1,0}-p^2g_{2,1}=  \bigg\langle P_{J}(\sqrt{1-4p^2/s}) \bigg\rangle_{++} +   \bigg\langle  (1+\frac{2p^2}{u}) P^{(0,2)}_{J-1}(\frac{u-p^2}{u+p^2}) \bigg\rangle_{+-}\, .
\label{forNCgenfunc}
\ee
By performing an expansion at small $p^2$, and matching the coefficients at each $p^2$-order  on both sides of this expression,  we   find sum rules for $g_{1,0}$ and $g_{2,1}$ as well as null constraints: 
\begin{align}
g_{1,0} = &   - \big\langle  1 \big\rangle_{++}  -  \big\langle  1 \big\rangle_{+-} \,, \label{c10expression} \\ 
\label{c21expression}
g_{2,1} = &    \big\langle \frac{\mathcal{J}^2}{s} \big\rangle_{++}  +  \big\langle \frac{\mathcal{J}^2-4}{u} \big\rangle_{+-} \,,\qquad \qquad\\
0 =  &  \big\langle \frac{\mathcal{J}^2(\mathcal{J}^2-6)}{s^2}\big\rangle_{++}  + \big\langle \frac{(\mathcal{J}^2-2)(\mathcal{J}^2-10)}{u^2} \big\rangle_{+-} \,,\\ 
0= &  \big\langle \frac{\mathcal{J}^2(\mathcal{J}^2-6)(\mathcal{J}^2-20)}{s^3}\big\rangle_{++}  +  \big\langle \frac{(\mathcal{J}^2-2)(\mathcal{J}^4-18 \mathcal{J}^2+36)}{u^3} \big\rangle_{+-} \,,\\
\vdots &  \nonumber
\end{align} 
where we have defined $\mathcal{J}^2= J(J+1)$.

An alternative way of thinking about these null constraints is via a generating function, which is determined  by  taking the second derivative with respect to $p^2$ of Eq.~\eqref{forNCgenfunc}:
\begin{equation}
0=\mathcal{A}^{\prime\prime}_{1}(p^2) =\frac{d^2}{d(p^2)^2}\left( \bigg\langle P_{J}(\sqrt{1-4p^2/s}) \bigg\rangle_{++} +   \bigg\langle  (1+\frac{2p^2}{u}) P^{(0,2)}_{J-1}(\frac{u-p^2}{u+p^2}) \bigg\rangle_{+-} \right)  \,. 
\end{equation}
This function is identically zero for all values of $p^2$, without having to necessarily expand in the forward limit. 

We can easily increase the number of subtractions. For $n=2$ we have 
\begin{equation}
-\left(g_{2,0}+ p^2 g_{3,1}+ p^4 g_{4,2}\right) =  \bigg\langle  \frac{P_{J}(\sqrt{1-4p^2/s})}{s} \bigg\rangle_{++}  -   \bigg\langle  \frac{(1+\frac{p^2}{u})(1+\frac{2p^2}{u}) P^{(0,2)}_{J-1}\big(\frac{u-p^2}{u+p^2}\big) }{u}
 \bigg\rangle_{+-}\, ,
\end{equation}
from which another set of sum rules and null constraints is obtained:
\begin{align}
g_{2,0}= &  - \big\langle\frac{1}{s} \big\rangle_{++}  +    \big\langle\frac{1}{u} \big\rangle_{+-} \,,\label{eq:c20dispersive} \\ 
\label{eq:c31dispersive}
g_{3,1}= &   \big\langle\frac{\mathcal{J}^2}{s^2} \big\rangle_{++}  -    \big\langle\frac{\mathcal{J}^2-5}{u^2} \big\rangle_{+-} \,, \\ 
g_{4,2}=&  - \big\langle\frac{(\mathcal{J}^2-6)\mathcal{J}^2}{4s^3} \big\rangle_{++}  +    \big\langle\frac{\mathcal{J}^4-16\mathcal{J}^2+36}{4u^3} \big\rangle_{+-} \,, \\
0 =&  \big\langle \frac{\mathcal{J}^2(\mathcal{J}^2-6)(\mathcal{J}^2-20)}{s^4}\big\rangle_{++}   -  \big\langle \frac{(\mathcal{J}^2-2)(\mathcal{J}^2-6)(\mathcal{J}^2-21)}{u^4}\big\rangle_{+-} \,,\\
 0 =&  \big\langle \frac{\mathcal{J}^2(\mathcal{J}^2-6)(\mathcal{J}^2-20)(\mathcal{J}^2-42)}{s^5}\big\rangle_{++}   -  \big\langle \frac{(\mathcal{J}^2-2)(\mathcal{J}^2-6)(\mathcal{J}^4-32\mathcal{J}^2+96)}{u^5}\big\rangle_{+-} \,,\\
& \vdots \nonumber
\end{align}
Finally, we can define the generating function for any subtraction
\begin{equation}
0= \mathcal{A}^{(n+1)}_{n}(p^2) = -\frac{d^{n+1}}{d(p^2)^{n+1}} \bigg(\bigg\langle \frac{P_J\big(\sqrt{1-4p^2/s}\big)}{s^{n-1}}\bigg \rangle_{++}+(-1)^{n+1}\bigg\langle \frac{\big(1+\frac{p^2}{u}\big)^{n-1} \big(1+\frac{2p^2}{u}\big)P_{J-1}^{(0,2)}\big(\frac{u-p^2}{u+p^2}\big)}{u^{n-1}}\bigg \rangle_{+-}   \bigg) \,,
\end{equation}
valid $\forall p^2$ in the interval $0 \leq p^2\leq M_s^2/4 $. 

\subsubsection{Inelastic Constraints}
There is another class of null constraints which are not captured in the description above. These are constructed doing dispersion relations (at fixed-$t$) of inelastic amplitudes, such as the one defined in Eq.~\eqref{inamp}. Following a similar logic to the one depicted in Fig.~\ref{arcdef}, we define the inelastic Arc
\begin{align}
\mathcal{A}^{+-+-}_{n}(t)&= \frac{1}{2\pi i} \oint \frac{ds}{s^n}\frac{1}{su}\frac{\mathcal{M}(1^{+} 2^{-} 3^{+} 4^{-})}{ [13] \langle 4 2 \rangle}= \\
&=  \frac{1}{2\pi i}\frac{1}{t} \int_{M^2}^\infty \frac{\text{Disc}\M^{+-+-}(s+i\epsilon,t)}{s^ns(s+t)}ds+
\frac{1}{2\pi i}\frac{1}{t} \int_{-\infty}^{-M^2-t} \frac{\text{Disc}\M^{+-+-}(s+i\epsilon,t)}{s^ns(s+t)}ds\, ,\nonumber
\end{align}
where for $n\geq1$ one can pass from the first to the second equation (under the assumption that the amplitude decays according to Froissart bound) disregarding the contours at infinity. Then, via a change of variables and using crossing symmetry in the form of the identity $\M(1^+2^-3^+4^-)(s+i\epsilon,t)=\M(1^+2^-3^+4^-)(-s-t-i\epsilon,t)$, we find  
\be
\mathcal{A}^{+-+-}_{n}(t)=  \frac{1}{2\pi i}\frac{1}{t} \int_{M^2}^\infty \frac{ds}{s(s+t)}\bigg(\frac{1}{s^n}+\frac{(-1)^{n-1}}{(s+t)^n}\bigg)\text{Disc}\M^{+-+-}(s+i\epsilon,t)\, .
\ee
The inelastic discontinuity can be turned into an elastic one using one final crossing identity
\be
\mathcal{M}(1^+\, 2^-\, 3^+\, 4^-)(s+i\epsilon, t) = \mathcal{M}(1^+\, 2^-\, 3^-\, 4^+)(s+i\epsilon, -s-t-i\epsilon)\, .
\ee
In the integration region we have $s>0$, $t<0$ fixed (and small) and $u=-s-t<0$,  the discontinuity can be expanded in partial waves with angular dependence
 \be
d_{1,1}^J( \cos \theta_u) = -t/s (-1)^{J-1} P_{J-1}^{(2,0)
}(1+2t/s)\xrightarrow[t\to0]{}(-t/s)(-1)^{J-1}\left[\frac{\mathcal{J}^2}{2}+\frac{t}{s}\frac{\mathcal{J}^2(\mathcal{J}^2-2)}{6}+\ldots \right]\, ,
 \ee
where  $t\to0$ corresponds to the backwards limit.
Plugging everything back in the inelastic Arc we get expressions of the form
 \begin{align}
\mathcal{A}^{+-+-}_{n=1}(t) & = -2\left\langle \frac{\mathcal{J}^2}{u} (-1)^{J-1} \right\rangle_{+-} - 2t \left\langle \frac{\mathcal{J}^2(2\mathcal{J}^2 - 13)}{6u^2} (-1)^{J-1} \right\rangle_{+-} + \ldots \nonumber \,,\\
\mathcal{A}^{+-+-}_{n=2}(t) & = -2t \left\langle \frac{\mathcal{J}^2}{2u^3} (-1)^{J-1} \right\rangle_{+-} + \ldots \,,\\
&\ \ \vdots \nonumber
\end{align}

The Arcs can also be calculated in the IR. Since we are not fixing a crossing symmetric variable, infinite Wilson coefficients will be projected out. Nonetheless, by matching the $t\to0$ expansion in the IR with the UV relations given above, we are still able to derive new sum rules, the first few being
\begin{align}
g_{2,1}  = &   -\big\langle \frac{\mathcal{J}^2}{u} (-1)^{J-1} \big\rangle_{+-} \nonumber \,,\\ 
g_{3,1}  = &  - \big\langle \frac{\mathcal{J}^2(2\mathcal{J}^2-13)}{6u^2} (-1)^{J-1} \big\rangle_{+-} \,,\\ 
   \vdots & \nonumber
\end{align}

We can equate these sum rules with those derived via crossing-symmetric dispersion relations, \textit{e.g.} Eq.~\eqref{c21expression} and Eq.~\eqref{eq:c31dispersive}, obtaining expressions such as
\begin{align}
 \big\langle\frac{\mathcal{J}^2}{s} \big\rangle_{++} &=  - \big\langle \frac{((-1)^{J-1}+1)\mathcal{J}^2-4}{u} \big\rangle_{+-}\,, \label{ncanalysis1} \\
 \big\langle\frac{\mathcal{J}^2}{s^2} \big\rangle_{++} &=   \big\langle \frac{\mathcal{J}^2-5}{u^2} +\frac{ (-1)^{J} \mathcal{J}^2(2\mathcal{J}^2-13)}{6u^2} \big\rangle_{+-}\, ,\\
&\ \nonumber  \vdots
\end{align}
yielding a new \textit{infinite} set of null constraints.

\subsection{General implications of Null Constraints}

By  looking at the null constraints it is possible to obtain information on the UV completions of a theory of Goldstinos. Let us consider Eq.~\eqref{ncanalysis1} as an example. If we make the sum over spins explicit, the expression becomes\footnote{Let us remind the reader that because of selection rules the $++$ sector is populated by even spin particles with $J\geq0$, while in the $+-$ sector all spins enter with $J\geq 1$.}
\be
\label{sumrule}
 \big\langle\frac{6}{s} \big\rangle_{++}^{J=2}+ \big\langle\frac{20}{s} \big\rangle_{++}^{J=4}+ \big\langle\frac{42}{s} \big\rangle_{++}^{J=6}+\cdots =  \big\langle\frac{4}{s} \big\rangle_{+-}^{J=2}- \big\langle\frac{20}{s} \big\rangle_{+-}^{J=3}+ \big\langle\frac{4}{s} \big\rangle_{+-}^{J=4}+\cdots\, , 
\ee
where $ \big\langle(...) \big\rangle_{++}^{J=j}$ refers to the contribution of spin $j$ states to the high energy integral (and the dots refer to higher spins states).
Three properties are immediately noticeable (and remain true for all null constraints):
\begin{enumerate}
\item $J=0$ states coupled to same-helicity Goldstinos ($++$) do not enter in the null constraints. 
\item $J=1$ states coupled to opposite-helicity Goldstinos ($+-$) do not enter in the null constraints.
\item If all $\big\langle...\big\rangle_{+-}$  are set to zero (e.g., no states couple to a pair of Goldstinos with helicity  $+-$), \eqref{sumrule} becomes $ \big\langle\frac{6}{s} \big\rangle_{++}^{J=2}+ \big\langle\frac{20}{s} \big\rangle_{++}^{J=4}+ \big\langle\frac{42}{s} \big\rangle_{++}^{J=6}+\cdots =0$, and since all $++$ contributions have the same sign, there is no way to satisfy the constraint unless the theory is trivial.
\end{enumerate}
These observations tell  us the following. Firstly,
we can have amplitudes mediated only by  $J=0$ or $J=1$ states  without the need of other states (we will provide these amplitudes in the next section). 
Secondly, amplitudes mediated by higher-spin states coupled to a pair of Goldstinos with helicity $++$, require also states coupled to a  $+-$  Goldstino pair. The converse is not true: all null constraints can be satisfied solely with states coupled to a $+-$ Goldstino pair.  
Looking at the full set of constraints, it becomes clear that as soon as a $J>1$ state enters into amplitude, states with all $J$  values must be present. 


\subsection{Bounds on Wilson Coefficients and possible UV Models}
Armed with the sum rules and the null constraints derived in the previous section, we are ready to bound the  Wilson coefficients and find the \textit{Goldstino EFT-hedron}. 

We first illustrate how these bounds work with a simple example. Let us recall the sum rules for the lowest order Wilson coefficients
\begin{equation}
    \begin{split}
        g_{1,0}&=-\langle 1\rangle_{++}-\langle 1 \rangle_{+-} \,,\\
        g_{2,0}&=-\langle 1/s \rangle_{++}+\langle 1/u \rangle_{+-} \,,
    \end{split}
\end{equation}
respectively from Eq.~\eqref{c10expression} and Eq.~\eqref{eq:c20dispersive}. Defining the dimensionless quantities
\be
\tilde{g}_{n,l}=\frac{g_{n,l}}{g_{1,0}}M^{2(n-1)},
\ee
where $M^2=\text{min}\{M_s^2,M_u^2\}$, one can easily derive the bound 
\be
-1 \leq  \tilde{g}_{2,0} \leq 1\, .
\ee

\begin{figure}[t]
\begin{centering}
    \includegraphics[width=0.8\textwidth]{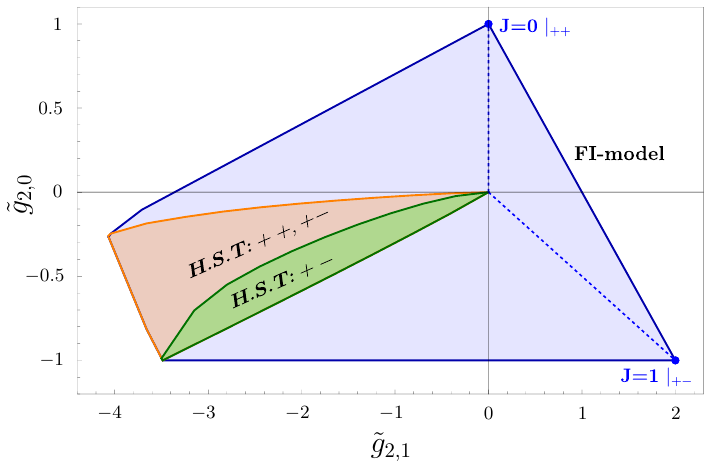}
    \caption{Allowed regions for the normalised Wilson coefficients $\tilde{g}_{2,0}$ and $\tilde{g}_{2,1}$. In blue the total allowed parameter space. In orange the allowed space assuming only higher-spin states ($J\geq2$) coupled to a pair of Goldstino with helicity  $++$ and $+-$, while in green higher-spin states coupled only to  $+-$.  The two kinks on the RHS of the plot corresponds to  spin-0 and spin-1 theories.}
    \label{posbound1}
    \end{centering}
\end{figure}

In general, semidefinite programming provides a systematic way of obtaining numerical bounds from the full set of sum rules and null constraints, and can be implemented using \texttt{SDPB} \cite{Simmons-Duffin:2015qma}\footnote{There are many references which give an overview of the numerical implementation. For further details regarding similar setups see for example \cite{Caron-Huot:2020cmc, Albert:2022oes,Fernandez:2022kzi}.} .
We show the results in Fig.~\ref{posbound1}  for the  normalized Wilson coefficients $\tilde{g}_{2,0}$ and $\tilde{g}_{2,1}$. 

The top and bottom kinks on the RHS of  Fig.~\ref{posbound1} can be identified with  the two  theories described above: a $J=0$ state with $++$ coupling and a $J=1$ state with $+-$ coupling. The Wilson coefficients for these theories can be directly read from the sum rules in Eq.~\eqref{c10expression}, Eq.~\eqref{c21expression} and Eq.~\eqref{eq:c20dispersive}, substituting in the UV the $J=0$ or $J=1$ state. We can  write down explicitly the amplitudes: 
\bea
\mathcal{M}^{J=0}\big|_{++}(s|t,u) &=& -\frac{g_{++}^2}{m^2_S}\frac{s^2}{s-m_S^2}\, , \\
\mathcal{M}^{J=1}\big|_{+-}(s|t,u) &=& -g_{+-}^2\frac{s}{m_V^2}\bigg(\frac{u}{u-m_V^2}  +\frac{t}{t-m_V^2}\bigg)\,.
\eea
These amplitudes clearly satisfy the criteria of positivity (partial wave decomposition manifests a single resonance with fixed positive coupling) and satisfy Froissart bound. When expanded at low energies, 
they predict $\tilde{g}_{2,0}$ and $\tilde{g}_{2,1}$ corresponding to   the two kinks in the RHS of Fig.~\ref{posbound1}.

These  two kinks corresponds to well-known models of supersymmetry-breaking. For example the upper kink corresponds to  UV theories where supersymmetry is broken by scalars such as the O'Raifeartaigh models \cite{ORaifeartaigh:1975nky}.
On the other hand,   the Fayet-Iliopoulos (FI) model \cite{Fayet:1974jb}
 lies in the blue line of  Fig.~\ref{posbound1}. joining the two kinks in the RHS.
This model corresponds to a supersymmetric $U(1)$ model with a pair of chiral supermultiplets $\Phi=(\psi,\phi)$ and $\bar\Phi=(\bar\psi,\bar\phi)$ of charge $+1$ and $-1$ respectively. The superpotential is given by $W=m\Phi\bar\Phi$, that leads to a potential for the scalars:
\be
V(\phi,\bar\phi)=m^2 (|\phi|^2+|\bar\phi|^2)+\frac{g^2}{2}(|\phi|^2-|\bar\phi|^2+\xi)^2\,,
\ee
where $\xi$ is the FI-term. 
For $m^2<g^2\xi$ the potential leads to $\langle\bar\phi\rangle\not=0$  breaking supersymmetry and the $U(1)$.
The massless Goldstino is a mixture of  $\psi$ and the gaugino,  and the four-Goldstino amplitude
is mediated by the scalar $\phi$ and the gauge boson  $V$. This model  predicts 
$\tilde{g}_{2,0}$ and $\tilde{g}_{2,1}$ 
lying  on the blue line of  Fig.~\ref{posbound1}  joining the two RHS kinks.
The two kinks corresponds to the two limits:
\begin{enumerate}
\item $m^2\to g^2\xi$ ($\langle \bar\phi\rangle\to0$): In this limit the mass of $V$ becomes much lighter than that of $\phi$  and the Goldstino  is mostly the gaugino (supersymmetry-breaking is dominated by the $D$-term). 
The four-Goldstino amplitude is dominated  by the vector exchange, corresponding to the lower RHS kink of  Fig.~\ref{posbound1}. 
\item $m^2\to0$:  
In this limit the mass of $\phi$  goes to zero and the Goldstino is mostly $\psi$ ($F_\phi$-term supersymmetry-breaking).
The four-Goldstino amplitude is dominated  by the $\phi$ exchange, corresponding to the upper  kink on the RHS of  Fig.~\ref{posbound1}. 
\end{enumerate}

\subsubsection*{Higher-Spin UV Models:}
\begin{figure}[t]
\begin{centering}
    \includegraphics[width=0.8\textwidth]{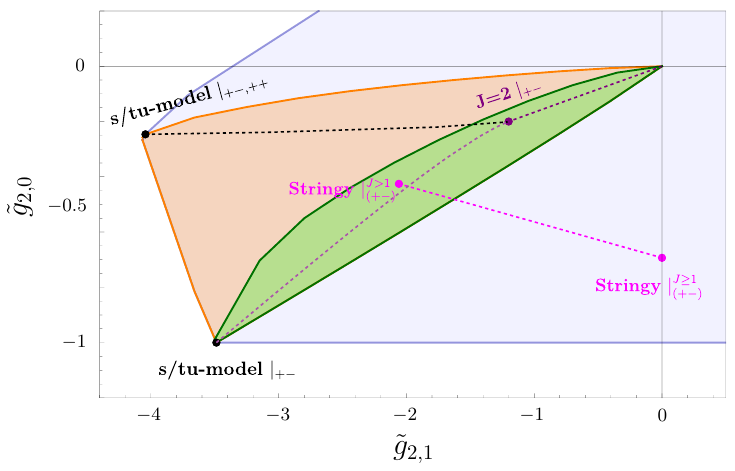}
    \caption{Zooming in the higher-spin UV model  portion of Fig.~\ref{posbound1}. The models depicted in the figure are discussed in the text.}
    \label{posbound2}
    \end{centering}
\end{figure}

We now turn our attention to the region of the plot associated with higher-spin UV models which lead to Goldstino amplitudes that fulfill all our assumptions about causality and unitarity.~\footnote{We refer to these constructions as “models” rather than “UV completions” because additional causality and unitarity constraints may arise when the extra states appear on external legs in scattering processes.} These solutions can be obtained numerically by excluding states with spin $J=0,1$  from the UV spectrum. The $+-$ sector, appearing from the exchange of states in the $t$ or $u$ channels, is always required. In contrast, the $++$ sector, corresponding to the exchange of states in the $s$-channel, can be either included or excluded. Including it results in the orange region, while excluding it yields the green region in Fig.~\ref{posbound1} (the latter being, correctly, contained in the first). 
In Fig.~\ref{posbound2}  we zoom in this region to highlight some interesting UV models.

The two kinks on the LHS of Fig.~\ref{posbound2} and the orange line that joins them correspond to  models where all higher-spin states  have equal  mass $M$. These type of models have appeared as extremal amplitudes in other setups \cite{Caron-Huot:2020cmc,Albert:2022oes,Fernandez:2022kzi}, we will call them $s/tu$-models. Enforcing the  $t$--$u$ crossing symmetry,  these amplitudes take the general form 
\be
\mathcal{M}^{s/tu-\text{model}}(s|t,u) = \frac{-s}{M^2} \bigg(\frac{M^4s + \alpha M^2 s^2 + \beta M^2 t u +\gamma s t u}{(s-M^2)(t-M^2)(u-M^2)}\bigg)\, ,
\ee
where $(\alpha,\beta,\gamma)$ are free parameters. We can fix two of these parameters by imposing that the $J=0$ $(++)$ state and the $J=1$ $(+-)$ states vanish.\footnote{These values are $\alpha= \frac{129+\log 2 (-319+336 \log 2)}{2466+\log 2(-6785+4656 \log 2)}$ and $\beta= \frac{1138+18\log 2 (-257+240 \log 2)}{2466+\log 2(-6785+4656 \log 2)}$.} This leaves us with an amplitude that depends on a single remaining parameter, which is bounded from both above and below to ensure positivity of all higher-spin residues. By varying this parameter within its allowed range, we can interpolate between the two  kinks shown in Fig.~\ref{posbound2}, moving along the orange line connecting them. The top left kink corresponds to the $s/tu$-model where the $s$-channel poles have the largest residue possible, while the bottom left kink corresponds to an amplitude without  $s$-channel poles.\footnote{The values of $\gamma$ corresponding to these two amplitudes are respectively
\be
\gamma=-\frac{16 (-2 + 3 \log 2 )}{-17 + 24 \log 2}\sim 3.48\quad \text{and}\quad \gamma= \frac{-1131+16\log 2 (181-114 \log 2)}{2466+\log 2(-6785+4656 \log 2)}\sim -0.55\, . \nonumber
\ee}

There is a way of deforming the $s/tu$-models above while maintaining $t$--$u$ symmetry, which corresponds to assuming that the $s$-channel states have mass $M_2$, while the $t/u$-channel states are at mass $M$. The amplitude becomes 
\be
\mathcal{M}_{M\neq M_2}^{s/tu-\text{model}}(s|t,u) = \frac{-s}{M_2^2} \bigg(\frac{M_2^4s + \alpha M_2^2 s^2 + \beta M_2^2 t u +\gamma s t u}{(s-M_2^2)(t-M^2)(u-M^2)}\bigg)\, .
\label{stumodel2}
\ee
As long as $M<M_2$, the residues of the poles  are positive (once again within some regime of the parameters $\alpha,\beta,\gamma$).
 We are therefore allowed to push $M_2$ larger and larger, effectively decoupling the $(++)$ states. Starting with an amplitude where $\alpha,\beta,\gamma$ are tuned to be at the top left corner (where the contribution of the $(++)$ states is maximised) and gradually increasing $M_2$, we move along the dashed black line in Fig.~\ref{posbound2}.\footnote{Note that the parameters $\alpha,\beta,\gamma$ will depend on the mass gap, and therefore will have to be adjusted as one moves along the dashed black line. } In the limit where $M_2/M\gg1$ (it can be chosen arbitrarily large) the amplitude \eqref{stumodel2} can be rewritten as
\be
\mathcal{M}_{M\ll M_2}^{s/tu-\text{model}}(s|t,u) \to \frac{-s}{M^2}\bigg(\frac{s-3t}{u-M^2} +\frac{s-3u}{t-M^2}\bigg)\frac{M_2^2}{M_2^2-s}\, ,
\ee
which corresponds to an amplitude with a $J=2$ state of mass $M$  in the $t$ and $u$ channel, with $(+-)$ coupling, and a tower of higher-spin states exchanged in the $s$-channel at $M_2\gg M$ with ($++$) coupling. This theory corresponds to the purple dot in Fig.~\ref{posbound2}. This point in parameter space can be found also from another equivalent amplitude, where the $(++)$ states do not enter at all, that is
\be
\mathcal{M}^{J=2}\big|_{+-}(s|t,u) = \frac{-s}{M^2}\bigg(\frac{s-3t}{u-M^2} \frac{M_2^2}{M_2^2-t}+\frac{s-3u}{t-M^2}\frac{M_2^2}{M_2^2-u}\bigg)\, .
\label{eqref:j2}
\ee
In this amplitude the states with mass $M_2$  also have $(+-)$ coupling. The necessity of additional higher-spin states at some scale (much larger than $M$)  is mandatory to make an  amplitude mediated by a $J=2$ state consistent with the Froissart bound. Without these extra states the amplitude's behavior at large s would be $\propto s^2$, therefore violating the Froissart bound. The presence of these additional states restores the correct large-$s$ behavior. Since a theory with an intermediate $J=2$  state alone does not satisfy the constraints, the purple dot in Fig.~\ref{posbound2} must be an extremal point of the green area. With the present number of constraints $\mathcal{O}(15)$ and due to the slow convergence this expectation is not yet fulfilled.
\

Let us emphasize again that the decoupling limit of the $s/tu$-model in Eq.~\eqref{stumodel2} is consistent only if we decouple the $s$-channel poles, aka the $(++)$ states. This was expected since, as we discussed previously, $(++)$ states alone do not consistently UV complete the theory, while $(+-)$ states alone can. 

One final amplitude we identify corresponds to a "stringy" UV model. For example one can take the Lovelace-Shapiro \cite{LOVELACE1968264,shap}\footnote{Used to characterize pion scattering, further studied in \cite{Albert:2022oes,Fernandez:2022kzi} in the context of Large-$N_c$ and in \cite{Bianchi:2020cfc} in the context of higher-point scattering.} amplitude in the $(+-)$ sector, and dressing it with the little group factor, the resulting amplitude
\be
\mathcal{M}^{string}\big|_{+-}(s|t,u) = \frac{-s}{M^2}\ \frac{\Gamma(1/2-t/2M^2) \Gamma(1/2-u/2M^2)}{\Gamma(-(t+u)/2M^2)}\, ,
\ee
satisfies all the consistency constraints. The Wilson coefficients corresponding to this amplitude are shown with the pink point in Fig.~\ref{posbound2}. Notice that this amplitude includes $J=1$ states, which as discussed, decouple from the other states and can therefore be subtracted while keeping the good properties of the amplitude. Performing this subtraction moves us along the dashed pink line towards the boundary of the green exclusion region. 

\section{Conclusions}
\label{sec:conclusions}

Positivity of scattering amplitudes, expressing the fundamental constraints of causality and unitarity, imposes remarkably stringent  conditions on the EFTs of massive higher-spin particles. In this work we have explored these constraints for the borderline case of spin $J = 3/2$, whose fate ---whether the theory admits a healthy low–energy description with a large separation from UV dynamics, or instead is inconsistent--- depends critically on the details of the theory, and ultimately on the very existence of gravity.

In Section~\ref{sec:MajoranaGeneral} we first analyzed the simplest illustrative example: a massive Majorana (neutral) spin–$3/2$ particle. We showed that its EFT fails to satisfy positivity bounds unless new states appear at $\Lambda \simeq 9 m$, where $m$ is the  mass of the spin–$3/2$ particle. In other words, a lone  spin–$3/2$ state cannot be consistently isolated: new dynamics must accompany it. Moreover, these new states are highly constrained: weakly coupled EFTs with a finite number of degrees of freedom must be gravitational and couple in a supersymmetric way. In short, the fundamental \textit{tenets} of causality and unitarity enforce the emergence of gravity and supersymmetry when a massive and light spin–$3/2$ particle sits at the bottom of the spectrum. As an example, the SUSY relation $F^2 = 3 m^2 \mpl^2$ emerges as one such necessary tuning.

The inevitable emergence of gravity (and supersymmetry) persists when we consider a massive Dirac spin–$3/2$ particle carrying a global $U(1)$ charge, as discussed in Section~\ref{sec:DiracGeneral}. In this case, gravity alone is insufficient for consistency: an accompanying $U(1)$ gauge field must also be present, representing an infrared avatar of the \textit{no global symmetry conjecture}. Even more is required — the gyromagnetic coupling must satisfy $g = 2$, and the electric charge must saturate the \textit{weak gravity conjecture}, unless further light degrees of freedom are introduced, as explored in Section~\ref{sec:DiracGeneral}.

In Section~\ref{sec:GoldstinoEFThedron} we have taken the limit in which the transverse modes of the spin-$3/2$ particle decouple, allowing
us to implement all positivity constraints on the EFT of the longitudinal components – the Goldstinos. 
We first developed a novel class of $t$--$u$ symmetric dispersion relations, which allow to elegantly group all Null Constraints in a closed form. These constraints bound the space of
Wilson coefficients, defining the Goldstino EFT-hedron. In the extremal regions,  we have identified  known supersymmetry-breaking models, such as the
Fayet-Iliopoulos and O' Raifeartaigh models, as well as simple amplitudes involving higher-spin particles.

Importantly, the consistency conditions we have derived --- such as the relation involving $F$ and $m$, the quantitative upper bound $\Lambda / m \lesssim 9$, the characterization of the requisite additional states and their couplings,  as e.g. captured by the EFT-hedron analysis--- are completely invisible to analyses that focus exclusively on the massless transverse helicities $\lambda = \pm 3/2$ (see \textit{e.g.} \cite{Grisaru:1977kk,Benincasa:2007xk} and references therein). It is precisely the dynamics of the longitudinal modes, and their interplay with the transverse polarizations, that encode these constraints and reveal the deeper structure of the theory.

While our results point to gravity as an inescapable requirement for the consistency of theories with a light spin–$3/2$ particle, it is intriguing that the converse — the necessity for a combined spin–$3/2$ and spin–$1/2$ state — has also been advocated in \cite{Dvali:2024dlb}. In this sense, our findings suggest that as soon as a light spin–$3/2$ particle appears in the spectrum (perhaps motivated by applications such as canceling topological susceptibilities \cite{Dvali:2024dlb}), a fully-fledged supergravity structure bootstraps itself from the demands of consistency.\\

We stress that our findings also have potential phenomenological implications: should helicity-$\lambda = \pm 3/2$ modes of a massive spin-$3/2$ particle be discovered with couplings larger than gravity, it would immediately imply that  new states must be nearby. We can be more quantitative: new physics must appear below a scale $\sim 9m$.  
Discovering the longitudinal helicity-$\lambda = \pm 1/2$ modes, and measuring its decay constant and mass,  one could also directly test whether it corresponds to the Goldstino component of a Gravitino, as opposed to something more mundane, such as a bound state in a confining theory. 
The mass of new states can also be directly tied to the departure from saturation of the Weak Gravity Conjecture by a Dirac spin-$3/2$ particle. Discovering such a particle would signal a new mass scale and provide concrete clues about the underlying gravitational dynamics of new physics.\\

Finally, it is possible to envision some future directions and open problems.
A natural extension is to apply our analysis to particles of higher spin. For example, while an isolated massive spin-$2$ ($J=2$) particle is known to be inconsistent with positivity constraints~\cite{Bellazzini:2023nqj}, one may ask whether this conclusion can be avoided by changing the EFT---e.g., by including additional light degrees of freedom such as a massless graviton. Interpreting the massive spin-$2$ state as the first Kaluza-Klein (KK) mode in a compactified extra-dimensional model, our framework could provide quantitative bounds on the mass gap to higher KK modes, thereby constraining properties of the compactification itself.

Another direction stems from the peculiar result that positivity bounds require all gravitational and electromagnetic multipoles to be naturally vanishing, for $J=3/2$. It would be interesting to explore whether a similar approach could provide insights on the case of \textit{classical} spinning bodies with $J\gg 1$, giving a first-principle criterion for the choice of the three-point amplitude, with relevant applications to the physics of gravitational waves and astrophysical objects.

\vskip1cm
{\bf Note added:}
We  recently became aware that similar results on the EFT parameters of massive spin-3/2 particles have been  obtained via on-shell recursive relations in  \cite{Gherghetta2025}.
\vskip1cm

\subsection*{Acknowledgments}
This research was supported in part by NSF grant PHY-2309135 to the Kavli Institute for Theoretical Physics (KITP). BB, AP, and MR are very grateful to KITP for its hospitality, and to the organizers of the ``What's Particle Theory'' program, where this work was initiated. 
AP and FS have been partly
supported by the research grants 2021-SGR-00649 and PID2023-146686NB-C31.
This research was also supported by the Munich Institute for Astro-, Particle and BioPhysics (MIAPbP) which is funded by the Deutsche Forschungsgemeinschaft (DFG, German Research Foundation) under Germany´s Excellence Strategy – EXC-2094 – 390783311.
We also thank Tony Gherghetta and Yael Shadmi for valuable discussions. MR thanks Stefano De Angelis for discussions. BB thanks Arkady Vainshtein for many lively---if occasionally spirited---yet always enlightening conversations about dispersion relations, along with other stimulating physics discussions, and very enjoyable tennis games. BB thanks Justin Berman and Rachel Rosen for interesting discussions.  BB also warmly thanks Nathaniel Craig and the KITP for making possible his and his family’s extended and wonderful stay in Santa Barbara, both at KITP and UCSB. He is also grateful to the staff of the Munger Physics Residence for their kind support throughout their time there.

\appendix

\section{Spinor-helicity conventions}\label{app:conventions}
We work in mostly-minus metric $g_{\mu\nu}=\text{diag}(1,-1,-1,-1)$. Amplitudes are expressed in terms of left-handed $\la_{\alpha}$ and right-handed $\lat_{\dot\alpha}$ spinors of $SL(2,\mathbb{C})_{\mathbb{C}}$. The invariant contractions are obtained with the totally antisymmetric tensor $\epsilon^{\alpha\beta}=\epsilon^{\dot\alpha\dot\beta}$ (with $\epsilon^{12}=1$) and are denoted as $\angmm{\la}{\rho}=\epsilon^{\alpha\beta}\la_{\beta}\rho_{\alpha}=\la^{\alpha}\rho_{\alpha}$ and $\sqrmm{\lat}{\rhot}=\epsilon^{\dot\beta\dot\alpha}\lat_{\dot\beta}\rhot_{\dot\alpha}=\lat_{\dot\alpha}\rhot^{\dot\alpha}$. For each particle, depending on the mass, we realize the Poincaré algebra as explained in the following.\\

\noindent \textbf{Massless particles.} The four-momentum is given by
\begin{equation}
    p_{\alpha\dot\alpha}=\la_{\alpha}\lat_{\dot\alpha} \,,
\end{equation}
defined up to little-group transformations $\la\to w\la$, $\lat\to w^{-1} \lat$. Given two particles of momenta $p_1$ and $p_2$ we use the shortcut notation $\angmm{1}{2}$ and $\sqrmm{1}{2}$, respectively for the contractions $\angmm{\la_{p_1}}{\la_{p_2}}$ and $\sqrmm{\lat_{p_1}}{\lat_{p_2}}$.\\

\noindent \textbf{Massive particles.} The four-momentum is given by
\begin{equation}
    p_{\alpha\dot\alpha}=\la_{\alpha}\lat_{\dot\alpha} +\frac{m^2}{\angmm{\la}{\rho}\sqrmm{\rhot}{\lat}}\rho_{\alpha}\rhot_{\dot\alpha}=\epsilon_{IJ}\la^{I}_{\alpha}\lat^J_{\dot\alpha}=\la^{I}_{\alpha}\lat_{I\,\dot\alpha}\,,
\end{equation}
where we defined the little-group $SU(2)$ spinors
\begin{equation}
    \la^{I}_{\alpha}=\begin{pmatrix}
        \frac{m}{\angmm{\rho}{\la}}\rho_{\alpha}\\
        \la_{\alpha}
    \end{pmatrix}\;,\;\;\lat^{I}_{\dot\alpha}=\begin{pmatrix}
        \lat_{\dot\alpha}\\
        \frac{m}{\sqrmm{\rhot}{\lat}}\rhot_{\dot\alpha}
    \end{pmatrix} \,.
\end{equation}
Following Ref. \cite{Arkani-Hamed:2017jhn}, we adopt the bold notation $\angMM{1}{2}$, $\sqrMM{1}{2}$ for spinor contractions, where the little-group indices are understood.\\

Throughout the paper we consider amplitudes both in \textit{incoming-outgoing} convention, denoted with in-- and out-- states on two separate lines $\M_{\la_1\la_2}^{\la_3\la_4}$, and \textit{all-incoming} convention, denoted with all states on the same footing, \textit{e.g.} $\M(1^{\la_1}2^{\la_2}3^{\la_3}...)$ or $\M^{\la_1\la_2\la_3\la_4}(s,t)$ when in the CoM frame. To flip a state from incoming to outgoing the replacement $\la^{I}\to-\la_{I}\;,\;\;\lat^{I}\to\lat_{I}$ is needed.

\section{Derivations of positivity bounds}\label{app:positivity_bounds}
\par In this section, after reviewing the necessary basics of $S$-matrix theory, we give a full derivation of the positivity bounds employed in the main text. We also show how the bounds are generalized at finite values of $m/\Lambda$.
\subsection{The analytic $S$-matrix}
The $2\to2$ scattering amplitude $\M_{\la_1\la_2}^{\la_3\la_4}(s,t)$ of four massive spin-$J$ particles enjoys notable analytic properties, descending from microcausality, Lorentz invariance and unitarity.\\
\\
\noindent \textit{a) Analyticity}\\

\noindent $\M_{\la_1\la_2}^{\la_3\la_4}(s,t)$ is an analytic function in $s$, at fixed $-\Lambda^2\leq t\leq0$,~\footnote{This condition is weaker than maximal analyticity, which holds also at $-t\gg m^2$. For theories that admit no separation of scales, such as $\Lambda<\mathcal{O}(10)m$ in the isolated Majorana case \eqref{eq:boundMajorana8}, the stated region of analyticity can be rigorously proven, see \textit{e.g.} \cite{Sommer:1970mr} for a review.} in the whole complex plane, except for the real axis. Singularities are either dynamical (single-particle simple poles, multi-particle branch cuts at $s\geq4m^2-t$ and $s\leq0$) or kinematical.\footnote{These are determined by angular momentum selection rules and have been classified in \cite{Cohen-Tannoudji:1968lnm}, see also \cite{Hebbar:2020ukp} for a modern review.}
The amplitude is \textit{hermitian analytic}, \textit{i.e.}
    \begin{equation}\label{eq:hermitian_analyticity}
        \left[\M_{\la_1\la_2}^{\la_3\la_4}(s,t)\right]^*=\M_{\la_3\la_4}^{\la_1\la_2}(s^*,t) \,.
    \end{equation}

\noindent \textit{b) Crossing symmetry}\\

\noindent Amplitudes for different physical scattering channels are related to the same function $\M_{\la_1\la_2}^{\la_3\la_4}(s,t)$ by analytic continuation. In particular $s$-$u$ crossing relates the processes $1\,2\to3\,4$ and $1\,\bar4\to3\,\bar2$ as
\begin{equation}
    \M_{\la_1\la_2}^{\la_3\la_4}(s-i\epsilon,t)=\sum_{\la'_i}X^{\la'_1\la'_2\la'_3\la'_4}_{\la_1\la_2\la_3\la_4}(s,t)\M_{\la'_1\bar\la'_4}^{\la'_3\bar\la'_2}(4m^2-s-t+i\epsilon,t) \,,
\end{equation}
 for $-s-t\geq-t\geq0$, where $\bar\la=-\la$. The crossing matrix is defined as
 \begin{equation}
     X^{\la'_1\la'_2\la'_3\la'_4}_{\la_1\la_2\la_3\la_4}(s,t)=(-1)^{\la_1-\la_3}d^{J}_{\la_1'\la_1}(\theta_u)d^{J}_{\bar\la_2'\la_2}(\theta_u-\pi)d^{J}_{\la_3'\la_3}(-\theta_u)d^{J}_{\bar\la_4'\la_4}(-\theta_u+\pi)\,,
 \end{equation}
 where $d^{J}_{\la'\la}(\theta)$ are Wigner $d$-matrices and
 \begin{equation}
     \cos\theta_u=\sqrt{\frac{s(s+t-4m^2)}{(s-4m^2)(s+t)}}\,,
 \end{equation}
Further details are discussed in the following subsections.\\
\\
\noindent \textit{c) Unitarity}\\

\noindent Unitarity of the $S$-matrix requires the discontinuity of the amplitude to be positive semi-definite
    \begin{equation}
        \text{Disc}\,\M_{\la_1\la_2}^{\la_3\la_4}(s,t)\succeq0 \,.
    \end{equation}
    via the optical theorem and hermitian analyticity. In the elastic and forward limit we have simply $\text{Disc}\,\M_{\la_1\la_2}^{\la_1\la_2}(s,0)>0$. The discontinuity admits a partial-waves decomposition
    \begin{equation}
    \text{Disc}\,\M_{\la_1\la_2}^{\la_3\la_4}(s,t)=\sum_{j}8(2j+1)a_{\la_1\la_2}^{\la_3\la_4}(s,j)d^j_{\la_{12}\la_{34}}(\theta(s,t)) \,,
    \end{equation}
    where $\la_{ij}=\la_i-\la_j$, and
    \begin{equation}
        a_{\la_1\la_2}^{\la_3\la_4}(s)\succeq0 \,.
    \end{equation}
\\
\noindent \textit{d) Polynomial boundedness}\\

\noindent The amplitude is polynomially bounded as $\abs{s}\to+\infty$. In particular, we assume
    \begin{equation}
    \frac{\M_{\la_1\la_2}^{\la_3\la_4}(s,t)}{s^2}\to0\;\;\text{as}\;\abs{s}\to+\infty\;,\;\;t\text{ fixed} \,.
    \end{equation}
    For gapped theories this property follows from the Froissart-Martin bound \cite{Froissart:1961ux,Martin:1962rt}. For results in presence of gravity see \cite{Haring:2022cyf}.

\subsection{Derivation of the bounds \eqref{eq:positivity_bounds0}}
As discussed in Section~\ref{subsec:positivitydecoupling}, we define the Arcs as the contour integrals
\begin{equation}
    \A_{\la_1\la_2}^{\la_3\la_4}(t,n)=\frac{1}{2}\oint_{C_{\text{low}}}\frac{ds}{2\pi i}\frac{\M_{\la_1\la_2}^{\la_3\la_4}(s,t)+\M_{\la_1\bar\la_4}^{\la_3\bar\la_2}(s,t)}{(s-2m^2+\frac{t}{2})^{3+n}}\,,
\end{equation}
with $n\geq0$. Leveraging analyticity, crossing symmetry, hermitian analyticity and polynomial boundedness, we can deform the integration contour, as shown in Fig. \ref{fig:dispersioncontours}, to provide a UV representation of the Arcs as an integral over the discontinuity
\begin{equation}\label{eq:UVarcsX}
\begin{split}
    \A_{\la_1\la_2}^{\la_3\la_4}(t,n)&=\frac{1}{2}\int_{\Lambda^2}^{+\infty}\frac{ds}{2\pi}\frac{\text{Disc}\,\M_{\la_1\la_2}^{\la_3\la_4}(s,t)+\text{Disc}\,\M_{\la_1\bar\la_4}^{\la_3\bar\la_2}(s,t)}{(s-2m^2+\frac{t}{2})^{3+n}}\\
    &+\frac{(-1)^n}{2}\int_{\Lambda^2}^{+\infty}\frac{ds}{2\pi}\frac{\left[X^{\la'_1\la'_2\la'_3\la'_4}_{\la_1\la_2\la_3\la_4}(s,t)+X^{\la'_1\la'_2\la'_3\la'_4}_{\la_1\bar\la_4\la_3\bar\la_2}(s,t)\right]\text{Disc}\,\M_{\la'_1\bar\la'_4}^{\la'_3\bar\la'_2}(s,t)}{(s-2m^2+\frac{t}{2})^{3+n}} \,.
\end{split}
\end{equation}
This expression is in general quite complicated to handle, because of the non-trivial crossing matrix. Nonetheless, we can exploit the assumption of large scale separation in order to simplify it. Indeed, the crossing matrix can be expanded in the regime $m^2\ll-t\ll\Lambda^2<s$ as
\begin{equation}
    X^{\la'_1\la'_2\la'_3\la'_4}_{\la_1\la_2\la_3\la_4}(s,t)=\delta_{\la_1\la'_1}\delta_{\la_2\la'_2}\delta_{\la_3\la'_3}\delta_{\la_4\la'_4}+\mathcal{O}\left(\frac{\sqrt{-t}\,m}{s}\right)\propto\prod_{i} \left(\frac{\sqrt{-t}\,m}{s}\right)^{|\la'_i-\la_i|}\,,
\end{equation}
where each higher order term is explicitly known in terms of products of Wigner-$d$ functions and their derivatives.  

Let us first focus on the leading contribution, in order to derive Eq.~\eqref{eq:positivity_bounds0}. 
In this limit we have simply
\begin{equation}\label{eq:UVarcs_simp}
    \A_{\la_1\la_2}^{\la_3\la_4}(t,n)=\frac{1+(-1)^n}{2}\int_{\Lambda^2}^{+\infty}\frac{ds}{2\pi}\frac{\text{Disc}\,\M_{\la_1\la_2}^{\la_3\la_4}(s,t)+\text{Disc}\,\M_{\la_1\bar\la_4}^{\la_3\bar\la_2}(s,t)}{(s-2m^2+\frac{t}{2})^{3+n}} \,.
\end{equation}
At this point, we can make use of the following property of the discontinuity:
\begin{equation}\label{eq:disc_inequality}
    0\leq2\abs{\text{Disc}\,\M_{\la_1\la_2}^{\la_3\la_4}(s,t)}\leq\text{Disc}\,\M_{\la_1\la_2}^{\la_1\la_2}(s,0)+\text{Disc}\,\M_{\la_3\la_4}^{\la_3\la_4}(s,0) \,,
\end{equation}
which simply follows from $|\hat{T}(\ket{\alpha}+\ket{\beta})|^2\geq0$ and the optical theorem. Applied to Eq.~\eqref{eq:UVarcs_simp}, it yields the following inequality
\begin{equation}\label{eq:positivitybounds0app}
\begin{split}
    0&\leq\abs{\A_{\la_1\la_2}^{\la_3\la_4}(t,n)\pm\A_{\la_1\bar\la_2}^{\la_3\bar\la_4}(t,n)}\leq\abs{\A_{\la_1\la_2}^{\la_3\la_4}(t,n)}+\abs{\A_{\la_1\bar\la_2}^{\la_3\bar\la_4}(t,n)}=\\
    &\leq\frac{1}{2}\frac{(\Lambda^2-2m^2)^3}{(\Lambda^2-2m^2+\frac{t}{2})^{n+3}}\left[\A_{\la_1\la_2}^{\la_1\la_2}(0,0)+\A_{\la_3\la_4}^{\la_3\la_4}(0,0)+\A_{\la_1\la_4}^{\la_1\la_4}(0,0)+\A_{\la_3\la_2}^{\la_3\la_2}(0,0)\right] \,.
\end{split}
\end{equation}

\subsection{Corrections from the crossing matrix}
\par The leading finite-mass effects are captured by Eq.~\eqref{eq:positivity_bounds0}. Before discussing how the bounds in the decoupling limit are deformed by the finite mass, let us show how also the subleading corrections, coming from the crossing matrix, can be bounded. As a warm up, let us consider the first-order correction, following the analysis of \cite{Bellazzini:2023nqj}. We expand the crossing matrix as
\begin{equation}
    X^{\la'_1\la'_2\la'_3\la'_4}_{\la_1\la_2\la_3\la_4}(s,t)=\delta_{\la_1\la'_1}\delta_{\la_2\la'_2}\delta_{\la_3\la'_3}\delta_{\la_4\la'_4}+\frac{\sqrt{-t}\,m}{s}c^{\la'_1\la'_2\la'_3\la'_4}_{\la_1\la_2\la_3\la_4}+\mathcal{O}\left(\frac{-t\,m^2}{s^2}\right) \,.
\end{equation}
When taking the absolute value  $|\A_{\la_1\la_2}^{\la_3\la_4}(t,n)\pm\A_{\la_1\bar\la_2}^{\la_3\bar\la_4}(t,n)|$, the delta-function contribution combines with the $s$-channel integral to give the RHS of Eq.~\eqref{eq:positivitybounds0app}, and we get a new term $\Delta$:
 \begin{equation}
\begin{split}
    0&\leq\abs{\A_{\la_1\la_2}^{\la_3\la_4}(t,n)\pm\A_{\la_1\bar\la_2}^{\la_3\bar\la_4}(t,n)}\leq\abs{\A_{\la_1\la_2}^{\la_3\la_4}(t,n)}+\abs{\A_{\la_1\bar\la_2}^{\la_3\bar\la_4}(t,n)}=\\
    &\leq\frac{1}{2}\frac{(\Lambda^2-2m^2)^3}{(\Lambda^2-2m^2+\frac{t}{2})^{n+3}}\left[\A_{\la_1\la_2}^{\la_1\la_2}(0,0)+\A_{\la_3\la_4}^{\la_3\la_4}(0,0)+\A_{\la_1\la_4}^{\la_1\la_4}(0,0)+\A_{\la_3\la_2}^{\la_3\la_2}(0,0)+\Delta_{\la_1\la_2}^{\la_3\la_4}\right] \,.
\end{split}
\end{equation}
Using symmetry properties of Wigner $d$-matrices and relabeling dummy indices which are summed over, the correction term can be brought in the form
 \begin{eqnarray}
 \nonumber
    \Delta_{\la_1\la_2}^{\la_3\la_4}=\frac{\sqrt{-t}\,m}{\Lambda^2}\abs{c^{\la'_1\la'_2\la'_3\la'_4}_{\la_1\la_2\la_3\la_4}} &\int_{\Lambda^2}^{+\infty}\frac{ds}{2\pi}\frac{1}{(s-2m^2+\frac{t}{2})^{3}}  \left[\abs{\text{Disc}\,\M_{\la'_1\la'_2}^{\la'_3\la'_4}(s,t)}+\abs{\text{Disc}\,\M_{\la'_1\bar\la'_4}^{\la'_3\bar\la'_2}(s,t)}\right.\\
    &\left.+\abs{\text{Disc}\,\M_{\la'_1\bar\la'_2}^{\la'_3\bar\la'_4}(s,t)}+\abs{\text{Disc}\,\M_{\la'_1\la'_4}^{\la'_3\la'_2}(s,t)}\right] \,.
\end{eqnarray}
Applying repeatedly Eq.~\eqref{eq:disc_inequality} yields 
\begin{equation}\label{eq:crossing_correction}
\begin{split}
    \Delta_{\la_1\la_2}^{\la_3\la_4}\leq\frac{\sqrt{-t}\,m}{\Lambda^2}\frac{\abs{c^{\la'_1\la'_2\la'_3\la'_4}_{\la_1\la_2\la_3\la_4}}}{2}\left[\A_{\la'_1\la'_2}^{\la'_1\la'_2}(0,0)+\A_{\la'_3\la'_4}^{\la'_3\la'_4}(0,0)+\A_{\la'_1\la'_4}^{\la'_1\la'_4}(0,0)+\A_{\la'_3\la'_2}^{\la'_3\la'_2}(0,0)\right] \,.
\end{split}
\end{equation}
This equation gives an explicit bound on the leading correction generated by crossing. Notice that $c$ is non-vanishing only for $8$ entries and its absolute value can be at most of order $J$, as can be easily seen by group-theoretical arguments.

\subsubsection{Bounding the full crossing matrix}
\par It is actually possible to give an exact bound, without expanding the crossing matrix. In this case it is convenient to keep separated the $s$- and $u$-channel cuts contributions:
 \begin{equation}
\begin{split}
    0&\leq\abs{\A_{\la_1\la_2}^{\la_3\la_4}(t,n)\pm\A_{\la_1\bar\la_2}^{\la_3\bar\la_4}(t,n)}\leq\abs{\A_{\la_1\la_2}^{\la_3\la_4}(t,n)}+\abs{\A_{\la_1\bar\la_2}^{\la_3\bar\la_4}(t,n)}=\\
    &\leq\frac{1}{4}\frac{(\Lambda^2-2m^2)^3}{(\Lambda^2-2m^2+\frac{t}{2})^{n+3}}\left[\A_{\la_1\la_2}^{\la_1\la_2}(0,0)+\A_{\la_3\la_4}^{\la_3\la_4}(0,0)+\A_{\la_1\la_4}^{\la_1\la_4}(0,0)+\A_{\la_3\la_2}^{\la_3\la_2}(0,0)+\tilde\Delta_{\la_1\la_2}^{\la_3\la_4}\right] \,,
\end{split}
\end{equation}
where the $s$-channel cut gives half the RHS of \eqref{eq:positivitybounds0app} and the $u$-channel cut gives
\begin{equation}
\begin{split}
    \tilde\Delta_{\la_1\la_2}^{\la_3\la_4}=\int_{\Lambda^2}^{+\infty}\frac{ds}{2\pi}\frac{\abs{X^{\la'_1\la'_2\la'_3\la'_4}_{\la_1\la_2\la_3\la_4}(s,t)}}{(s-2m^2+\frac{t}{2})^{3}}&\left[\abs{\text{Disc}\,\M_{\la'_1\la'_2}^{\la'_3\la'_4}(s,t)}+\abs{\text{Disc}\,\M_{\la'_1\bar\la'_4}^{\la'_3\bar\la'_2}(s,t)}\right.\\
    &\left.+\abs{\text{Disc}\,\M_{\la'_1\bar\la'_2}^{\la'_3\bar\la'_4}(s,t)}+\abs{\text{Disc}\,\M_{\la'_1\la'_4}^{\la'_3\la'_2}(s,t)}\right] \,.
\end{split}
\end{equation}
Let us now observe that given a set of functions $f^{\la'_1\la'_2\la'_3\la'_4}_{\la_1\la_2\la_3\la_4}\left(\frac{-t}{s},\frac{m^2}{s}\right)$ satisfying the properties
\begin{subequations}
\begin{align}    \abs{X^{\la'_1\la'_2\la'_3\la'_4}_{\la_1\la_2\la_3\la_4}(s,t)}&\leq f^{\la'_1\la'_2\la'_3\la'_4}_{\la_1\la_2\la_3\la_4}\left(\frac{-t}{s},\frac{m^2}{s}\right) \,,\\
\label{subeq:flimit}
f^{\la'_1\la'_2\la'_3\la'_4}_{\la_1\la_2\la_3\la_4}\left(\frac{-t}{s},\frac{m^2}{s}\right)&\to\delta_{\la_1\la'_1}\delta_{\la_2\la'_2}\delta_{\la_3\la'_3}\delta_{\la_4\la'_4}\;\;\text{as}\;\;t\to0\;\;\text{or}\;\;m\to0 \,,\\
\frac{\partial}{\partial s}f^{\la'_1\la'_2\la'_3\la'_4}_{\la_1\la_2\la_3\la_4}\left(\frac{-t}{s},\frac{m^2}{s}\right)&\leq0 \,.
\end{align}
\end{subequations}
we have 
\begin{equation}
\begin{split}
    \tilde\Delta_{\la_1\la_2}^{\la_3\la_4}\leq f^{\la'_1\la'_2\la'_3\la'_4}_{\la_1\la_2\la_3\la_4}\left(\frac{-t}{\Lambda^2},\frac{m^2}{\Lambda^2}\right) \int_{\Lambda^2}^{+\infty}\frac{ds}{2\pi}\frac{1}{(s-2m^2+\frac{t}{2})^{3}}&\left[\abs{\text{Disc}\,\M_{\la'_1\la'_2}^{\la'_3\la'_4}(s,t)}+\abs{\text{Disc}\,\M_{\la'_1\bar\la'_4}^{\la'_3\bar\la'_2}(s,t)}\right.\\
    &\left.+\abs{\text{Disc}\,\M_{\la'_1\bar\la'_2}^{\la'_3\bar\la'_4}(s,t)}+\abs{\text{Disc}\,\M_{\la'_1\la'_4}^{\la'_3\la'_2}(s,t)}\right] \,.
\end{split}
\end{equation}
Eventually, applying again Eq.~\eqref{eq:disc_inequality}, we get
 \begin{equation}\label{eq:fullcrossing_bound}
\begin{split}
    0&\leq\abs{\A_{\la_1\la_2}^{\la_3\la_4}(t,n)\pm\A_{\la_1\bar\la_2}^{\la_3\bar\la_4}(t,n)}\leq\abs{\A_{\la_1\la_2}^{\la_3\la_4}(t,n)}+\abs{\A_{\la_1\bar\la_2}^{\la_3\bar\la_4}(t,n)}\\
    &\leq\frac{1}{4}\frac{(\Lambda^2-2m^2)^3}{(\Lambda^2-2m^2+\frac{t}{2})^{n+3}}\left[\A_{\la_1\la_2}^{\la_1\la_2}(0,0)+\A_{\la_3\la_4}^{\la_3\la_4}(0,0)+\A_{\la_1\la_4}^{\la_1\la_4}(0,0)+\A_{\la_3\la_2}^{\la_3\la_2}(0,0)\right.\\
    &\left.+\sum_{\la_i'}f^{\la'_1\la'_2\la'_3\la'_4}_{\la_1\la_2\la_3\la_4}\left(\frac{-t}{\Lambda^2},\frac{m^2}{\Lambda^2}\right) \left(\A_{\la'_1\la'_2}^{\la'_1\la'_2}(0,0)+\A_{\la'_3\la'_4}^{\la'_3\la'_4}(0,0)+\A_{\la'_1\la'_4}^{\la'_1\la'_4}(0,0)+\A_{\la'_3\la'_2}^{\la'_3\la'_2}(0,0)\right)\right] \,.
\end{split}
\end{equation}
Notice that Eq.~\eqref{subeq:flimit} ensures that in the limit $m^2\ll-t\ll\Lambda^2$ Eq.~\eqref{eq:positivitybounds0app} is recovered.
\par The take-home message is that finite-mass corrections coming from the non-trivial crossing are indeed small in the regime $m^2\ll-t\ll\Lambda^2$ and can be bounded explicitly.
\par Exploiting bounds on Wigner $d$-matrices from the mathematical literature \cite{haagerup2014inequalities}, we can provide an explicit set of $f$:
\begin{equation}
    f^{\la'_1\la'_2\la'_3\la'_4}_{\la_1\la_2\la_3\la_4}\left(\frac{-t}{s},\frac{m^2}{s}\right)=h^{J}_{\la_1'\la_1}(\theta_u)h^{J}_{\bar\la_2'\la_2}(\theta_u-\pi)h^{J}_{\la_3'\la_3}(-\theta_u)h^{J}_{\bar\la_4'\la_4}(-\theta_u+\pi) \,,
\end{equation}
with
\begin{equation}
    h^{J}_{\la\la'}(\theta)=\min\Biggl\{
        \left[\frac{(n+1)(n+\alpha+\beta+1)}{(n+\alpha+1)(n+\beta+1)}\right]^{\frac{1}{4}}\;,\;\;
        \binom{n+\alpha}{n}\sqrt{\frac{\Gamma(n+1)\Gamma(n+\alpha+\beta+1)}{\Gamma(n+\alpha+1)\Gamma(n+\beta+1)}}\left(\frac{1-\cos\theta}{2}\right)^{\frac{\alpha}{2}}\Biggr\} \,,
\end{equation}
and
\begin{equation}
    \begin{cases}
        \alpha=\abs{\lambda-\lambda'}\\
        \beta=\abs{\lambda+\lambda'}\\
        n=J-\max{(\abs{\lambda},\abs{\lambda'})} \,.
    \end{cases}
\end{equation}

\subsection{Finite-mass effects in the isolated Majorana spin-3/2 EFT}

\begin{figure}[h!!]
    \begin{subfigure}[h]{0.33\linewidth}
    \includegraphics[width=\textwidth]{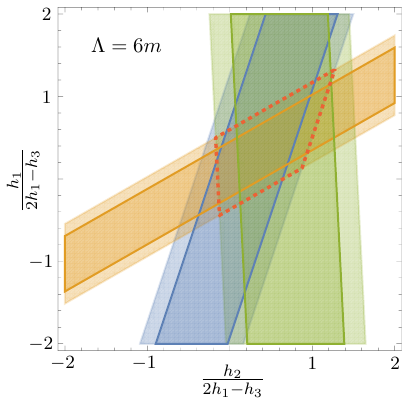}
    \end{subfigure}
    \begin{subfigure}[h]{0.33\linewidth}
    \includegraphics[width=\textwidth]{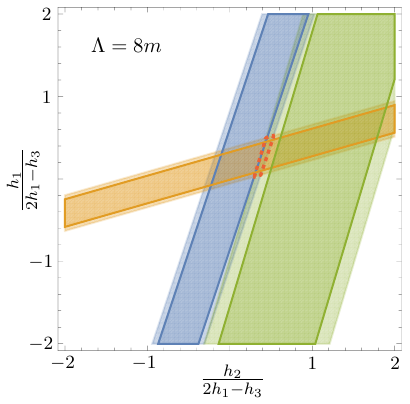}
    \end{subfigure}
    \begin{subfigure}[h]{0.33\linewidth}
    \includegraphics[width=\textwidth]{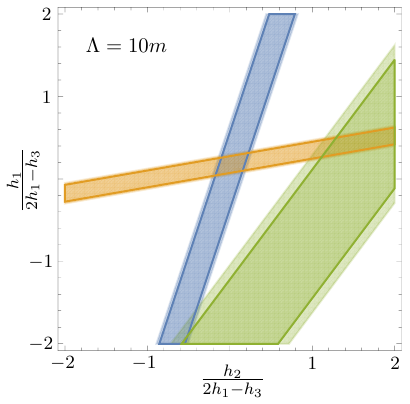}
    \end{subfigure}
    \caption{Region of parameter space allowed by the constraints in Eq.~\eqref{eq:positivity_isolatenmajorana_finitem} (solid bands, corresponding to leading-order crossing effects) and Eq.~\eqref{eq:positivitybounds0app} (light bands, full crossing effects), evaluated at $t = -\frac{\Lambda^2}{10}$ for increasing values of $\Lambda = 6m, 8m, 10m$ (from left to right). The blue, orange and green bands correspond to bounding respectively $|\A_{-\frac{1}{2}-\frac{1}{2}}^{-\frac{1}{2}-\frac{1}{2}}(t,0)|$, $|\A_{-\frac{1}{2}-\frac{1}{2}}^{+\frac{1}{2}+\frac{1}{2}}(t,0)+\A_{-\frac{1}{2}+\frac{1}{2}}^{+\frac{1}{2}-\frac{1}{2}}(t,0)|$ and $|\A_{-\frac{1}{2}-\frac{1}{2}}^{+\frac{1}{2}+\frac{1}{2}}(t,0)-\A_{-\frac{1}{2}+\frac{1}{2}}^{+\frac{1}{2}-\frac{1}{2}}(t,0)|$. The intersection region, when present, is indicated by a red dashed contour.}\label{fig:finitemapp}
\end{figure}

\begin{figure}[h!!]
    \centering
    \includegraphics[width=0.6\linewidth]{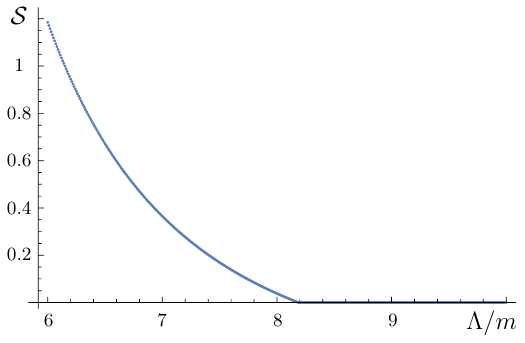}
    \caption{Plot of the area $\mathcal{S}$ of the allowed region in parameter space as a function of $\Lambda/m$, with $t = -\Lambda^2/10$.}
    \label{fig:finitmboundsintersectionarea}
\end{figure}

We now apply the results of the previous sections to the case of an isolated Majorana spin-$3/2$ particle. This scenario was analyzed in the decoupling limit in Section~\ref{sec:isolated_majorana}; we now extend those bounds to the case of finite mass. The relevant Arcs are 
\begin{equation}
\begin{split}
\A_{-\frac{1}{2}-\frac{1}{2}}^{-\frac{1}{2}-\frac{1}{2}}(t,0)&= \frac{t(6h_2-h_3)-8m^2(h_3-2h_1)}{36M^6} \,,\\
\A_{-\frac{1}{2}-\frac{1}{2}}^{+\frac{1}{2}+\frac{1}{2}}(t,0)&=-\frac{3th_1+2m^2(h_3+4h_2)}{9M^6} \,,\\
\A_{-\frac{1}{2}+\frac{1}{2}}^{+\frac{1}{2}-\frac{1}{2}}(t,0)&=-t\frac{h_3}{9M^6} \,.
\end{split}
\end{equation}
In particular the forward elastic Arc is
\begin{equation}
\begin{split}
\A_{-\frac{1}{2}-\frac{1}{2}}^{-\frac{1}{2}-\frac{1}{2}}(0,0)&= \frac{2m^2(2h_1-h_3)}{9M^6}\,.
\end{split}
\end{equation}
The leading bounds at finite mass, neglecting corrections from crossing, are given by Eq.~\eqref{eq:positivitybounds0app} and read

\begin{equation}\label{eq:positivity_isolatenmajorana_finitem}
\begin{split}
\abs{\frac{t(6h_2-h_3)-8m^2(h_3-2h_1)}{36M^6}}&\leq \frac{2m^2(2h_1-h_3)}{9M^6}\frac{(\Lambda^2-2m^2)^3}{(\Lambda^2-2m^2+\frac{t}{2})^{3}} \,,\\
\abs{-\frac{3th_1+2m^2(h_3+4h_2)}{9M^6}\mp t\frac{h_3}{9M^6}}&\leq\frac{4m^2(2h_1-h_3)}{9M^6}\frac{(\Lambda^2-2m^2)^3}{(\Lambda^2-2m^2+\frac{t}{2})^{3}}\,.
\end{split}
\end{equation}
%
%
Since the forward Arc vanishes when $2h_1 - h_3 = 0$, Eq.~\eqref{eq:positivity_isolatenmajorana_finitem} would in that case simply enforce $h_1 = h_2 = h_3 = 0$. Therefore, we can directly focus on the case $2h_1 - h_3 > 0$, and analyze the parameter space in terms of the ratios $\frac{h_2}{2h_1 - h_3}$ and $\frac{h_1}{2h_1 - h_3}$. The constraints imposed by Eq.~\eqref{eq:positivity_isolatenmajorana_finitem} are shown as colored bands in Fig.~\ref{fig:finitemapp} for $t = -\frac{\Lambda^2}{10}$ and increasing values of $\Lambda = 6m, 8m, 10m$, along with those obtained from the full bounds in Eq.~\eqref{eq:fullcrossing_bound}. Already at $\Lambda = 10m$, no intersection---no solution to the constraints---is found. Performing a fine-grained scan of $\Lambda/m$ values (see Fig.~\ref{fig:finitmboundsintersectionarea}) we find the precise bound
\begin{equation}
    \Lambda<8.2m \,.
\end{equation}
Furthermore, the plots show explicitly how the corrections to Eq.~\eqref{eq:positivitybounds0app} coming from the non-trivial crossing matrix are indeed small as soon as $m^2\ll-t\ll\Lambda^2$. Numerically they modify the bounds respectively of $21\%$, $15\%$ and $11\%$ for $\Lambda = 6m, 8m, 10m$.

\section{The spin-$\mathbf{3/2}$ EFT on-shell}\label{app:3_2eft}
\subsection{Majorana case}
We list here all the $CP$-invariant on-shell three-point amplitudes involving  two Majorana spin-$3/2$ particles and a third particle, which can be a scalar, a massive vector or a massless graviton:

\begin{itemize}
    \item \underline{Scalar and pseudoscalar}:
    \begin{equation}
        \begin{split}
            \M(\mathbf{1}\mathbf{2}\mathbf{3}^{S_{+}})&=\frac{a_1^{+}}{M_S^2}(\angMM{1}{2}^3+\sqrMM{1}{2}^3)+\frac{a_2^{+}}{M_S^2}\angMM{1}{2}\sqrMM{1}{2}(\angMM{1}{2}+\sqrMM{1}{2})\,,\\
            \M(\mathbf{1}\mathbf{2}\mathbf{3}^{S_{-}})&=i\frac{a_1^{-}}{M_S^2}(\angMM{1}{2}^3-\sqrMM{1}{2}^3)+i\frac{a_2^{-}}{M_S^2}\angMM{1}{2}\sqrMM{1}{2}(\angMM{1}{2}-\sqrMM{1}{2})\,,
        \end{split}
    \end{equation}
    where $M_S$ controls the mass dimension of the three point vertex.
    %
    \item \underline{Vector and pseudovector}:
    \begin{equation}
        \begin{split}
            \M(\mathbf{1}\mathbf{2}\mathbf{3}^{V_{+}})&=  i\frac{b_1^{+}}{m_{V_+} M_V^2}(\angMM{1}{3}\sqrMM{2}{3}-\angMM{2}{3}\sqrMM{1}{3})(\angMM{1}{2}^2-\sqrMM{1}{2}^2)\,,\\
            \M(\mathbf{1}\mathbf{2}\mathbf{3}^{V_{-}})&=\frac{b_1^{-}}{m_{V_-} M_V^2}(\angMM{1}{3}\sqrMM{2}{3}-\angMM{2}{3}\sqrMM{1}{3})(\angMM{1}{2}^2+\sqrMM{1}{2}^2)\\
            &+\frac{b_2^{-}}{m_{V_-} M_V^2}(\angMM{1}{3}\sqrMM{2}{3}-\angMM{2}{3}\sqrMM{1}{3})\angMM{1}{2}\sqrMM{1}{2} \,,
        \end{split}
    \end{equation}
    where  $m_{V_\pm}$ is the mass of the vector, and $M_V$ the scale controlling the overall mass dimension of the vertex.
     \item \underline{Graviton}:
    \begin{equation}
        \begin{split}
            \M(\mathbf{1}\mathbf{2}3^{++})&=\frac{x_{12}^2}{\mpl^2}\left[\angMM{2}{1}^3+\frac{g_{4}}{m^2}\angMM{2}{1}\angsqrMM{1}{3}{2}^2+\frac{g_{8}}{m^3}\angsqrMM{1}{3}{2}^3\right]\,,\\
            \M(\mathbf{1}\mathbf{2}3^{--})&=\frac{x_{12}^{-2}}{\mpl^2}\left[\sqrMM{2}{1}^3+\frac{g_{4}}{m^2}\sqrMM{2}{1}\angsqrMM{2}{3}{1}^2+\frac{g_{8}}{m^3}\angsqrMM{2}{3}{1}^3\right] \,,
        \end{split}
    \end{equation}
    where $\mpl$ is the Planck mass and the three terms correspond, respectively, to minimal coupling, the gravitational quadrupole ($g_4$), and the octupole ($g_8$).
\end{itemize}

\subsection{Dirac case}
\par We list here all the parity and charge-conjugation invariant on-shell three point amplitudes of two Dirac spin-$3/2$ particles with $U(1)$ charges and a third particle, which can be a scalar, a massive vector, a massless photon or a massless graviton:

\begin{itemize}
    \item \underline{Scalar and pseudoscalar:}
   \begin{equation}
        \begin{split}
            \M(\mathbf{1}_q\mathbf{2}_{-q}\mathbf{3}^{S_{+,+}})&=\frac{a_1^{+,+}}{M_S^2}(\angMM{1}{2}^3+\sqrMM{1}{2}^3)+\frac{a_2^{+,+}}{M_S^2}\angMM{1}{2}\sqrMM{1}{2}(\angMM{1}{2}+\sqrMM{1}{2}) \,,\\
            \M(\mathbf{1}_q\mathbf{2}_{-q}\mathbf{3}^{S_{-,+}})&=i\frac{a_1^{-,+}}{M_S^2}(\angMM{1}{2}^3-\sqrMM{1}{2}^3)+i\frac{a_2^{-,+}}{M_S^2}\angMM{1}{2}\sqrMM{1}{2}(\angMM{1}{2}-\sqrMM{1}{2}) \,.
        \end{split}
    \end{equation}
    \item \underline{Vector and pseudovector:}
    \begin{equation}
        \begin{split}
            \M(\mathbf{1}_q\mathbf{2}_{-q}\mathbf{3}^{V_{-,+}})&=i\frac{b_1^{-,+}}{m_V M_V^2}(\angMM{2}{3}\sqrMM{1}{3}-\angMM{1}{3}\sqrMM{2}{3})(\sqrMM{1}{2}^2-\angMM{1}{2}^2) \,,\\
            \M(\mathbf{1}_q\mathbf{2}_{-q}\mathbf{3}^{V_{+,+}})&=\frac{b_1^{+,+}}{m_V M_V^2}(\angMM{2}{3}\sqrMM{1}{3}-\angMM{1}{3}\sqrMM{2}{3})(\angMM{1}{2}^2+\sqrMM{1}{2}^2)\\
            &+2\frac{b_2^{+,+}}{m_V M_V^2}(\angMM{2}{3}\sqrMM{1}{3}-\angMM{1}{3}\sqrMM{2}{3})\sqrMM{1}{2}\angMM{1}{2}\\
            \M(\mathbf{1}_q\mathbf{2}_{-q}\mathbf{3}^{V_{-,-}})&=\frac{b_1^{-,-}}{M_V^3}(\sqrMM{1}{3}\sqrMM{2}{3}\sqrMM{1}{2}^2+\angMM{1}{3}\angMM{2}{3}\angMM{1}{2}^2)\\
            &+\frac{b_2^{-,-}}{M_V^3}(\sqrMM{1}{3}\sqrMM{2}{3}\angMM{1}{2}^2+\angMM{1}{3}\angMM{2}{3}\sqrMM{1}{2}^2)\\
            &+\frac{b_3^{-,-}}{M_V^3}(\sqrMM{1}{3}\sqrMM{2}{3}+\angMM{1}{3}\angMM{2}{3})\sqrMM{1}{2}\angMM{1}{2}\\
            &+\frac{b_4^{-,-}}{m_VM_V^2}(\angMM{1}{3}\sqrMM{2}{3}+\angMM{2}{3}\sqrMM{1}{3})\sqrMM{1}{2}\angMM{1}{2}\\
            \M(\mathbf{1}_q\mathbf{2}_{-q}\mathbf{3}^{V_{+,-}})&=i\frac{b_1^{+,-}}{M_V^3}(\sqrMM{1}{3}\sqrMM{2}{3}\sqrMM{1}{2}^2-\angMM{1}{3}\angMM{2}{3}\angMM{1}{2}^2)\\
            &+i\frac{b_2^{+,-}}{M_V^3}(\sqrMM{1}{3}\sqrMM{2}{3}\angMM{1}{2}^2-\angMM{1}{3}\angMM{2}{3}\sqrMM{1}{2}^2)\\
            &+i\frac{b_3^{+,-}}{M_V^3}(\sqrMM{1}{3}\sqrMM{2}{3}-\angMM{1}{3}\angMM{2}{3})\sqrMM{1}{2}\angMM{1}{2}\,.
        \end{split}
    \end{equation}
    \item \underline{Photon:}
    \begin{equation}
        \begin{split}
            \M(\mathbf{1}_q\mathbf{2}_{-q}3^{+})&=\sqrt{2}qe\frac{x_{12}}{m^2}\left[\angMM{2}{1}^3+\frac{3c_{2}}{2m}\angMM{2}{1}^2\angsqrMM{1}{3}{2}+\frac{c_{4}}{m^2}\angMM{2}{1}\angsqrMM{1}{3}{2}^2+\frac{c_{8}}{m^3}\angsqrMM{1}{3}{2}^3\right]\\
            \M(\mathbf{1}_q\mathbf{2}_{-q}3^{-})&=\sqrt{2}qe\frac{x_{12}^{-1}}{m^2}\left[\sqrMM{2}{1}^3+\frac{3c_{2}}{2m}\sqrMM{2}{1}^2\angsqrMM{2}{3}{1}+\frac{c_{4}}{m^2}\sqrMM{2}{1}\angsqrMM{2}{3}{1}^2+\frac{c_{8}}{m^3}\angsqrMM{2}{3}{1}^3\right] \,.
        \end{split}
    \end{equation}
    \item \underline{Graviton:}
    \begin{equation}
        \begin{split}
            \M(\mathbf{1}_q\mathbf{2}_{-q}3^{++})&=\frac{x_{12}^2}{\mpl^2}\left[\angMM{2}{1}^3+\frac{g_{4}}{m^2}\angMM{2}{1}\angsqrMM{1}{3}{2}^2+\frac{g_{8}}{m^3}\angsqrMM{1}{3}{2}^3\right]\\
            \M(\mathbf{1}_q\mathbf{2}_{-q}3^{--})&=\frac{x_{12}^{-2}}{\mpl^2}\left[\sqrMM{2}{1}^3+\frac{g_{4}}{m^2}\sqrMM{2}{1}\angsqrMM{2}{3}{1}^2+\frac{g_{8}}{m^3}\angsqrMM{2}{3}{1}^3\right] \,.
        \end{split}
    \end{equation}
    \item \underline{Charged scalar and pseudoscalar:}
    \begin{equation}
        \begin{split}
            \M(\mathbf{1}_q\mathbf{2}_{q}\mathbf{3}^{S_{-,+}}_{-2q})&=\M(\mathbf{1}_{-q}\mathbf{2}_{-q}\mathbf{3}^{\bar S_{-,+}}_{2q})=\frac{\tilde a_1^{-,+}}{M_S^2}(\angMM{1}{2}^3+\sqrMM{1}{2}^3)+\frac{\tilde a_2^{-,+}}{M_S^2}\angMM{1}{2}\sqrMM{1}{2}(\angMM{1}{2}+\sqrMM{1}{2}) \,,\\
            \M(\mathbf{1}_q\mathbf{2}_{q}\mathbf{3}^{S_{+,+}}_{-2q})&=\M(\mathbf{1}_{-q}\mathbf{2}_{-q}\mathbf{3}^{\bar S_{+,+}}_{2q})=i\frac{\tilde a_1^{+,+}}{M_S^2}(\angMM{1}{2}^3-\sqrMM{1}{2}^3)+i\frac{\tilde a_2^{+,+}}{M_S^2}\angMM{1}{2}\sqrMM{1}{2}(\angMM{1}{2}-\sqrMM{1}{2}) \,.
        \end{split}
    \end{equation}
    \item \underline{Charged vector and pseudovector:}
    \begin{equation}
        \begin{split}
            \M(\mathbf{1}_q\mathbf{2}_{q}\mathbf{3}^{V_{+,-}}_{-2q})=\M(\mathbf{1}_{-q}\mathbf{2}_{-q}\mathbf{3}^{\bar V_{+,-}}_{2q})&=i\frac{\tilde b_1^{+,-}}{m_V M_V^2}(\angMM{2}{3}\sqrMM{1}{3}-\angMM{1}{3}\sqrMM{2}{3})(\sqrMM{1}{2}^2-\angMM{1}{2}^2) \,,\\
            \M(\mathbf{1}_q\mathbf{2}_{q}\mathbf{3}^{V_{-,-}}_{-2q})=\M(\mathbf{1}_{-q}\mathbf{2}_{-q}\mathbf{3}^{\bar V_{-,-}}_{2q})&=\frac{\tilde b_1^{-,-}}{m_V M_V^2}(\angMM{2}{3}\sqrMM{1}{3}-\angMM{1}{3}\sqrMM{2}{3})(\angMM{1}{2}^2+\sqrMM{1}{2}^2)\\
            &+2\frac{\tilde b_2^{-,-}}{m_V M_V^2}(\angMM{2}{3}\sqrMM{1}{3}-\angMM{1}{3}\sqrMM{2}{3})\sqrMM{1}{2}\angMM{1}{2} \,.
        \end{split}
    \end{equation}
\end{itemize}

\section{Details of Section~\ref{subsec:diracscalarvector}}\label{app:diracscalarvector}

We provide here some additional details on the scenarios considered in Section~\ref{subsec:diracscalarvector}.

\subsubsection*{Spontaneously broken $U(1)$ symmetry:}
The on-shell three-point amplitudes for a massive vector $V_{-,-}$ are:
\begin{equation}\label{eq:threepoint_massivephoton}
        \begin{split}
            \M(\mathbf{1}_q\mathbf{2}_{-q}\mathbf{3}^{V_{-,-}})&=\frac{b_1^{-,-}}{M_V^3}(\sqrMM{1}{3}\sqrMM{2}{3}\sqrMM{1}{2}^2+\angMM{1}{3}\angMM{2}{3}\angMM{1}{2}^2)\\
            &+\frac{b_2^{-,-}}{M_V^3}(\sqrMM{1}{3}\sqrMM{2}{3}\angMM{1}{2}^2+\angMM{1}{3}\angMM{2}{3}\sqrMM{1}{2}^2)\\
            &+\frac{b_3^{-,-}}{M_V^3}(\sqrMM{1}{3}\sqrMM{2}{3}+\angMM{1}{3}\angMM{2}{3})\sqrMM{1}{2}\angMM{1}{2}\\
            &+\frac{b_4^{-,-}}{m_VM_V^2}(\angMM{1}{3}\sqrMM{2}{3}+\angMM{2}{3}\sqrMM{1}{3})\sqrMM{1}{2}\angMM{1}{2} \,.
        \end{split}
    \end{equation}
We can match the notation to the massless photon amplitudes by the following replacements
\begin{equation}\label{eq:relabeling_massivephoton}
\begin{split}
    \frac{b_1^{-,-}}{M_V^3}&=-\sqrt{2}\frac{q e}{m^3}c_8 \,,\\
    \frac{b_2^{-,-}}{M_V^3}&=-\frac{q e}{\sqrt{2}m^3}[3 c_2 + 
    2 (1 +  c_8 +  c_{4})] \,,\\
    \frac{b_3^{-,-}}{M_V^3}&=\sqrt{2}\frac{q  e}{m^3}(2c_8+ c_{4}) \,,\\
    \frac{b_4^{-,-}}{m_V M_V^2}&=-\sqrt{2}\frac{q e}{m^2 m_V} \,.
\end{split}
\end{equation}
In the limit $m_V\to0$ the four-point amplitudes match the massless photon ones up to contact terms.
\par In the decoupling limit analysis we do not find any qualitative difference with the case in which the photon is massless. We arrive indeed to the same final result
\begin{equation}\label{eq:tuningdirac_massive}
\begin{split}
    g_8&=g_4=0 \,,\\
     c_8&= c_4= c_2=0 \,,\\
    \frac{q^2 e^2}{m^2}&=\frac{1}{2\mpl^2} \,.
\end{split}
\end{equation}
Interestingly, from positivity of elastic Arcs in the longitudinal sector we also have
\begin{equation}
    \A_{-\frac{1}{2}+\frac{1}{2}}^{-\frac{1}{2}+\frac{1}{2}}(0,0)=\frac{1}{3m^2\mpl^2}-\frac{m_V^2}{m^2}\frac{1}{18m^2\mpl^2}\geq0 \,,
\end{equation}
which turns into the bound on the photon mass of Eq.~\eqref{eq:massbound}.

\subsubsection*{Massless photon and massive $V_{-,-}$ vector:}

In this scenario we consider, in addition to the graviton and massless photon, also a massive photon ($V_{-,-})$ with couplings given by Eqs.~\eqref{eq:threepoint_massivephoton},~\eqref{eq:relabeling_massivephoton}. As discussed in the main text, we can take linear combinations of the two vectors so that only one of the two has a minimal coupling. Hence without loss of generality we rescale $q\tilde e\tilde c_{i}\to \tilde c_i$ and set $\tilde e\to0$, where we denote with the tilde the couplings of the massive vector.

\par Imposing positivity, starting from the order $E^8$, we find that the constraints of Eq.~\eqref{eq:tuningdiracE6} are qualitatively unchanged, and in addition we get
\begin{equation}
\tilde c_8=0\,.
\end{equation}
Once the tuning is performed, we can go to the decoupling limit that selects the leading $E^5$ contributions. The constraints found in Eq.~\eqref{eq:constraintsdiracE5} are deformed into
\begin{equation}\label{eq:constraintsdiracE5twovect}
    \begin{split}
        -\frac{3\tilde c_2(3\tilde c_2+4\tilde c_4)}{m^5}&=\frac{q^2e^2}{m^5}(2+3c_2)(3c_2+4c_4) \,,\\
        -\frac{27\tilde c_2^2-30\tilde c_2\tilde c_4+4\tilde c_4^2}{3m^5}+\frac{8}{3}\frac{h_{21}m^3}{M^8}&=\frac{q^2e^2}{m^5}\left[2+\frac{16}{3}c_4+c_2(8+9c_2+10c_4)\right] \,,\\
        \frac{1}{\mpl^2m^3}&=\frac{q^2e^2}{m^5}(2+3c_2)\,.
    \end{split}
\end{equation}
This system admits multiple solutions. However, it is straightforward to verify that all those in which gravity and the minimal electromagnetic coupling are switched off eventually violate the positivity of the elastic Arcs. The remaining solution is
\begin{equation}
\begin{split}
    \frac{h_{21}m^2}{M^8}&=\frac{1}{16q^2e^2m^2\mpl^4}+\frac{1}{4m^4\mpl^2}+\frac{3q^2e^2\mpl^2}{8m^4}\tilde c_2(3\tilde c_2+4\tilde c_4)+\frac{9\tilde c_2^2-8\tilde c_4^2}{16m^8} \,,\\
    \frac{q^2e^2c_2}{m^2}&=\frac{1}{3\mpl^2}-\frac{2}{3}\frac{q^2e^2}{m^2} \,,\\
    \frac{q^2e^2c_4}{m^2}&=-\frac{3}{4}\frac{q^2e^2c_2}{m^2}-\frac{3q^2e^2\mpl^2}{4m^4}\tilde c_2(3\tilde c_2+4\tilde c_4)\,.
\end{split}
\end{equation}
At this point, the amplitude is reduced to have a leading energy growth of order $E^4$. The simplest constraints coming from elastic Arcs positivity are
\begin{equation}
    \begin{split}
        \A_{-\frac{3}{2}-\frac{1}{2}}^{-\frac{3}{2}-\frac{1}{2}}(0,0)&=-\frac{(m^2-2q^2e^2\mpl^2)^2}{24q^2e^2m^4\mpl^2}-\frac{3\tilde c_2^2}{8m^8}[2m^4+q^2e^2\mpl^2(9\mpl^2\tilde c_2^2-4m^2)]\\
        &-\frac{\tilde c_2\tilde c_4}{m^8}[m^4+q^2e^2\mpl^2(9\mpl^2\tilde c_2^2-2m^2)]-\frac{2\tilde c_4^2}{3m^8}(m^4+9q^2e^2\mpl^4\tilde c_2^2)\geq0 \,,\\
        \A_{-\frac{1}{2}+\frac{1}{2}}^{-\frac{1}{2}+\frac{1}{2}}(0,0)&=\frac{1}{3\mpl^2m^2}-\frac{1}{18m^8}(m^4+9q^2e^2\mpl^4\tilde c_2^2)(3\tilde c_2+4\tilde c_4)^2\geq0 \,.
    \end{split}
\end{equation}
The solution to this system of inequalities is
\begin{equation}
\label{eq:D9}
\begin{split}
    \frac{1}{2\mpl^2}&\leq \frac{q^2e^2}{m^2}\leq\frac{2}{\mpl^2} \,,\\
    \frac{q^2e^2c_2}{m^2}&=\frac{1}{3\mpl^2}-\frac{2}{3}\frac{q^2e^2}{m^2} \,,\\
    \frac{\tilde c_2}{m^3}&=\pm\frac{1}{3m^2\mpl^2}\sqrt{2\mpl^2-\frac{m^2}{q^2e^2}} \,,\\
    \tilde c_4&=0 \,,\\
    c_4&=0  \,. 
\end{split}
\end{equation}
Additionally, positivity of the elastic Arcs in the longitudinal sector provides the bound on the vector mass of Eq.~\eqref{eq:doublevector_massbound}.

\subsubsection*{Massless photon and massive $V_{+,-}$ pseudovector:}\label{sec:axialpseudovector}

\par The on-shell three-point amplitudes for a massive vector $V_{+,-}$ are
\begin{equation}\label{eq:3ptaxialpseudovector}
    \begin{split}
        \M(\mathbf{1}_q\mathbf{2}_{-q}\mathbf{3}^{V_{+,-}})&=i\frac{b_1^{+,-}}{M_V^3}(\sqrMM{1}{3}\sqrMM{2}{3}\sqrMM{1}{2}^2-\angMM{1}{3}\angMM{2}{3}\angMM{1}{2}^2)\\
            &+i\frac{b_2^{+,-}}{M_V^3}(\sqrMM{1}{3}\sqrMM{2}{3}\angMM{1}{2}^2-\angMM{1}{3}\angMM{2}{3}\sqrMM{1}{2}^2)\\
            &+i\frac{b_3^{+,-}}{M_V^3}(\sqrMM{1}{3}\sqrMM{2}{3}-\angMM{1}{3}\angMM{2}{3})\sqrMM{1}{2}\angMM{1}{2} \,.
    \end{split}
\end{equation}
\par The leading contribution from the pseudovector exchanges enters at order $E^6$. Going to the decoupling limit, the constraints of Eq.~\eqref{eq:tuningdiracE6} are qualitatively unchanged, and in addition we get
\begin{equation}
b_1^{+,-}=0\,.
\end{equation}
Once the tuning is performed, we can go to the decoupling limit that selects the now leading $E^5$ contributions. The constraints found in Eq.~\eqref{eq:constraintsdiracE5} are deformed into
\begin{equation}\label{eq:constraintsdiracE5axial}
    \begin{split}
        2m\frac{(b_3^{+,-})^2-(b_2^{+,-})^2}{M_V^6}&=\frac{q^2e^2}{m^5}(2+3c_2)(3c_2+4c_4) \,,\\
        \frac{8}{3}\frac{h_{21}m^3}{M^8}&=\frac{q^2e^2}{m^5}\left[2+\frac{16}{3}c_4+c_2(8+9c_2+10c_4)\right] \,,\\
        \frac{1}{\mpl^2m^3}&=\frac{q^2e^2}{m^5}(2+3c_2) \,.
    \end{split}
\end{equation}
This system admits multiple solutions. However, it is straightforward to verify that all those in which gravity and the minimal electromagnetic coupling are switched off eventually violate the positivity of the elastic Arcs. The remaining solution is
\begin{equation}
\begin{split}
    \frac{h_{21}m^2}{M^8}&=\frac{1}{16q^2e^2m^2\mpl^4}+\frac{1}{4m^4\mpl^2}+\frac{b_2^{+,-}b_3^{+,-}}{4M_V^6}-\frac{(b_3^{+,-})^2-(b_2^{+,-})^2}{8M_V^6m^2}(m^2+2q^2e^2\mpl^2) \,,\\
    \frac{q^2e^2c_2}{m^2}&=\frac{1}{3\mpl^2}-\frac{2}{3}\frac{q^2e^2}{m^2} \,,\\
    \frac{q^2e^2c_4}{m^2}&=-\frac{3}{4}\frac{q^2e^2c_2}{m^2}+ q^2e^2m^2 \mpl^2\frac{(b_3^{+,-})^2-(b_2^{+,-})^2}{M_V^6}\,.
\end{split}
\end{equation}
At this point, the amplitude is reduced to have a leading energy growth of order $E^4$. The simplest constraints coming from elastic Arcs positivity are
\begin{equation}
    \begin{split}
        \A_{-\frac{3}{2}-\frac{1}{2}}^{-\frac{3}{2}-\frac{1}{2}}(0,0)&=-\frac{(m^2-2q^2e^2\mpl^2)^2}{24q^2e^2m^4\mpl^2}-q^2e^2m^4\mpl^4\frac{[(b_3^{+,-})^2-(b_2^{+,-})^2]^2}{6M_V^{12}}\\
        &-q^2e^2\mpl^2\frac{(b_3^{+,-})^2-(b_2^{+,-})^2}{3M_V^6}-m^2\frac{(b_3^{+,-})^2+(b_2^{+,-})^2}{6M_V^6}\geq0 \,,\\
        \A_{-\frac{1}{2}+\frac{1}{2}}^{-\frac{1}{2}+\frac{1}{2}}(0,0)&=\frac{1}{3\mpl^2m^2}-2q^2e^2m^4\mpl^4\frac{[(b_3^{+,-})^2-(b_2^{+,-})^2]^2}{9M_V^{12}}\\
        &-m^2\frac{(b_3^{+,-}-b_2^{+,-})^2}{9M_V^6}\geq0 \,.
    \end{split}
\end{equation}
The solution to this system of inequalities is
\begin{equation}
\begin{split}
    \frac{1}{2\mpl^2}&\leq \frac{q^2e^2}{m^2}\leq\frac{2}{\mpl^2} \,,\\
    \frac{b_2^{+,-}}{M_V^3}&=\pm\frac{1}{m^2\mpl^2}\sqrt{2\mpl^2-\frac{m^2}{q^2e^2}} \,,\\
    b_3^{+,-}&=0 \,,
\end{split}
\end{equation}
which retrospectively fixes $c_4=0$. In this case as well positivity of the elastic Arcs in the longitudinal sector provides the bound on the pseudovector mass of Eq.~\eqref{eq:pseudovector_massbound}.

\subsection*{Spin-0 and the other spin-1 states:}
The on-shell three-point amplitudes for the $S_{\pm,+}$ scalars and $V_{\pm,+}$ massive vectors are:
\begin{itemize}
    \item \underline{Neutral scalar and pseudoscalar:}
    \begin{equation}\label{eq:diracneutralscalar}
        \begin{split}
            \M(\mathbf{1}_q\mathbf{2}_{-q}\mathbf{3}^{S_{+,+}})&=\frac{a_1^{+,+}}{M_S^2}(\angMM{1}{2}^3+\sqrMM{1}{2}^3)+\frac{a_2^{+,+}}{M_S^2}\angMM{1}{2}\sqrMM{1}{2}(\angMM{1}{2}+\sqrMM{1}{2}) \,,\\
            \M(\mathbf{1}_q\mathbf{2}_{-q}\mathbf{3}^{S_{-,+}})&=i\frac{a_1^{-,+}}{M_S^2}(\angMM{1}{2}^3-\sqrMM{1}{2}^3)+i\frac{a_2^{-,+}}{M_S^2}\angMM{1}{2}\sqrMM{1}{2}(\angMM{1}{2}-\sqrMM{1}{2}) \,.
        \end{split}
    \end{equation}
    \item \underline{Neutral vector and pseudovector:}
    \begin{equation}
    \label{eq:diracneutralvectorgb}
        \begin{split}
            \M(\mathbf{1}_q\mathbf{2}_{-q}\mathbf{3}^{V_{-,+}})&=i\frac{b_1^{-,+}}{m_V M_V^2}(\angMM{2}{3}\sqrMM{1}{3}-\angMM{1}{3}\sqrMM{2}{3})(\sqrMM{1}{2}^2-\angMM{1}{2}^2) \,,\\
            \M(\mathbf{1}_q\mathbf{2}_{-q}\mathbf{3}^{V_{+,+}})&=\frac{b_1^{+,+}}{m_V M_V^2}(\angMM{2}{3}\sqrMM{1}{3}-\angMM{1}{3}\sqrMM{2}{3})(\angMM{1}{2}^2+\sqrMM{1}{2}^2)\\
            &+2\frac{b_2^{+,+}}{m_V M_V^2}(\angMM{2}{3}\sqrMM{1}{3}-\angMM{1}{3}\sqrMM{2}{3})\sqrMM{1}{2}\angMM{1}{2} \,,
        \end{split}
    \end{equation}
\end{itemize}
while for the charged states we have\footnote{The charge-conjugation of scalars and vectors can always be chosen to be $+1$ for scalars and $-1$ for vectors, as it can be redefined by a $U(1)$ transformation. Indeed, the amplitudes turn out to be independent of this choice.}
\begin{itemize}
    \item \underline{Charged scalar and pseudoscalar}
    \begin{equation}\label{eq:3ptdiracchargedscalars}
        \begin{split}
            \M(\mathbf{1}_q\mathbf{2}_{q}\mathbf{3}^{S_{-,+}}_{-2q})&=\M(\mathbf{1}_{-q}\mathbf{2}_{-q}\mathbf{3}^{\bar S_{-,+}}_{2q})=\frac{\tilde a_1^{-,+}}{M_S^2}(\angMM{1}{2}^3+\sqrMM{1}{2}^3)+\frac{\tilde a_2^{-,+}}{M_S^2}\angMM{1}{2}\sqrMM{1}{2}(\angMM{1}{2}+\sqrMM{1}{2}) \,,\\
            \M(\mathbf{1}_q\mathbf{2}_{q}\mathbf{3}^{S_{+,+}}_{-2q})&=\M(\mathbf{1}_{-q}\mathbf{2}_{-q}\mathbf{3}^{\bar S_{+,+}}_{2q})=i\frac{\tilde a_1^{+,+}}{M_S^2}(\angMM{1}{2}^3-\sqrMM{1}{2}^3)+i\frac{\tilde a_2^{+,+}}{M_S^2}\angMM{1}{2}\sqrMM{1}{2}(\angMM{1}{2}-\sqrMM{1}{2}) \,.
        \end{split}
    \end{equation}
    \item \underline{Charged vector and pseudovector}
    \begin{equation}\label{eq:3ptdiracchargedvector}
        \begin{split}
            \M(\mathbf{1}_q\mathbf{2}_{q}\mathbf{3}^{V_{+,-}}_{-2q})=\M(\mathbf{1}_{-q}\mathbf{2}_{-q}\mathbf{3}^{\bar V_{+,-}}_{2q})&=i\frac{\tilde b_1^{+,-}}{m_V M_V^2}(\angMM{2}{3}\sqrMM{1}{3}-\angMM{1}{3}\sqrMM{2}{3})(\sqrMM{1}{2}^2-\angMM{1}{2}^2) \,,\\
            \M(\mathbf{1}_q\mathbf{2}_{q}\mathbf{3}^{V_{-,-}}_{-2q})=\M(\mathbf{1}_{-q}\mathbf{2}_{-q}\mathbf{3}^{\bar V_{-,-}}_{2q})&=\frac{\tilde b_1^{-,-}}{m_V M_V^2}(\angMM{2}{3}\sqrMM{1}{3}-\angMM{1}{3}\sqrMM{2}{3})(\angMM{1}{2}^2+\sqrMM{1}{2}^2)\\
            &+2\frac{\tilde b_2^{-,-}}{m_V M_V^2}(\angMM{2}{3}\sqrMM{1}{3}-\angMM{1}{3}\sqrMM{2}{3})\sqrMM{1}{2}\angMM{1}{2} \,.
        \end{split}
    \end{equation}
\end{itemize}
The mass of various scalars $m_S$ and vectors $m_V$ can be \textit{a priori} distinguished, as in Section~\ref{sec:MajoranaGeneral}, but here for simplicity we do not keep track of this.
\par The constraints derived from Eq.~\eqref{eq:positivity_bounds_dec}, which effectively require the cancellation of terms growing more than $E^4$ toghether with positivity of fully elastic Arcs, do not qualitatively differ from those of Section~\ref{subsec:diracgravitonphoton}, still yielding
\begin{equation}
\begin{split}
    g_8&=g_4=0 \,,\\
    c_8&=c_4=c_2=0 \,,\\
    \frac{q^2e^2}{m^2}&=\frac{1}{2\mpl^2} \,.
\end{split}
\end{equation}

In addition, at $E^4$ order, we obtain the following constraint from fully elastic Arcs positivity in the transverse sector:
\begin{equation}\label{eq:boundsdirac01nc}
\begin{split}
    \A_{-\frac{3}{2}-\frac{3}{2}}^{-\frac{3}{2}-\frac{3}{2}}(0,0)&\to-\frac{(a_1^{+,+})^2 +(a_1^{-,+})^2}{M_S^4}-\frac{(b_1^{-,+})^2 +(b_1^{+,+})^2}{\mu_V^2M_V^4}\geq0 \,,\\
    \A_{-\frac{3}{2}+\frac{1}{2}}^{-\frac{3}{2}+\frac{1}{2}}(0,0)&\to-2\frac{(\tilde b_1^{-,-})^2 +(\tilde b_1^{+,-})^2}{3M_V^4}\geq0 \,,\\
    \A_{-\frac{3}{2}+\frac{3}{2}}^{-\frac{3}{2}+\frac{3}{2}}(0,0)&\to-2\frac{(\tilde a_1^{+,+})^2 +(\tilde a_1^{-,+})^2}{M_S^4}-4\frac{(\tilde b_1^{-,-})^2 +(\tilde b_1^{+,-})^2}{\mu_V^2M_V^4}\geq0 \,,\\
\end{split}
\end{equation}
where we are keeping $\mu_V=\frac{m_V}{m}$ fixed. This implies 
\begin{equation}
a_1^{+,+}=a_1^{-,+}=b_1^{-,+}=b_1^{+,+}=\tilde a^{+,+}_1=\tilde a^{-,+}_1=\tilde b^{-,-}_1=\tilde b^{+,-}_1=0\,.
\end{equation}
Moreover, analogously to the Majorana case, the remaining scalar and vector couplings are bounded. From the longitudinal modes elastic amplitude
\begin{equation}
\begin{split}
\M_{-\frac{1}{2}-\frac{1}{2}}^{-\frac{1}{2}-\frac{1}{2}}(s,t)&\to\frac{s^2}{F^2}-s^2\frac{(a_2^{+,+})^2 +(a_2^{-,+})^2}{9M_S^4}-8s^2\frac{(b_2^{+,+})^2}{9\mu_V^2M_V^4}-2s^2\frac{(\tilde b_2^{-,-})^2}{9 M_V^4} \,,\\
\M_{-\frac{1}{2}+\frac{1}{2}}^{-\frac{1}{2}+\frac{1}{2}}(s,t)&\to\frac{s^2}{F^2}-4s^2\frac{(b_2^{+,+})^2}{9M_V^4}-2s^2\frac{(\tilde a_2^{+,+})^2+(\tilde a_2^{-,+})^2}{9M_S^4}-4s^2\frac{(\tilde b_2^{-,-})^2}{9\mu_V^2 M_V^4} \,,
\end{split}
\end{equation}
we extract the bounds
\begin{equation}\label{eq:scalarvectbound_dirac}
	\begin{split}
		\frac{(a_2^{+,+})^2 +(a_2^{-,+})^2}{M_S^4}+\frac{8(b_2^{+,+})^2}{\mu_V^2 M_V^4}+2\frac{(\tilde b_2^{-,-})^2}{ M_V^4}&\leq\frac{9}{F^2} \,,\\
        \frac{4(b_2^{+,+})^2}{M_V^4}+2\frac{(\tilde a_2^{+,+})^2+(\tilde a_2^{-,+})^2}{M_S^4}+4\frac{(\tilde b_2^{-,-})^2}{\mu_V^2 M_V^4}&\leq\frac{9}{F^2} \,.
	\end{split}
\end{equation}
Further bounds, relating the various couplings, can be obtained from inelastic Arcs.

\bibliography{biblio} 
\bibliographystyle{utphys}

\end{document}